\documentclass[letterpaper,twocolumn,10pt]{article}
\usepackage{usenix-2020-09}

\usepackage{tikz}
\usepackage{amsmath}
\usepackage{subfigure}
\usepackage{enumerate}

\usepackage{paralist}

\usepackage{times}
\usepackage{epsfig}
\usepackage{graphicx}
\usepackage{amssymb}
\usepackage{booktabs}
\usepackage{multirow}
\usepackage{array}
\usepackage{amsfonts}
\usepackage{amsmath}
\usepackage{bbm}
\usepackage{balance}
\usepackage{bm}
\usepackage{algorithm}
\usepackage{algorithmic}
\usepackage{makecell}
\usepackage{textcomp}

\usepackage{siunitx}
\usepackage{amssymb}

\usepackage{filecontents}

\begin{document}

\newcommand{\etal}{\textit{et al}.}
\newcommand{\etc}{\textit{etc}}
\newcommand{\ie}{\textit{i}.\textit{e}.}
\newcommand{\eg}{\textit{e}.\textit{g}.}

\date{}

\title{\Large \bf Optimizing Replacement Policies for Content Delivery Network Caching:\\
  Beyond Belady to Attain A Seemingly Unattainable Byte Miss Ratio }

\author{
{\rm Peng Wang}\\
WNLO, HUST
\and
{\rm Yu Liu}\\
WNLO, HUST
} 
\maketitle

\begin{abstract}
When facing objects/files of differing sizes in content delivery networks (CDNs) caches, pursuing an optimal object miss ratio (OMR) by approximating Belady no longer ensures an optimal byte miss ratio (BMR), creating confusion about how to achieve a superior BMR in CDNs. To address this issue, we experimentally observe that there exists a time window to delay the eviction of the object with the longest reuse distance to improve BMR without increasing OMR. As a result, we introduce a deep reinforcement learning (RL) model to capture this time window by dynamically monitoring the changes in OMR and BMR, and implementing a BMR-friendly policy in the time window. Based on this policy, we propose a Belady and Size Eviction (LRU-BaSE) algorithm, reducing BMR while maintaining OMR. To make LRU-BaSE efficient and practical, we address the feedback delay problem of RL with a two-pronged approach. On the one hand, our observation of a rear section of the LRU cache queue containing most of the eviction candidates allows LRU-BaSE to shorten the decision region. On the other hand, the request distribution on CDNs makes it feasible to divide the learning region into multiple sub-regions that are each learned with reduced time and increased accuracy. In real CDN systems, compared to LRU, LRU-BaSE can reduce "backing to OS" traffic and access latency by 30.05\% and 17.07\%, respectively, on average. The results on the simulator confirm that LRU-BaSE outperforms the state-of-the-art cache replacement policies, where LRU-BaSE's BMR is 0.63\% and 0.33\% less than that of Belady and Practical Flow-based Offline Optimal (PFOO), respectively, on average. In addition, compared to Learning
Relaxed Belady (LRB), LRU-BaSE can yield relatively stable performance when facing workload drift.
\end{abstract}

\section{Introduction}
The optimal cache replacement policy is to evict the objects with the longest reuse distance if these requests and their orders are known a priori, also known as Belady’s algorithm~\cite{belady1966study,jain2018rethinking}. In practice, because it is impossible to know requests’ reuse distance and their order in advance, one can only make educated guesses about the longest reuse distance. Least Recently Used (LRU)~\cite{huang2013analysis,eytan2020s}, for example, tries to reserve objects based on the intuitive assumption that objects recently accessed have a high probability of being accessed in the future, and approximates the object left at the end of the queue as the object with the longest reuse distance to be evicted. Least Frequently Used (LFU)~\cite{huang2013analysis} believes that the assumption should be based on the frequency of access. Other methods follow and/or fuse these two assumptions for Belady approximation, but improve on the structure of queue (\eg, segmentation)~\cite{megiddo2003arc,jiang2002lirs,huang2013analysis}, the approximation pattern (\eg, learning models)~\cite{vietri2018driving,berger2018towards}~\etc. These techniques have resulted in a low object miss ratio (OMR) and are widely employed.

\begin{figure}[t]
	\begin{center}
		\includegraphics[width=0.92\linewidth]{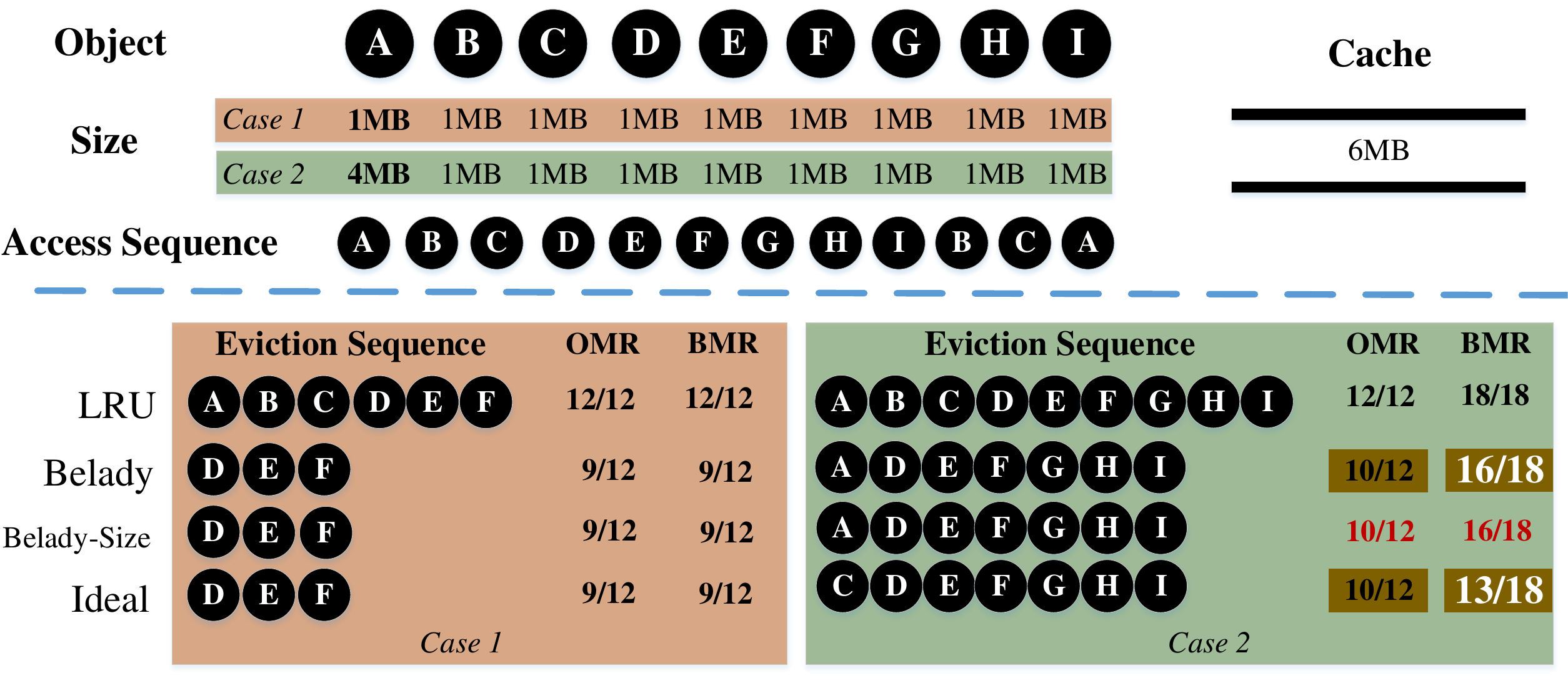}
	\end{center}
	\vspace{-0.3cm}
	\caption{Toy example - \textit{the phenomenon of Belady's inability to guarantee the minimum BMR when object sizes are different.}}
	\vspace{-0.3cm}
	\label{fig:toy}
\end{figure}

Since the data traffic is generally paid in proportion to the amount of data transmitted in web services, vendors desire the lowest byte miss ratio (BMR) in Content Delivery Network (CDN) caching. Conventionally, typical cache heuristics are effective for both BMR and OMR so long as objects are of similar size because both of them benefit from the data locality~\cite{PeledWE20, PeledMWE15, abrams1995caching}. However, the CDNs~\cite{maggs2015algorithmic,mukerjee2016impact,berg2020cachelib,berger2017adaptsize}, delivering files of different sizes through a network of caching servers to users, render impossible the rule that the optimal BMR should coincide with the optimal OMR because of (highly) different object sizes. We use a toy example to illustrate why this is the case, as follows. Assume an access sequence shown in Figure~\ref{fig:toy}, where the sizes of object-A are 1MB and 4MB in Case 1 and Case 2, respectively, and the size of all other objects is 1MB in both cases. On a 6GB cache, we use LRU, Belady, and Belady-Size~\cite{berger2018practical} to obtain the eviction sequences, OMRs, and BMRs, where the Belady-Size policy prioritizes the eviction of the object with the biggest product of the reuse distance and the object size. Comparing the BMR and OMR results of Belady's and Belady-Size's BMRs and OMRs with those of the ideal case, we conclude that, while Belady attains the minimum OMR, it fails to guarantee the minimum BMR when object sizes are different.

We also observe this phenomenon in tests on real-world CDN traces (See $\S$~\ref{sec:tph}). This phenomenon sparked our interest in whether there was a way to minimize BMR by considering object size when using Belady. Note that previous researchers attempted to reduce BMR by incorporating object size into heuristic approaches~\cite{abrams1995caching,abrams1996removal,wooster1997proxy,cao1997cost,aggarwal1999caching,rizzo2000replacement,JinB01,bahn2002efficient,li2019beating} where the replacement policies prioritize a pre-defined size threshold, followed by the Belady approximation, resulting in a trade-off between OMR and BMR. However, we believe that there is room to reduce BMR while maintaining OMR since we have observed that the eviction of objects within an appropriate reuse distance range does not increase the OMR (See $\S$~\ref{sec:win-win}). In other words, the replacement policy only increases the OMR if the object with the longest reuse distance cannot be evicted within a time window. In this time window, there is an opportunity to emphasize the reduction of BMR and achieve a win-win performance for both OMR and BMR. Nevertheless, it is hard to capture this window by traditional methods because it varies substantially depending on the traces and cache sizes.  

To accurately capture the aforementioned time window, we employ the reinforcement learning (RL) model~\cite{vietri2018driving,mnih2015human,sethumurugan2021designing} and propose a Belady and Size Eviction algorithm based on LRU (LRU-BaSE). The RL's Markov process can monitor the performance change caused by adjacent replacement behaviors, where the reward function rewards the behavior that maintains OMR while reducing BMR, implicitly capturing the time window to achieve a win-win performance for OMR and BMR. To tackle the problem of feedback delay~\cite{berger2018towards} in RL, we implement LRU-BaSE by a two-pronged approach. First, we observe that the LRU algorithm forces data with a relatively long reuse distance to be concentrated in a rear section of the queue (See $\S$~\ref{sec:rear}). LRU-BaSE shortens the decision region from the entire queue to this section, reducing its decision time. Second, we leverage data locality to divide the learning time region into four sub-regions based on experimental outcomes, then learn models for each of these sub-regions (See $\S$~\ref{sec:mono}). As a result, LRU-BaSE narrows the learning range for each model, lowering the training time with accurate decisions. These tricks, combined with the lightweight network, increase accuracy and efficiency.

Our evaluation of LRU-BaSE in real CDN systems show that, in comparison to the default cache algorithm, \ie, LRU, LRU-BaSE can reduce "backing to OS" traffic by 30.05\%, which can save \$795,000 per year in bandwidth costs. In addition, the average access latency and the request tail latency at the $99.9^{th}$ percentile are decreased by 17.07\% and 66.90\%, respectively. When compared to the state-of-the-art cache replacement algorithms on the simulator, LRU-BaSE outperforms the competition in terms of OMR and BMR on the public CDN traces. The BMR generated by LRU-BaSE is on average 0.63\% and 0.33\% lower than the lower bound of BMR, \ie, that of Belady and Practical Flow-based Offline Optimal (PFOO)~\cite{berger2018practical}, respectively. Finally, we simulate the workload drift~\cite{CaiLZZZLLCYX22} scene and demonstrate that LRU-BaSE is more robust for obtaining a superior BMR than Learning Relaxed Belady (LRB)~\cite{song2020learning}.


\section{Background and Motivation}
\subsection{XXPhoto and "Backing to OS" Traffic}
\label{sec:tph}
XXPhoto is company \textit{T}'s cloud service product and a special case of e-album business in the context of CDN. 
The cache layer of XXPhoto consists of the \textit{data center cache} (DC) and the \textit{outside cache} (OC). OC is closer, \eg, local, to clients in terms of access latency and helps improve the quality of user experience. DC, on the other hand, is located within the data center to alleviate the traffic burden on its storage system. 
In practice, when a request fails to hit in the cache, the request will be sent back to the data center, forming the so-called "backing to OS (Original Server)" traffic. This traffic consumes the costly data center resources and increases performance pressures (\eg, WAN bandwidth for OC and storage system I/O bandwidth for DC, and increased latencies, \etc.) and thus the LRU algorithm is widely deployed in the cache layer to minimize it. 
For the XXPhoto services during the month of August of 2019, the daily download traffic is 214.846Gbps and the daily "backing to OS" traffic is 73.166Gbps, which implies a BMR of 34.06\%. To further reduce the "backing to OS" traffic, we tested S4LRU~\cite{huang2013analysis} which is a cache replacement algorithm proven theoretically better than LRU, but the daily traffic increased to 88.197Gbps instead. Meanwhile, we found that OMR decreased from 32.81\% to 31.77\%, where changes in OMR and BMR are not always consistent because the sum of the sizes of objects hit is not always determined by the number of objects hit. This counterintuitive phenomenon inspired us to pay attention to the OMR-and-BMR relationship and how to reduce the "backing to OS" traffic in CDNs by improving cache replacement algorithms.

\subsection{Belady is no Panacea for BMR}
\label{sec:bnp}
\begin{figure}[t]
	\centering
	\includegraphics[width=0.4\textwidth]{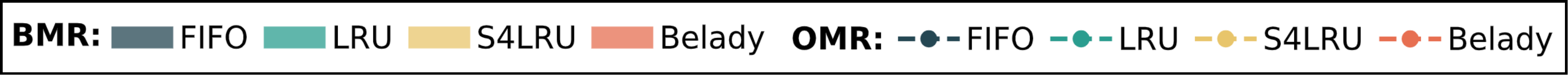}
	\vspace{-0.1cm}
	\\
	\subfigure[Trace-\textit{T}]{
	    \includegraphics[width=0.45\linewidth]{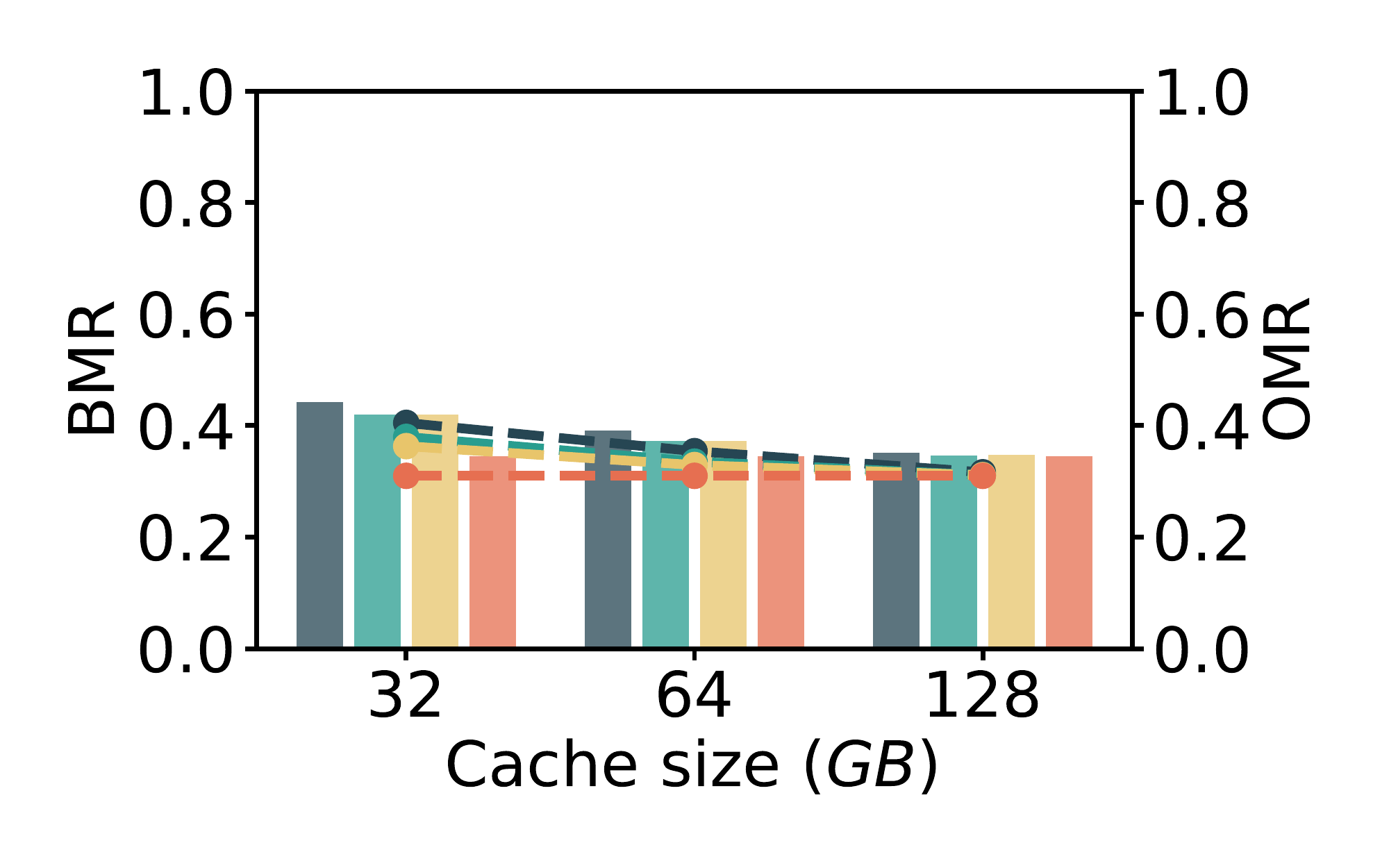}
	    \label{fig:bmr_omr_t}
	}
	\subfigure[Wikipedia]{
	    \includegraphics[width=0.45\linewidth]{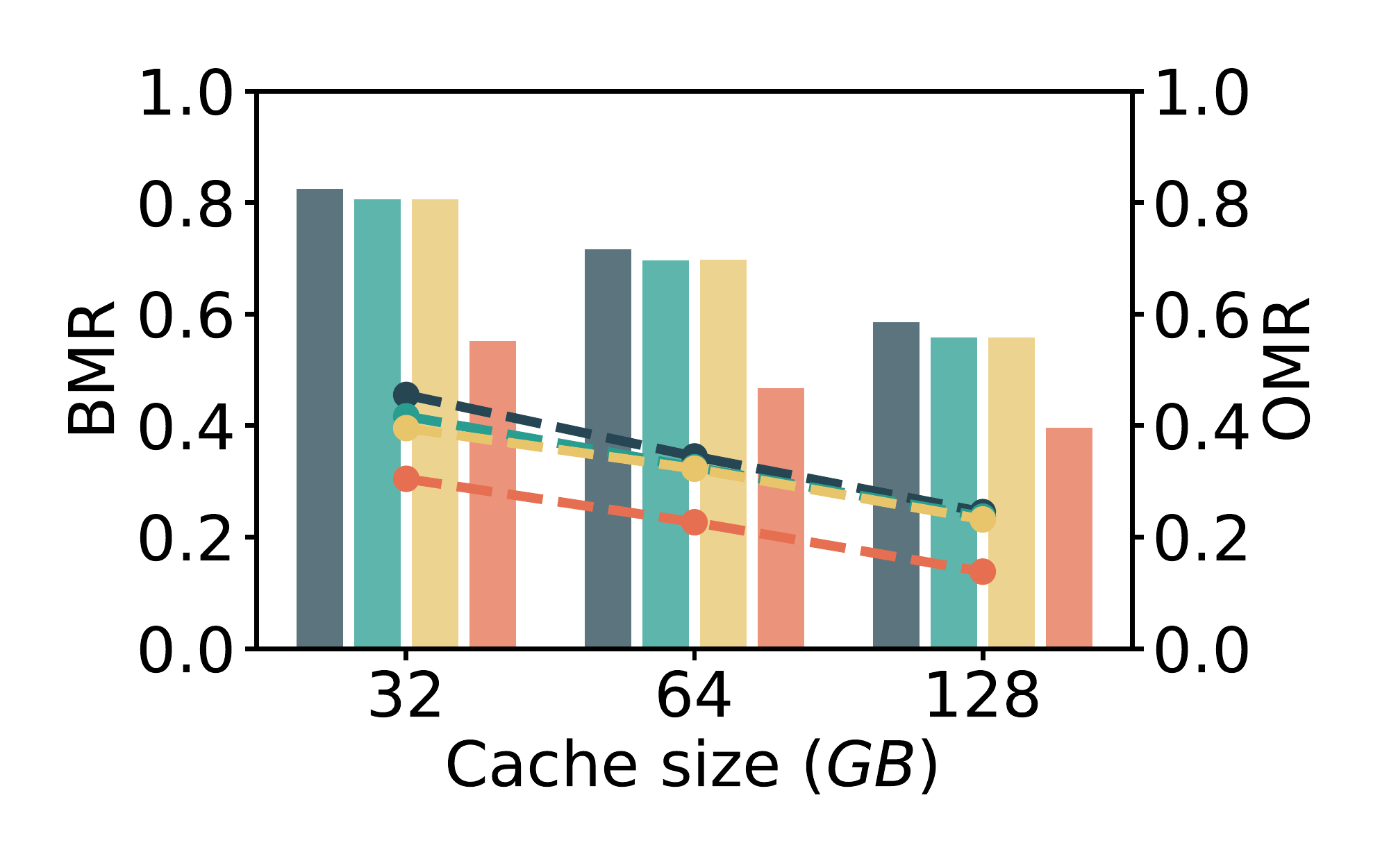}
	    \label{fig:bmr_omr_w}
	}
	\caption{BMR and OMR on Trace-\textit{T} and Wikipedia under different cache sizes and cache algorithms.}
	\label{fig:bmr_omr_gap}
\end{figure}

\begin{figure}[t]
	\centering
	\subfigure[32GB on Trace-\textit{T}]{
	    \includegraphics[width=0.3\linewidth]{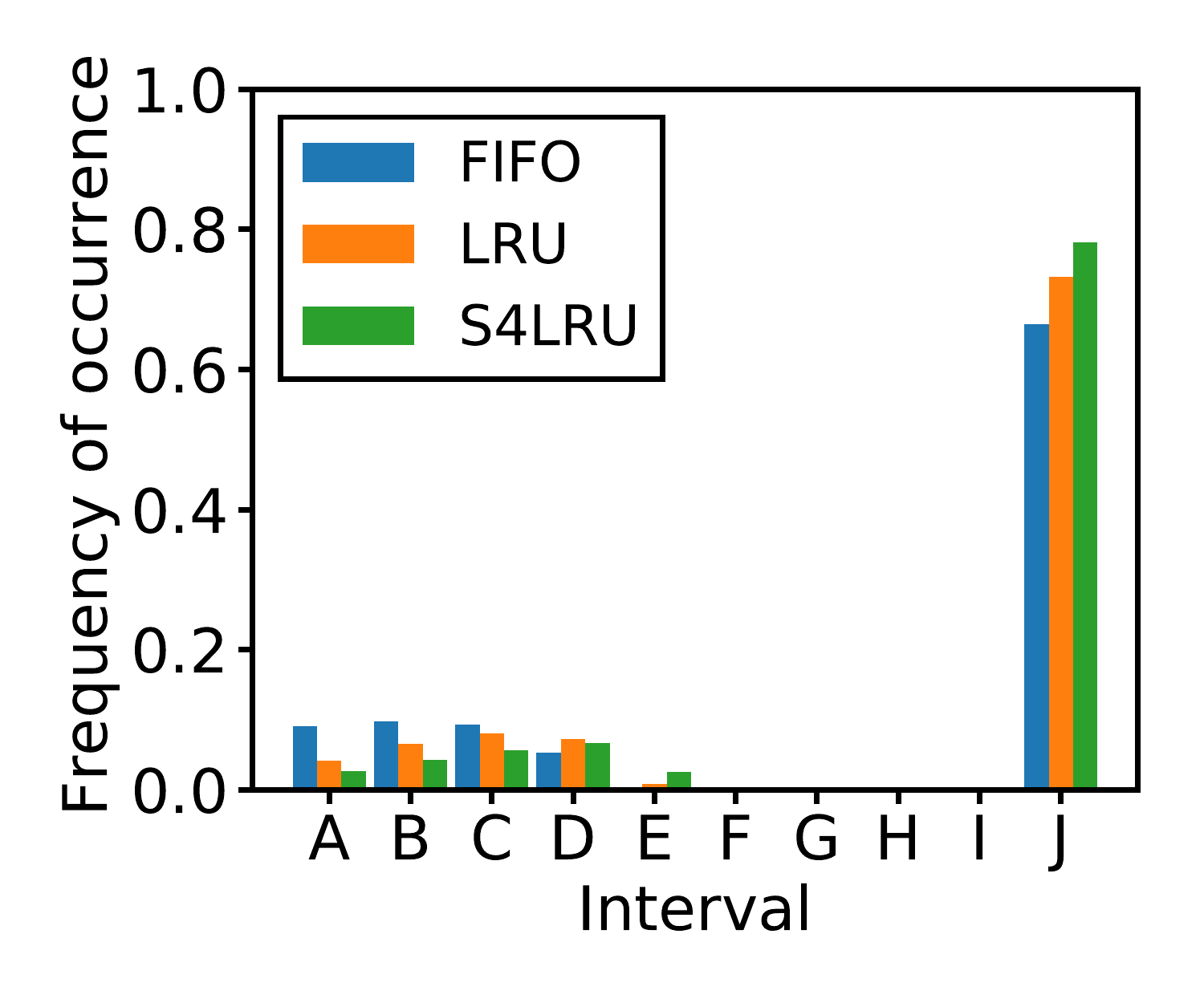}
	    \label{fig:rd32_t}
	}
	\subfigure[64GB on Trace-\textit{T}]{
	    \includegraphics[width=0.3\linewidth]{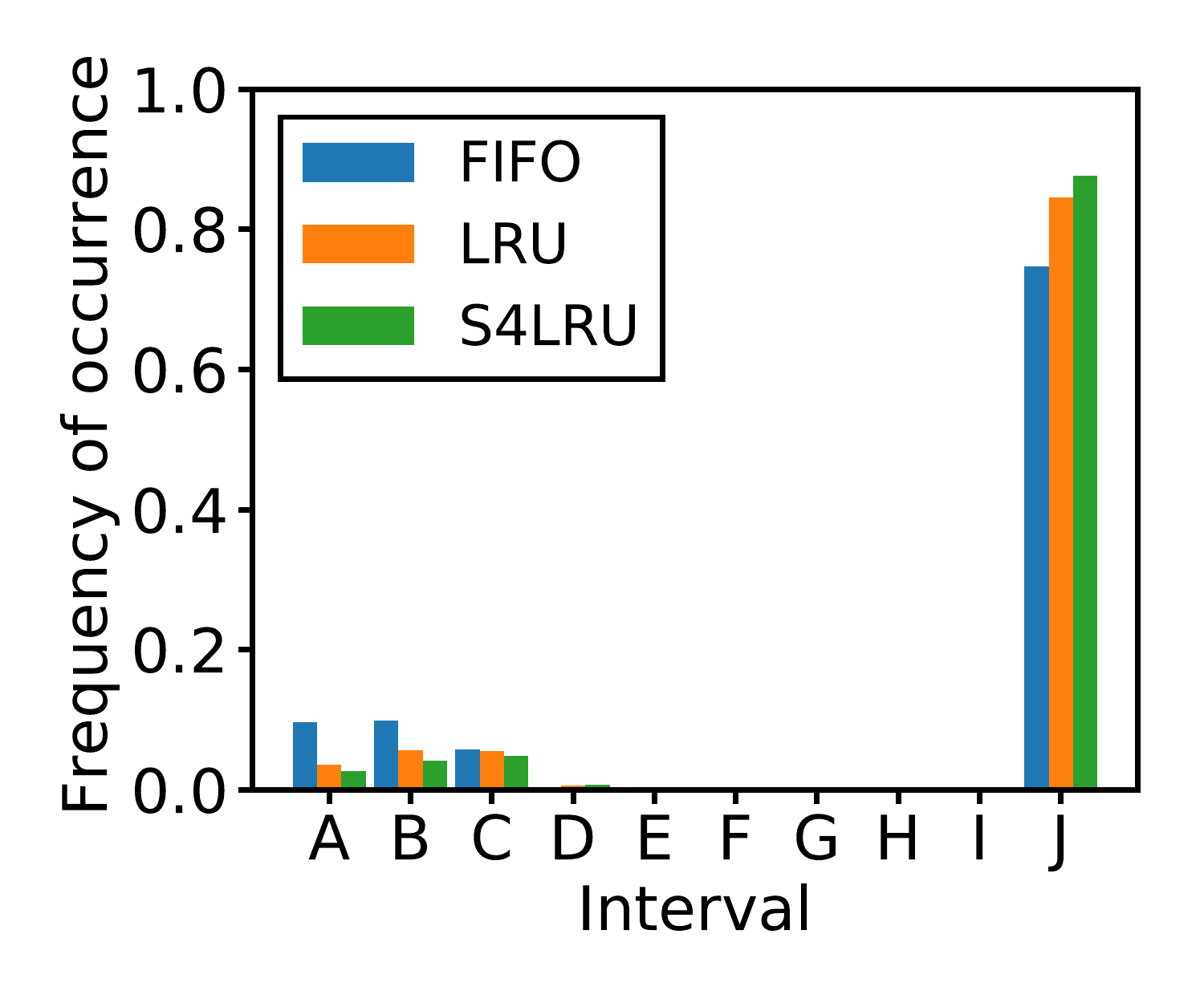}
	    \label{fig:rd64_t}
	}
	\subfigure[128GB on Trace-\textit{T}]{
	    \includegraphics[width=0.3\linewidth]{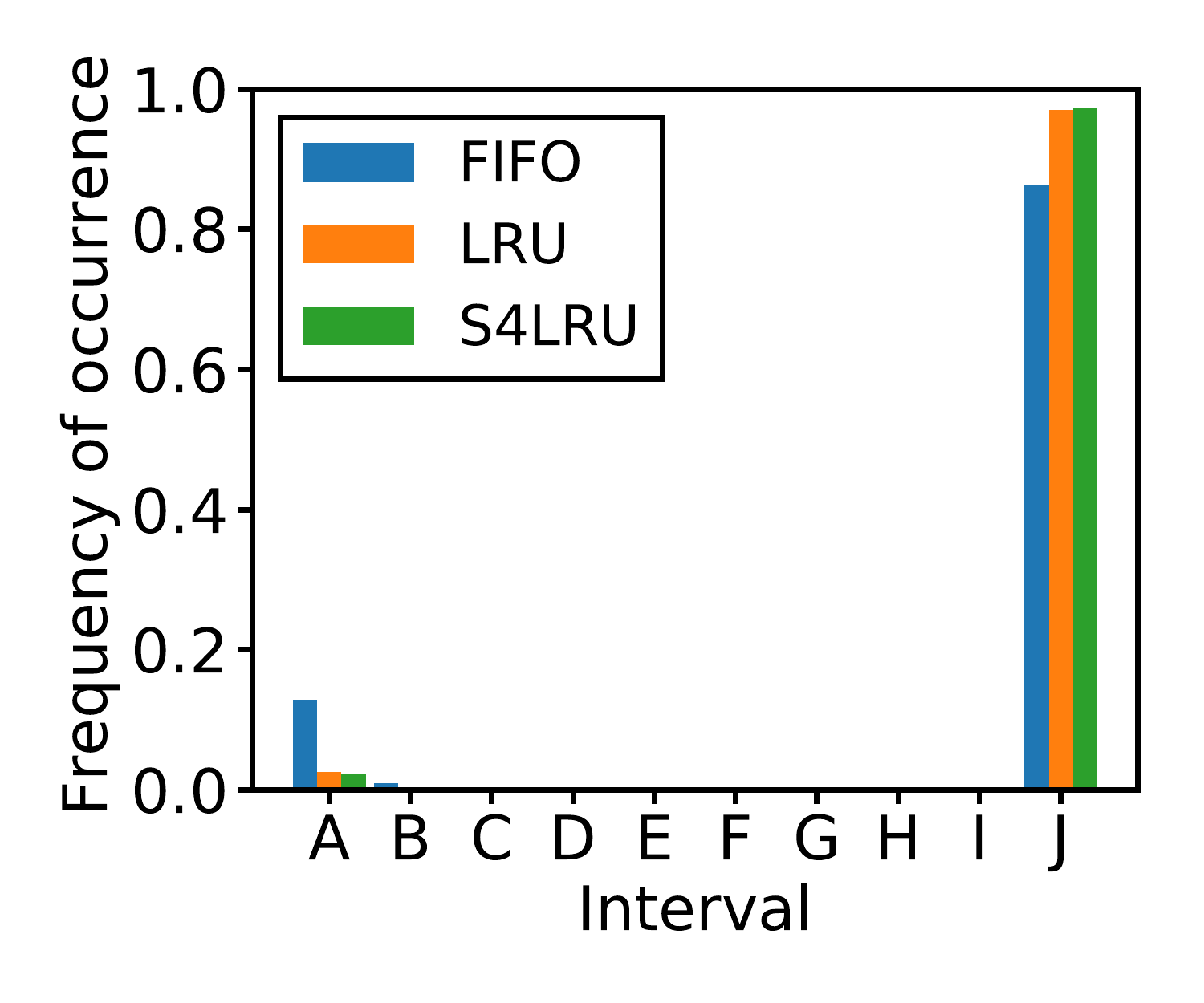}
	    \label{fig:rd128_t}
	}
	\\
	\subfigure[32GB on Wikipedia]{
	    \includegraphics[width=0.3\linewidth]{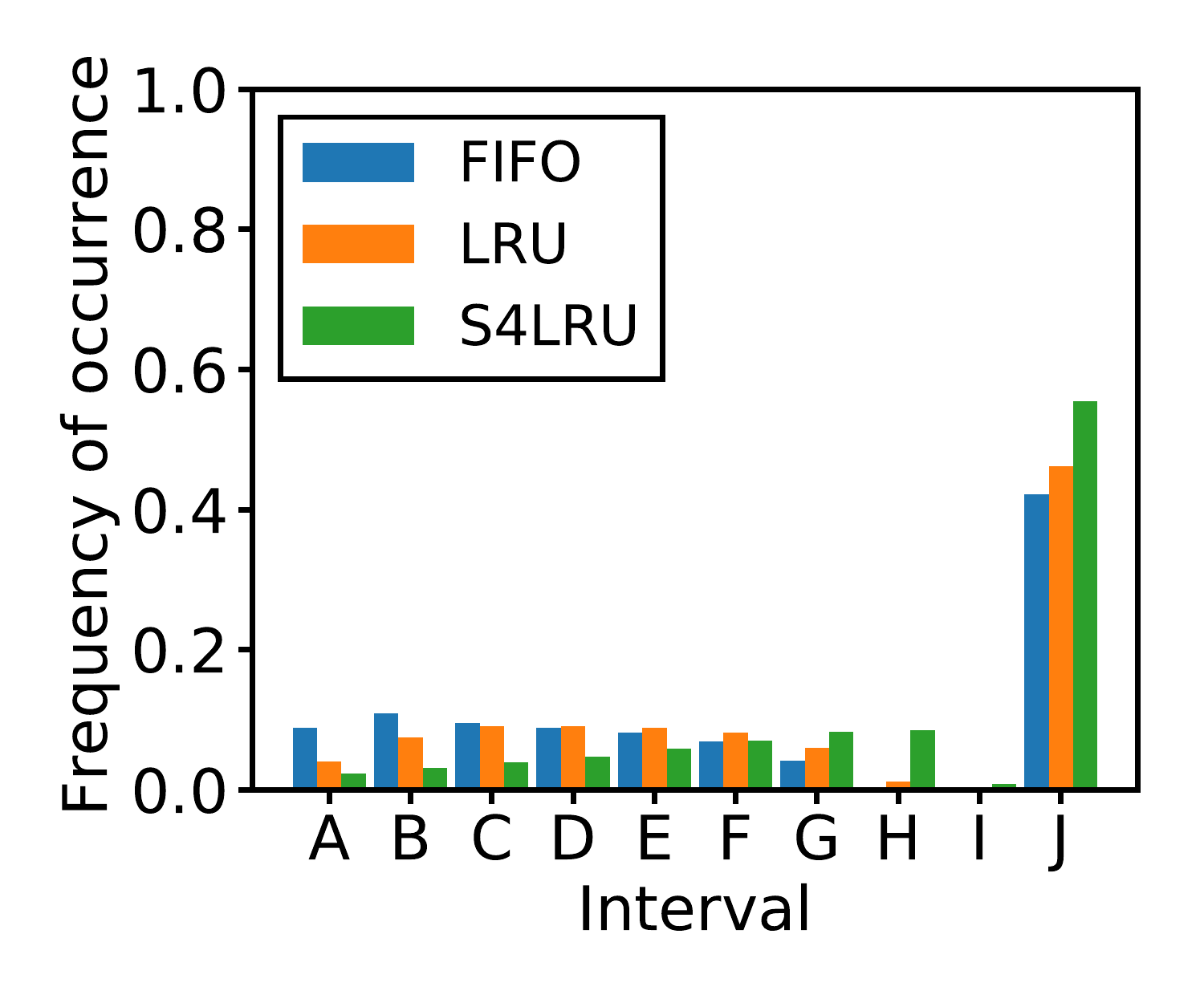}
	    \label{fig:rd32_w}
	}
	\subfigure[64GB on Wikipedia]{
	    \includegraphics[width=0.3\linewidth]{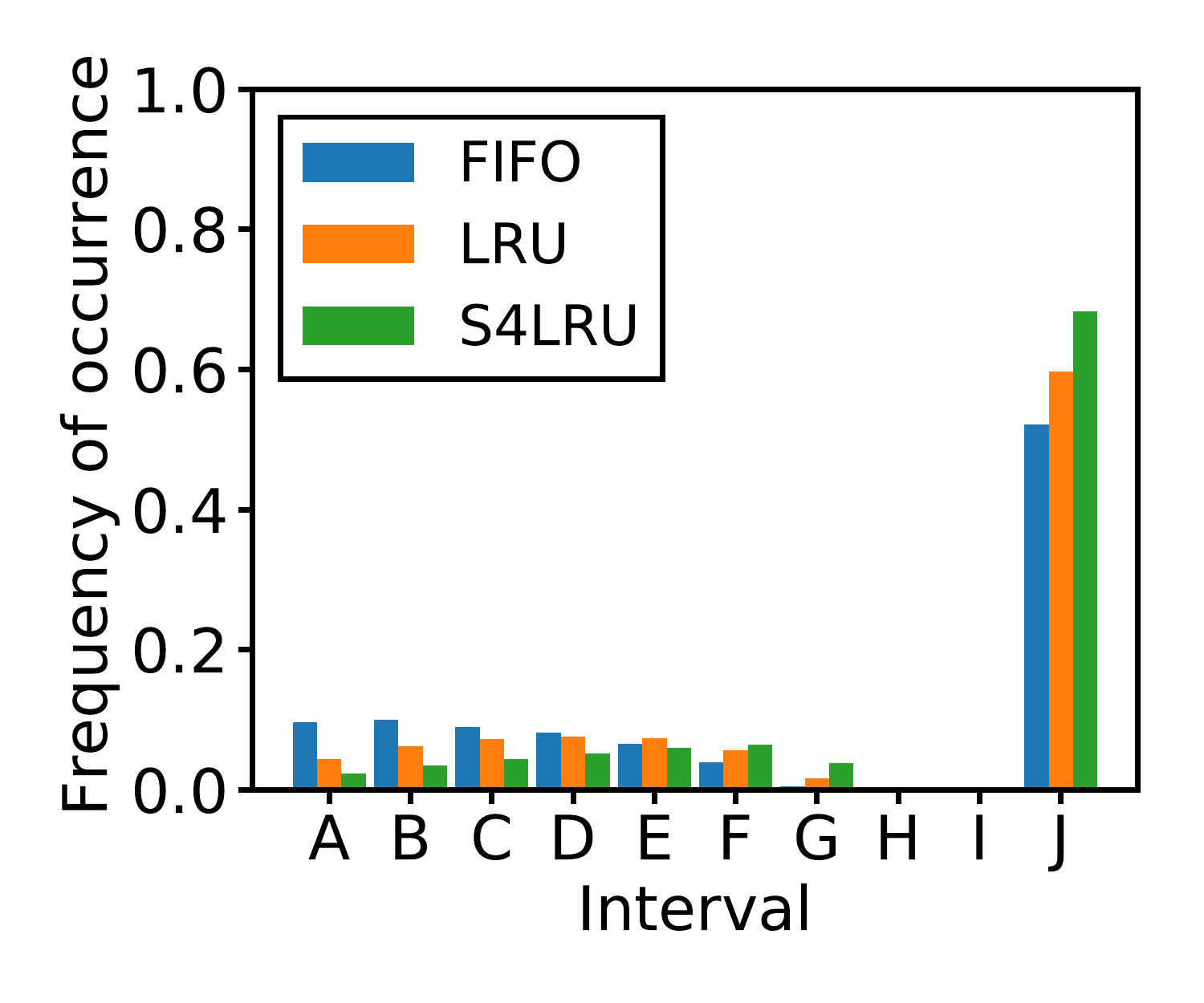}
	    \label{fig:rd64_w}
	}
	\subfigure[128GB on Wikipedia]{
	    \includegraphics[width=0.3\linewidth]{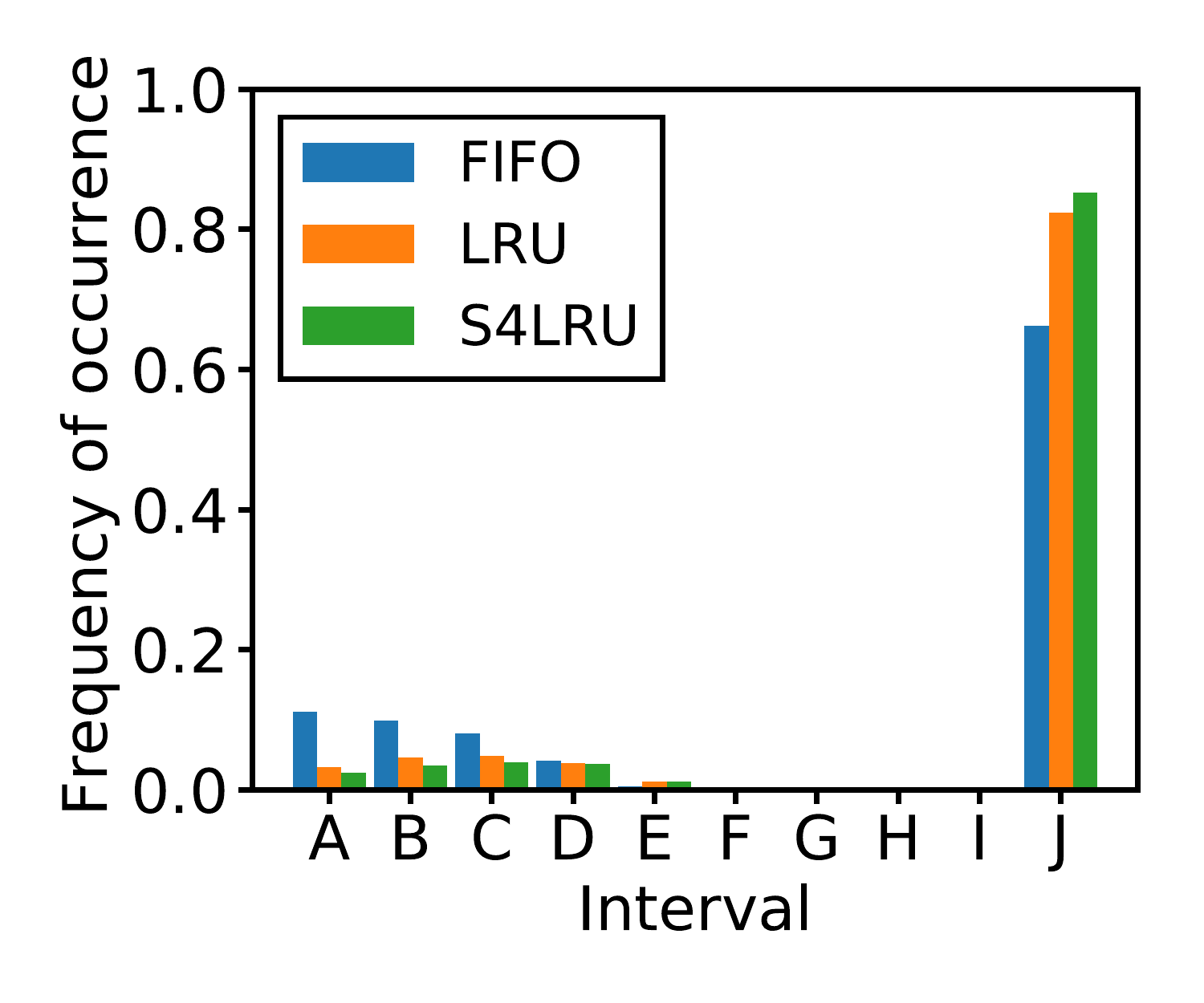}
	    \label{fig:rd128_w}
	}
	\vspace{-0.3cm}
	\caption{Reuse distance distribution of evicted objects on Trace-\textit{T} and Wikipedia under different cache sizes. A to J respectively represent intervals, \ie, subranges [0, 0.1), [0.1, 0.2), [0.2, 0.3), [0.3, 0.4), [0.4, 0.5), [0.5, 0.6), [0.6, 0.7), [0.7, 0.8), [0.8, 0.9), and [0.9, 1].}
	\vspace{-0.3cm}
	\label{fig:rd}
\end{figure}

To demonstrate S4LRU's~\cite{huang2013analysis} improvement in OMR over LRU by the former's accurate approximation of Belady, we present OMR, BMR, and the distribution of reuse distances of evicted objects at different cache sizes in Figure~\ref{fig:bmr_omr_gap} and Figure~\ref{fig:rd}. The FIFO algorithm is used to display the order of requests in the trace.
On the Trace-\textit{T} gathered from the workload described in $\S$~\ref{sec:tph}, OMR and BMR at different cache sizes are shown in Figure~\ref{fig:bmr_omr_t}. We equally divide the longest reuse distance into 10 subranges, \eg, $[0.9,1]$ for the subrange of 90\% to 100\% of the longest reuse distance. From Figure~\ref{fig:rd32_t} to Figure~\ref{fig:rd128_t}, we show distributions of reuse distances of evicted objects, where the height of a bar indicates the frequency of occurrence of evicted objects in the corresponding subrange. As shown in Figure~\ref{fig:bmr_omr_t}, S4LRU consistently outperforms LRU in OMR and BMR in all cache sizes except for 128GB, where an anomaly occurs in which S4LRU's BMR is slightly higher than LRU's BMR by 0.03\%, while the former maintains a 0.1\% advantage over the latter in OMR. Referring to these distributions, even though the number of objects in $[0.9, 1]$ tends to be close, S4LRU consistently evicts more objects with the longest reuse distances than LRU. After further analysis of the statistics, we find that the reuse distances of objects in $[0.9, 1]$ are all the longest. As a result, we believe that, while S4LRU does more accurately approximate Belady, it fails to produce an optimal BMR. To avoid bias from the results on a single trace, we perform the same test on the public trace, \ie, Wikipedia~\cite{song2020learning}. As shown in Figure~\ref{fig:bmr_omr_w}, the relative performances of S4LRU and LRU on Wikipedia seem different from those on Trace-\textit{T}. The LRU algorithm outperforms S4LRU in BMR and OMR at all cache sizes except for 32GB. Although we can attribute S4LRU's inferiority to LRU to the under-utilization of space resources in S4LRU~\cite{GastH15, huang2013analysis}, the algorithm that evicts the most objects with the longest reuse distance is still S4LRU. This yet again demonstrates that chasing OMR by approximating Belady no longer ensures an optimal BMR. A possible explanation for this phenomenon is that the Belady algorithm evicts objects with relatively large sizes in exchange for space to absorb more objects with relatively small sizes, albeit this is not necessarily beneficial for BMR~\cite{abrams1995caching}. 
As a result, we believe that approximating Belady can be the basis for achieving a satisfying BMR, but for a superior BMR, one should resort to the decisions taking object size into consideration. 

\section{Rationale of "Belady with Size"}
\label{sec:win-win}
Our analysis of existing heuristic-based cache replacement policies informed by both the Belady algorithm and object size~\cite{abrams1996removal,einziger2021lightweight,BahnKMN02} suggests that they generally follow a strategy that evicts the object with relatively large size to make room for caching more objects, treating the OMR and BMR as independent or adversarial objectives. The results in Figure~\ref{fig:toy} confirm that this strategy is not ideal for BMR. In addition, the learning-based replacement policies~\cite{berger2018towards,song2020learning,yan2021learning} learn by labels crafted by reuse distances, albeit with features containing object size, largely resulting in only OMR being affected by the object size. Clearly, ignoring BMR results in higher costs for webservice providers; whereas, minimizing BMR by compromising OMR gives rise to lower quality of user experience~\cite{berger2018robinhood,zhu2017workloadcompactor,AtreSWB20}.

\begin{figure}[t]
    \centering
	\subfigure[Trace-\textit{T}]{
	    \includegraphics[width=0.45\linewidth]{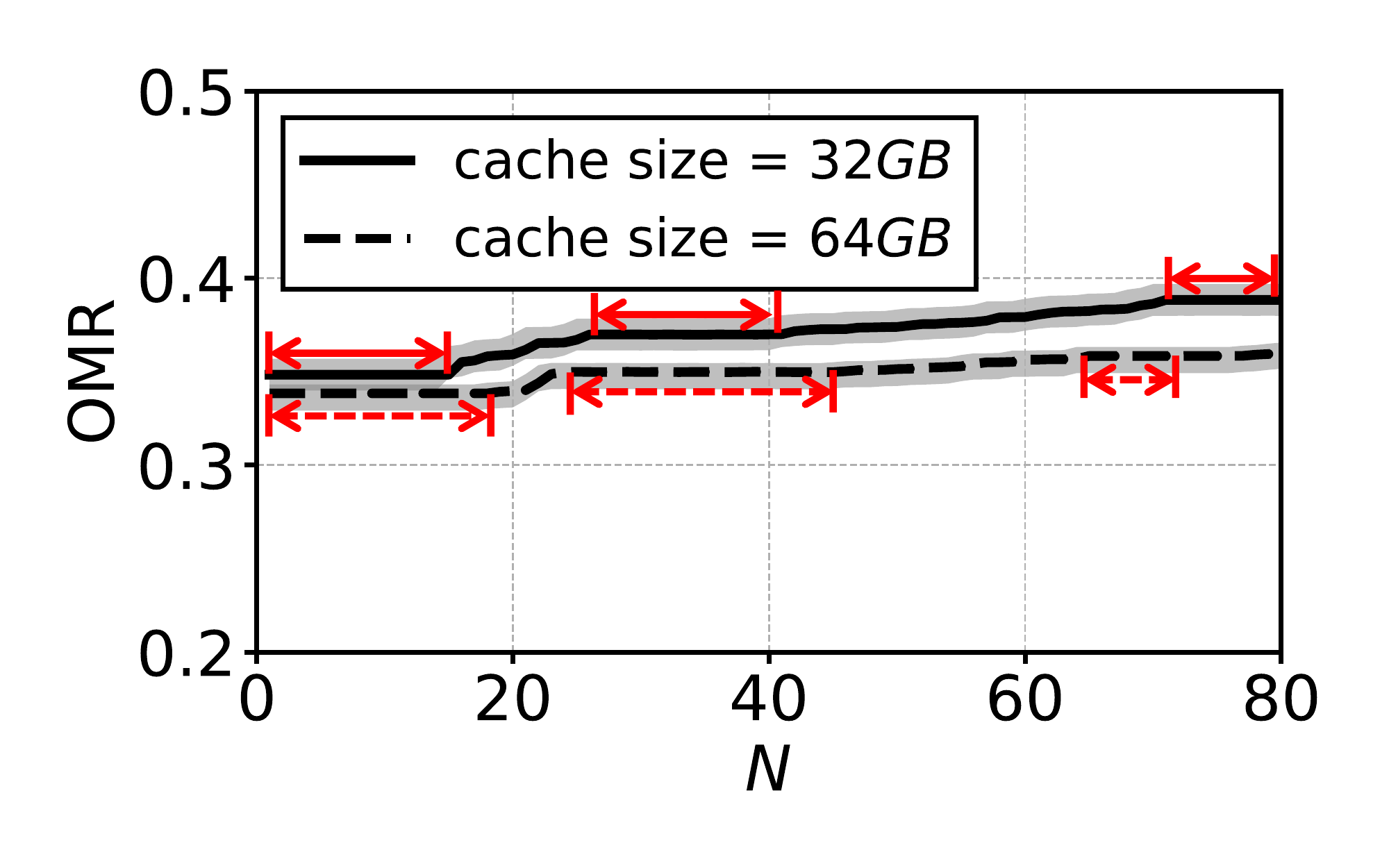}
	    \label{fig:bmr_omr_t_c}
	}
	\subfigure[Wikipedia]{
	    \includegraphics[width=0.45\linewidth]{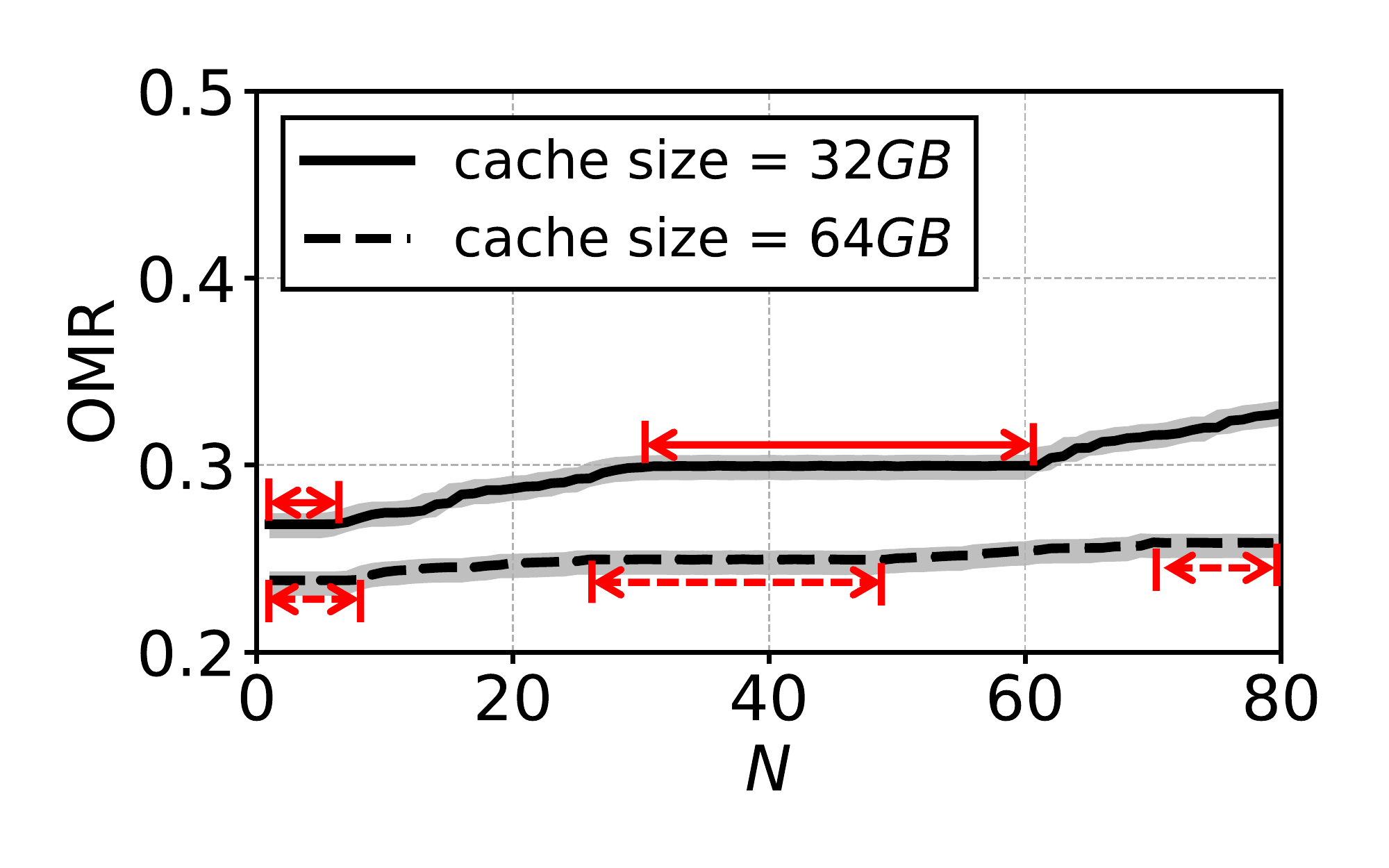}
	    \label{fig:bmr_omr_w_c}
	}
	\caption{OMR by evicting any one of the top-\textit{N} largest-reuse-distance objects. This experiment is implemented under a warmed-up cache. The redline-marked segments of the curves are the "flat" regions.}
	\label{fig:2.3}
\end{figure}

We believe that there is room for BMR to fall while keeping OMR. Because of the continuity of eviction behaviors, the object with the longest reuse distance may not be replaced upon the current request but could be replaced in the next one or one of the new few, which may possibly have no effect on OMR. For the example shown in Figure~\ref{fig:toy}, in Case 2, although the reuse distance of object-A is longer than that of object-C, prioritizing the eviction of object-C does not increase OMR. Based on this insight, we randomly evict one of the top-\textit{N} objects with the longest reuse distance when the current request misses. The OMR curves on Trace-\textit{T} and Wikipedia at different cache sizes are shown in Figure~\ref{fig:2.3}, where the shadow represents the upper and lower bound values under 500 repetitions of the test. In addition to the trend of curves growing as \textit{N} increases, we observed some "flat" regions in the curves (\eg, $N=16$ and $N=4$ for Trace-\textit{T} and Wikipedia respectively at the cache size of 32GB). The existence of these regions confirms our insight that evicting the longest-reuse-distance object upon each request miss is not necessarily an absolute requirement for the optimization of OMR. As a result, we believe that BMR and OMR have a chance to achieve a "win-win" outcome by judiciously evicting an object that can lower BMR within a specific time window without increasing OMR, just as we evicted object-C in Case 2 of Figure~\ref{fig:toy}. This approach may result in a lower BMR than the result yielded by Belady. However, since the "flat" region significantly changes with cache sizes or traces, as shown in Figure~\ref{fig:2.3}, it is hard to capture this time window manually. Note that LRB~\cite{song2020learning} attempted to capture this window by relaxing Belady's range through an analysis of limited data, but it is difficult for the pre-defined boundary to accommodate possible workload drift~\cite{CaiLZZZLLCYX22}. We show the experimental results to confirm this in $\S$~\ref{sec:robust}.


\begin{figure}[t]
	\begin{center}
		\includegraphics[width=0.7\linewidth,height=4cm]{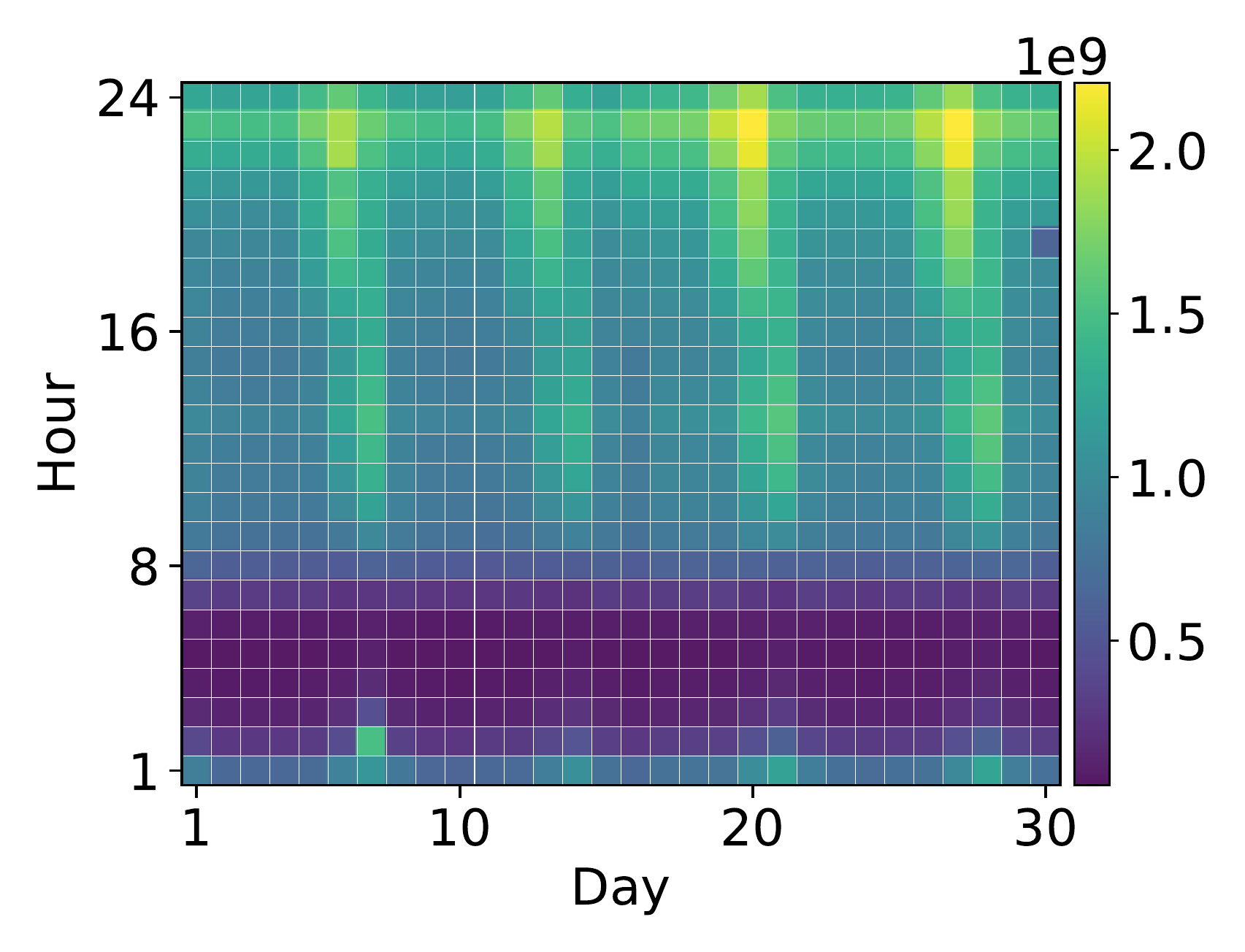}
	\end{center}
	\vspace{-0.7cm}
	\caption{Trace-\textit{T} - \textit{the number of user requests during the 24 hours, for the 30 days of the observed mouth.}}
	\vspace{-0.3cm}
	\label{fig:periodicity}
\end{figure}

\section{Methodology}
\label{sec:learnforcdn}
We believe that a key to capturing the window of opportunity for a win-win BMR and OMR performance is to dynamically identify the aforementioned "flat" regions by monitoring the request behaviors. To this end, we observe that Trace-\textit{T} has a stable daily periodicity in user request rate throughout the given traced month. This is visualized in the matrix of Figure~\ref{fig:periodicity}, where each column corresponds to a day of the month, each row represents a particular hour of the days, and the color on a given hour of a given day signifies the average request rate during that hour. The darker the color, the lower the request rate. With such observed patterns of request daily behaviors, which are consistent with our intuitive understanding of general user behaviors of webservices at large, we plan to use a learning model to "perceive" the distribution of requests and thus identify the "flat" regions. This, we expect, will help determine a time window of opportunity for proactive evictions of objects from the CDN caches based on a joint consideration of Belady and object sizes, while reducing overhead step by step and improving performance.

\subsection{Learning by the RL Model}
Because traditional classification models are insensitive to temporal changes, some lightweight learning models~\cite{ke2017lightgbm,Jerome2001} may not be sufficiently competent to identify this window. To the best of our knowledge, Long Short-Term Memory (LSTM)~\cite{hochreiter1997long}, Transformer~\cite{VaswaniSPUJGKP17}, and Reinforcement Learning (RL) models~\cite{vietri2018driving,mnih2015human} can achieve the purpose of monitoring in the temporal domain. Since it is a NP-hard problem~\cite{berger2018practical}, we cannot establish criteria that relate to BMR in a similar way to how the reuse distance relates to OMR, resulting in a lack of labels to learn by LSTM and Transformer. RL, on the other hand, can use BMR as a learning target directly. For example, based on the change of OMR, LeCaR~\cite{vietri2018driving} and CACHEUS~\cite{rodriguez2021learning} train the decisions to switch between two replacement policies. Nevertheless, RL models have a major problem with feedback delay and high training overhead. To solve this problem, in addition to deploying the lightweight model structure, we narrow the decision scope and learning range via the distribution of longest-reuse-distance objects.   

\subsubsection{Making Decision on the Rear Section}
\label{sec:rear}
\begin{figure}[t]
	\centering
	\includegraphics[height=0.35\linewidth]{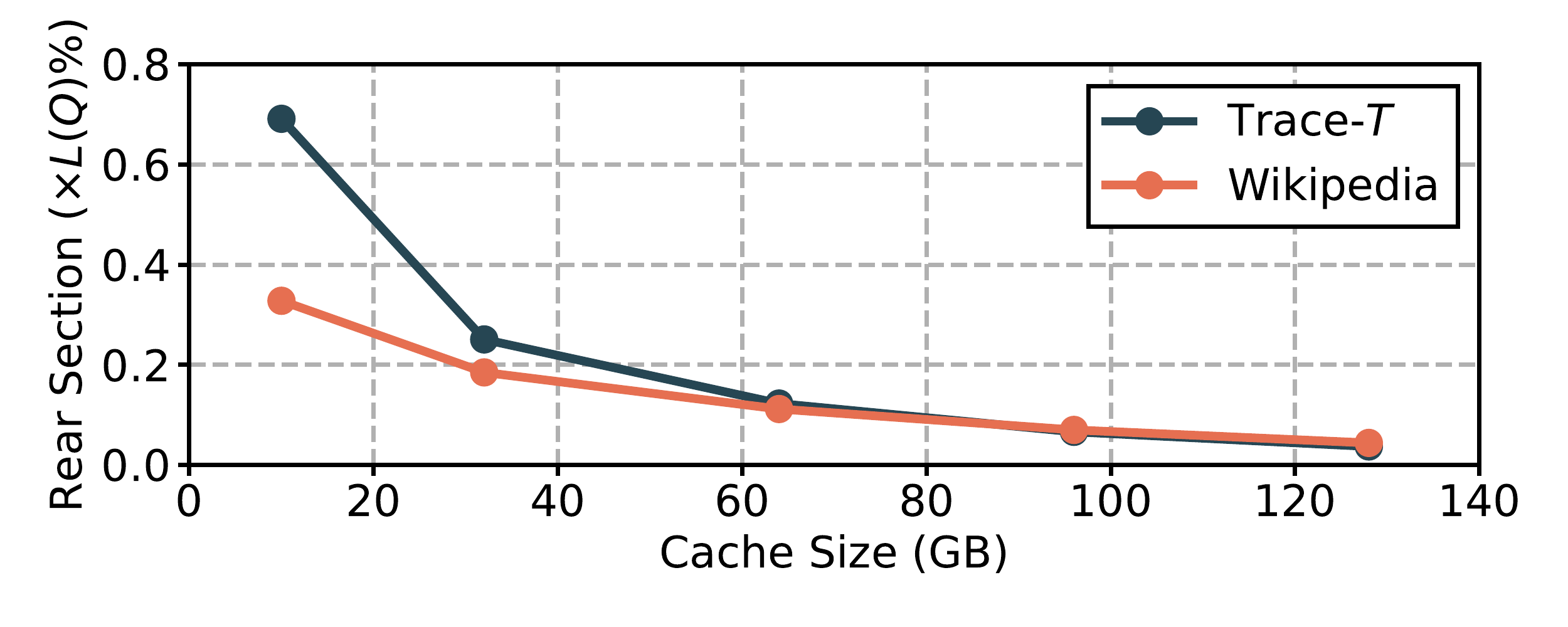}
	\caption{Curve fitted from ECDF on Trace-\textit{T} and Wikipedia.}
	\vspace{-0.1cm}
	\label{fig:evict}
\end{figure}

\begin{figure}[t]
	\centering
	\includegraphics[width=0.4\textwidth]{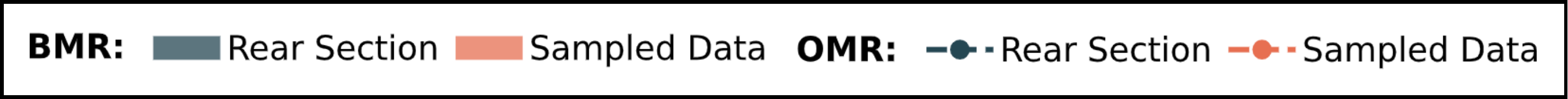}
	\vspace{-0.1cm}
	\subfigure[Trace-\textit{T}]{
	    \includegraphics[width=0.46\linewidth]{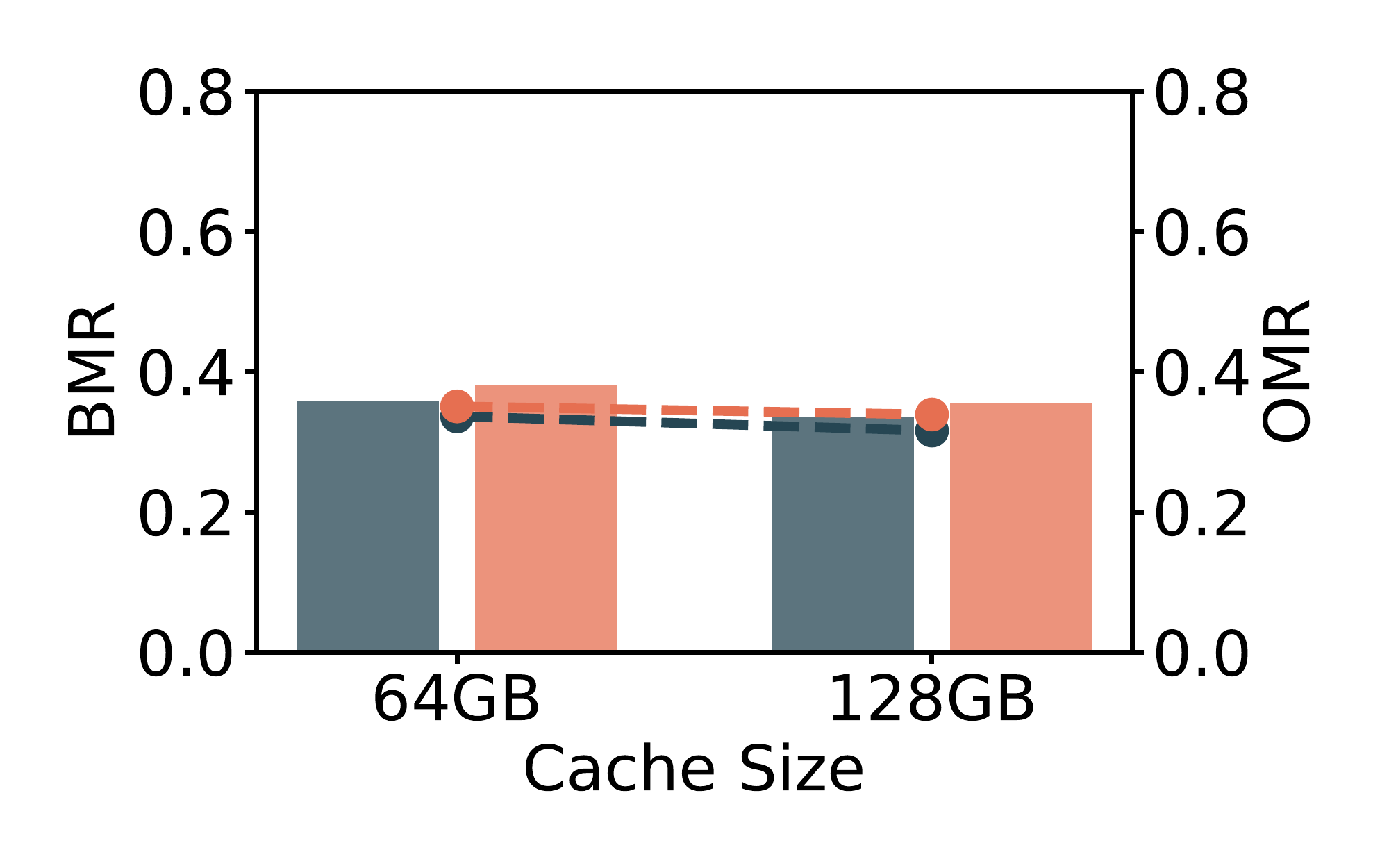}
	    \label{fig:rear_section}
	}
	\subfigure[Wikipedia]{
	    \includegraphics[width=0.46\linewidth]{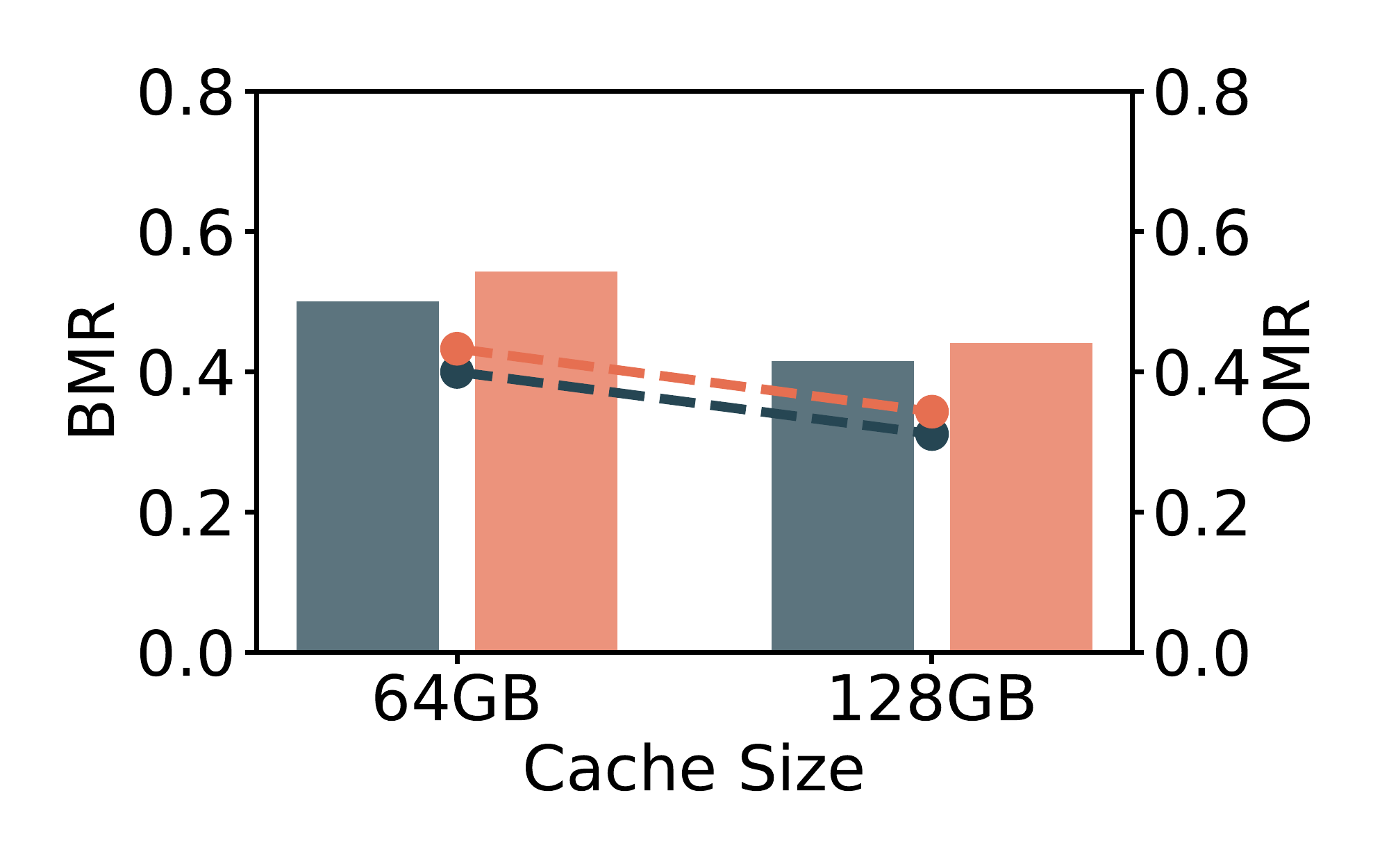}
	    \label{fig:sampled_data}
	}
	\caption{Decision results on the rear section and on the sampled data in the entire queue.}
	\label{fig:rs_sd}
\end{figure}

Proverbially, the LRU algorithm will promote the object that is hit to the head of the queue, leaving objects that have not been hit for a long time and generally have long reuse distances to loiter at the rear section of the queue. Recall that our experimental analysis in $\S$~\ref{sec:bnp} reveals that such a rear section is relatively very narrow.
If this rear section can be accurately identified and estimated from the queue, the spatial domain input to the RL model can be minimized, narrowing down the decision range while lowering the OMR and the overhead. To do so, we need to compute the cumulative distribution of the position of the longest-reuse-distance objects under different cache sizes. First, we define the eviction cumulative distribution function for candidate objects (ECDF) as: 
$$F_{X}(x) = P(X \leq x), (0 \leq x \leq 1)$$ 
where $X = D(O_d) / L(Q)$, $O_d$ denotes a candidate object, $Q$ represents the cache queue, $D(O_d)$ denotes the distance between the location of $O_d$ and the end of $Q$, $L(Q)=N$ represents the size of $Q$ (\ie, total number of objects $Q$ can hold). According to the ECDF values of $F_{X}(x)<0.99999$ over LRU at varying cache sizes from 10GB to 128GB, we fit the curve using the logistic regression algorithm and show the curve in Figure~\ref{fig:evict}. On both Trace-\textit{T} and Wikipedia, it is evident that as the cache size grows, the size of the conjectured rear section shrinks in proportion to the entire queue. There is no doubt that recording and learning data in a short rear section will greatly reduce the memory footprint compared to in the entire queue. Furthermore, Table~\ref{table:rs_overhead} shows the training and decision time on the rear section and the entire queue using the Deep Q-Learning (DQN)~\cite{mnih2013playing} model at 64GB and 128GB cache sizes. Obviously, the lower the dimensionality of the input, the shorter the training and decision time. Consequently, we randomly sample the same amount of data from the entire queue as in the rear section. Using the DQN model, the BMR and OMR created by the decision over the sampled data are compared to their counterparts over the data in the rear section. As shown in Figure~\ref{fig:rs_sd}, the decision on the rear section yields better performance than that on the sampled data. We attribute this result to more "right" objects being evicted from the rear section than from the sampled data. Furthermore, we can train and make decisions on the data sampled in the rear section, trading off between efficiency and precision in a time-sensitive environment. 

\begin{table}[t]
	\centering
	\scalebox{0.8}{
		\begin{tabular}{cccccc}
			\toprule[1.5pt]
			\multirow{2}{*}{Cache Size} & \multirow{2}{*}{} & \multicolumn{2}{c}{Trace-\textit{T}} & \multicolumn{2}{c}{Wikipedia} \\
			\cmidrule(lr){3-4} \cmidrule(lr){5-6} 
			& & RS & EQ & RS & EQ \\
			\midrule[1.2pt]
			\multirow{2}{*}{64GB} & TT & 24.55 h & 66.49h & 21.35 h  & 45.21 h \\
			&                       DT & 818.44$\mu$s & 1720.69$\mu$s & 720.46$\mu$s & 1365.90$\mu$s \\
			\midrule
			\multirow{2}{*}{128GB} & TT & 27.78 h & 71.12 h & 22.97 h & 47.01 h \\
			&                       DT & 935.08$\mu$s & 2041.76$\mu$s & 840.45$\mu$s & 1503.76$\mu$s \\
			\bottomrule
		\end{tabular}
	}
	\caption{Training time (TT) and decision time (DT) on the rear section (RS) and the entire queue (EQ) at 64GB and 128GB.}
	\label{table:rs_overhead}
\end{table}

\subsubsection{Learning in the Proper Time Regions}
\label{sec:mono}
\begin{figure}[t]
	\centering
	\subfigure[BMR]{
	    \includegraphics[width=0.46\linewidth]{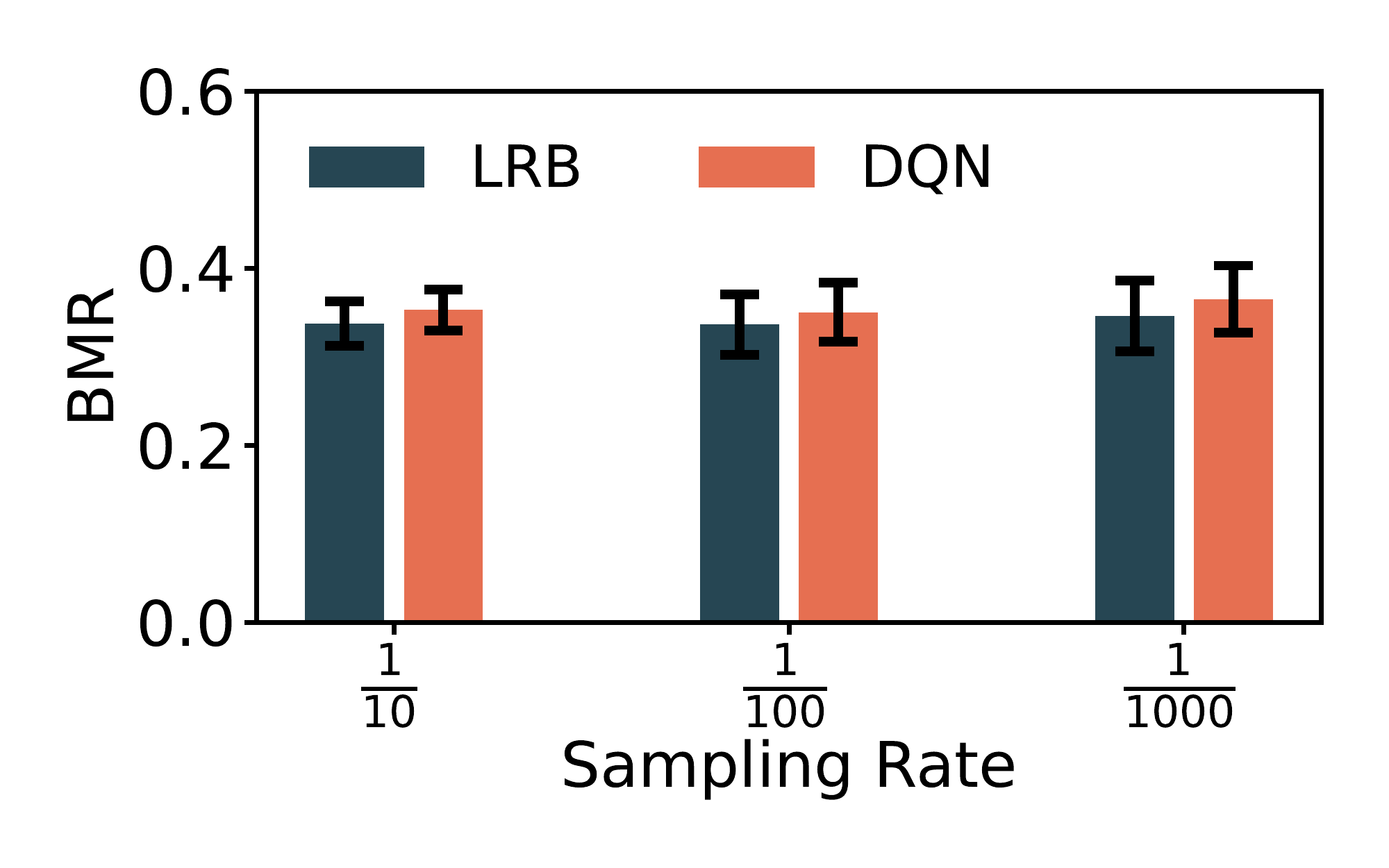}
	    \label{fig:lrb_dqn_bmr}
	}
	\subfigure[OMR]{
	    \includegraphics[width=0.46\linewidth]{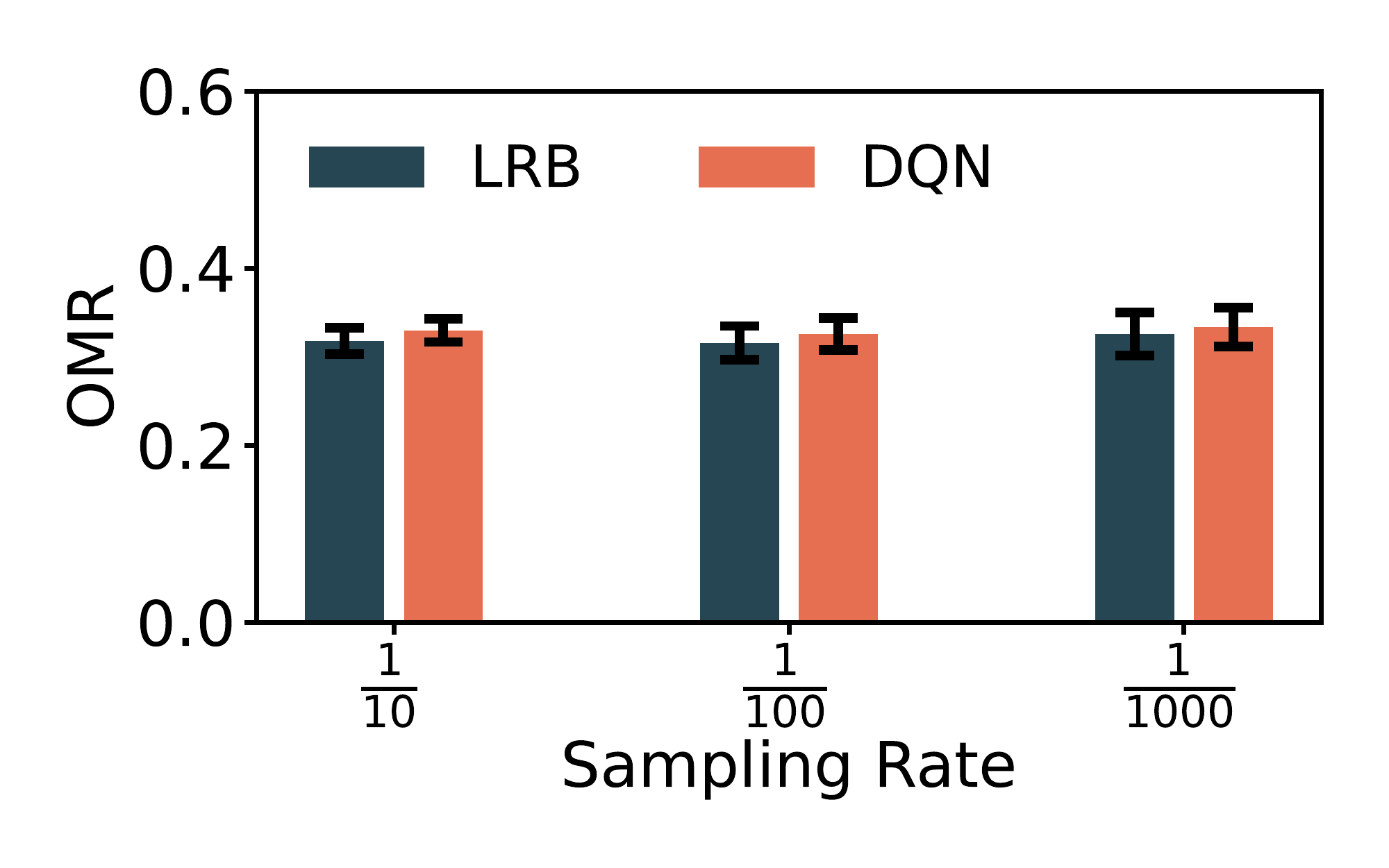}
	    \label{fig:lrb_dqn_mr}
	}
	\vspace{-0.3cm}
	\caption{Learning effects with different sampling rates.}
	\vspace{-0.3cm}
	\label{fig:lrb_dqn}
\end{figure}

Too wide a learning range will dilute the proportion of valid samples, resulting in low cognitive ability for the model. For example, with different sampling rates, we independently sample 1000 times from the data of Day 3 shown in Figure~\ref{fig:periodicity}. Then, we learn the sampled data by the DQN and LRB models, respectively, and estimate the requests for Day 4 on the 128GB cache. As shown in Figure~\ref{fig:lrb_dqn}, the lower the sampling rate, the larger the variances of both OMR and BMR. The average values are also inaccurate, especially for the DQN model. Statistically, the number of requests in a day of Trace-\textit{T} is over 1 billion. Exploring the large-scale data by an RL model deploying a few parameters will not likely yield accurate results.

\begin{table*}[t]
	\centering
	\scalebox{0.83}{
		\begin{tabular}{ccccccccc}
			\toprule[1.5pt]
			\multirow{2}{*}{} & \multicolumn{8}{c}{Time Span (beginning at 0 a.m.) } \\
			\cmidrule(lr){2-9}
			& 1h & 2h & 3h & 4h & 6h & 8h & 12h & 24h \\
			\midrule[1.2pt]
			TCT (h) & 3.91 & 4.20 & 4.78 & 5.17 & 6.78 & 10.36 & 14.89 & 27.78  \\ 
			\midrule
			BMR (\%) & \textbf{33.49$\pm$0.00} & 33.61$\pm$0.00 & 33.75$\pm$0.01 & 33.81$\pm$0.01 & 33.94$\pm$0.02 & 34.15$\pm$0.02 & 34.48$\pm$0.03 & 35.05$\pm$0.03  \\
			\midrule[1.2pt]
			TT (h) & 0.98 & 1.89 & 2.94 & 3.97 & 5.93 & 7.96 & 11.99 & 23.91  \\
			\midrule
			BMR (\%) & 34.55$\pm$0.04 & 34.28$\pm$0.03 & 34.17$\pm$0.03 & 34.18$\pm$0.02 & \textbf{34.12$\pm$0.02} & 34.33$\pm$0.03 & 34.71$\pm$0.04 & 35.09$\pm$0.04  \\
			\bottomrule[1.5pt]
		\end{tabular}
	}
	\caption{Time span tests on Trace-\textit{T} by the DQN model. TCT and TT represent the training convergence time and training time using CPU respectively.}
	\vspace{-0.1cm}
	\label{table:time_div}
\end{table*}

\begin{table}[t]
    \centering
    \scalebox{0.55}{
        \begin{tabular}{ccccccc}
            \toprule
            \multirow{2}{*}{} & \multicolumn{6}{c}{Beginning Time (a.m.)} \\
            \cmidrule(lr){2-7}
            & 0 & 1 & 2 & 3 & 4 & 5  \\
            \midrule
            BMR (\%) & 34.12$\pm$0.02 & 33.79$\pm$0.01 & \textbf{33.75$\pm$0.01} & 33.81$\pm$0.02 & 33.89$\pm$0.01 & 33.93$\pm$0.02 \\
            OMR (\%)  & 32.26$\pm$0.02 & 31.77$\pm$0.01 & \textbf{31.68$\pm$0.01} & 31.82$\pm$0.02 & 31.94$\pm$0.01 & 32.05$\pm$0.02 \\
            \bottomrule
        \end{tabular}
    }
    \caption{BMR and OMR with different beginning time.}
    \vspace{-0.1cm}
    \label{table:starting_time}
\end{table}

As a result, we explore the solution that narrows the learning scope, \eg, by dividing a day into numerous sub-regions to learn. We try to test the average BMR of many models at different time spans, assuming that the temporal regions are related to the data locality. The spatial overhead from additional models can be kept low by transferring the parameters using only two models (See $\S$\ref{sec:workflow}). Based on this idea, under 100 samplings with a sampling rate of $\frac{1}{100}$ (the best average values in Figure~\ref{fig:lrb_dqn}), we slice the Day 3 region into multiple non-overlapping sub-regions with the same time span and train the model for each sub-region by the DQN model. The 3$^{rd}$ and 4$^{th}$ rows of Table~\ref{table:time_div} show that the shorter the time span, the shorter the average training time, and the smaller variance of BMR. However, we cannot tolerate overlapping model training, \ie, the model's training duration is longer than the learning range's time span, due to limited computing resources used for update(\ie, online training). Consequently, we re-train each model within the learning range's time span. As shown in the 6$^{th}$ row of Table~\ref{table:time_div}, a 6-hour time span is the optimal scheme. Note that we also test the time spans for the following 6 days and obtain similar results. In the interest of space, we do not show them.

To further determine the proper time regions, we adjust the beginning time of the time span based on the periodicity and show the results in terms of BMR and OMR in Table~\ref{table:starting_time}. Based on these results, we confirm that 2 a.m. is the best beginning time. Reviewing Figure~\ref{fig:periodicity}, it's clear that the number of requests per hour in each time region is close (small chromatism) using this beginning time. Hence we prefer to train four models for each day, with each model's training data representing requests in 6 hours and beginning at 2 a.m. 

\subsection{Representative Learning Data}
\label{sec:rpd}

\begin{figure}[t]
	\centering
	\includegraphics[width=0.8\linewidth]{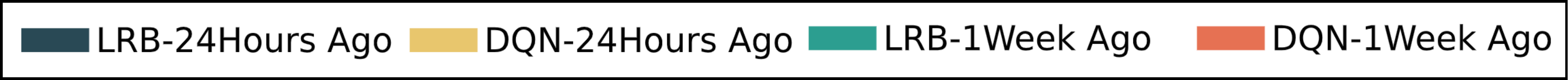}
	\vspace{-0.2cm}
	\\
	\subfigure[BMR]{
	    \includegraphics[width=0.46\linewidth]{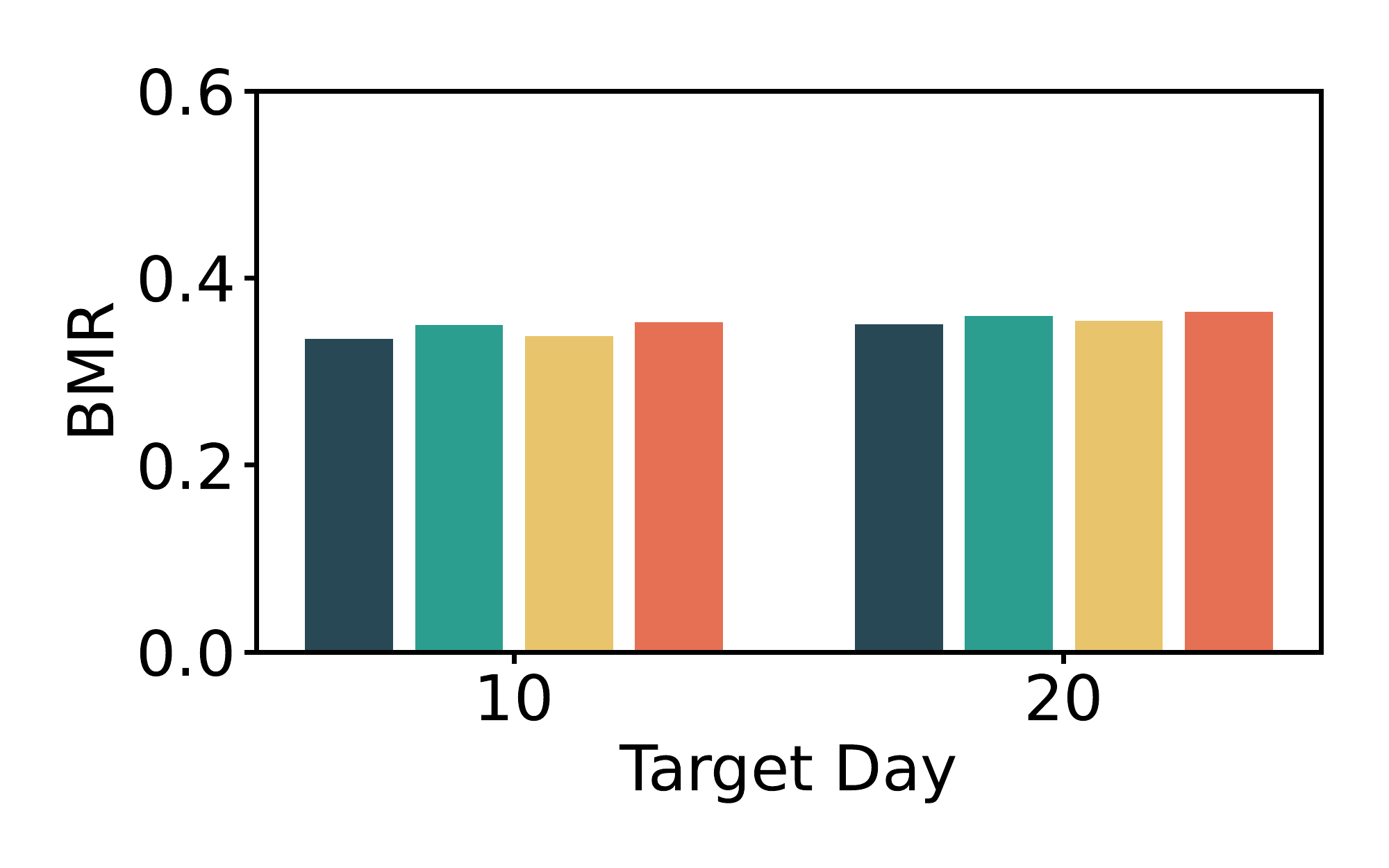}
	    \label{fig:lrb_dqn_bmr2}
	}
	\subfigure[OMR]{
	    \includegraphics[width=0.46\linewidth]{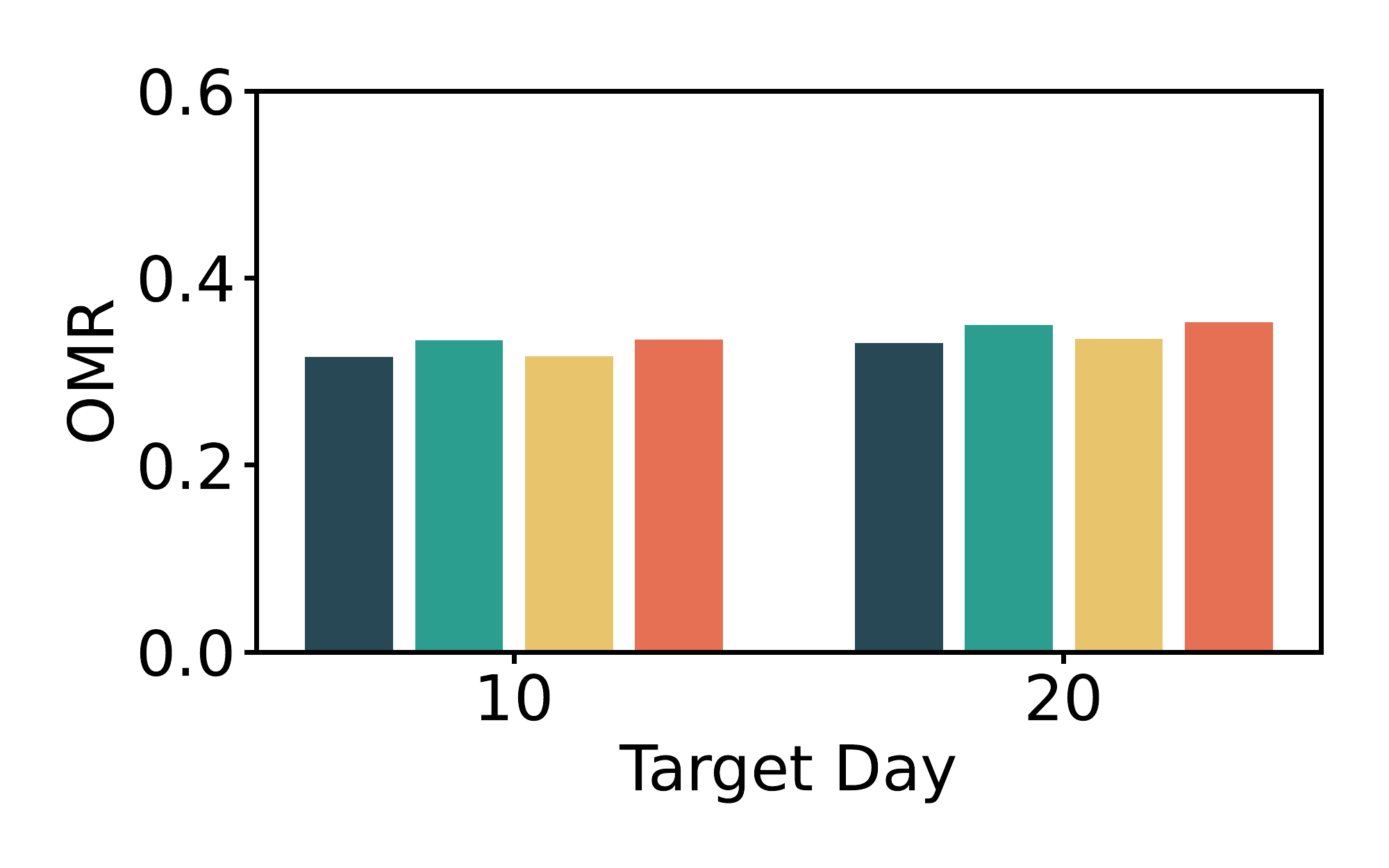}
	    \label{fig:lrb_dqn_mr2}
	}
	\vspace{-0.3cm}
	\caption{Learning results over different periodic data.}
	\vspace{-0.3cm}
	\label{fig:lrb_dqn2}
\end{figure}

The final RL issue is determining which periodic data is the most representative. We select the data from 24 hours ago as training data in the above verification based on the data locality. However, observing the request distribution in Figure~\ref{fig:periodicity}, we find that the data from a week ago more precisely matches the target data. As a result, based on DQN and LRB, the learning effects using the data from 24 hours ago and a week ago are compared on Trace-\textit{T}. We train the data of Day 3 and Day 9 to make decisions for the requests of Day 10. In addition, we train the data of Day 13 and Day 19 to make decisions for the requests of Day 20 that differ significantly from their counterpart of Day 10. As shown in Figure~\ref{fig:lrb_dqn2}, using 24-hour-old data as training data yields better results in both BMR and OMR. We believe that data locality plays a key role. Furthermore, using 24-hour-old data can reduce the space overhead required to save 1-week old data. 

\section{Design of LRU-BaSE}
\label{sec:design}
Based on the observations and analysis presented in previous sections, we employ the RL model and propose an LRU-based Belady algorithm with Size Eviction, \ie, LRU-BaSE. When a request hits its target object in the cache, the caching system uses the LRU algorithm based on the assumption of recency to promote the object. If the request misses instead, the caching system uses the learned model's policy to complete the eviction and replacement processes. 

\subsection{Learning Features and Training Data}
\label{sec:ll}
For the learning features, we select \textbf{frequency}, \textbf{reuse distance}, \textbf{reuse time} and \textbf{object size}~\cite{beckmann2018lhd,berger2018practical} from the content of the past request information. Frequency counts times of occurrence, while reuse distance and reuse time are measured in terms of the number of objects and time, respectively, between two consecutive accesses to an object. Note that the reuse time is important for perceiving the reuse behavior in the time dimension. Furthermore, we add object size to the set of learning features because photos have different sizes.

In the implementation, the training data is extracted from log files using the reservoir sampling approach~\cite{vitter1985random,zhou2018demystifying} and transformed into vectors. Specifically, we extract all unique IDs and sample out $\frac{1}{100}$ of the IDs to form an ID set. Based on this ID set, we obtain the final training data by sequential extraction. To fill the elements of training data vectors, in addition to the object size, we tally the number of times an object in the queue is hit and label this number to the object as frequency, while reuse distance and reuse time are calculated and labeled by request order and time stamp respectively. Note that we will control the object size distribution of the training data in this procedure to approach the object size distribution of the original trace.

\subsection{LRU-BaSE Architecture}
\begin{figure}[t]
	\begin{center}
		\includegraphics[width=0.86\linewidth]{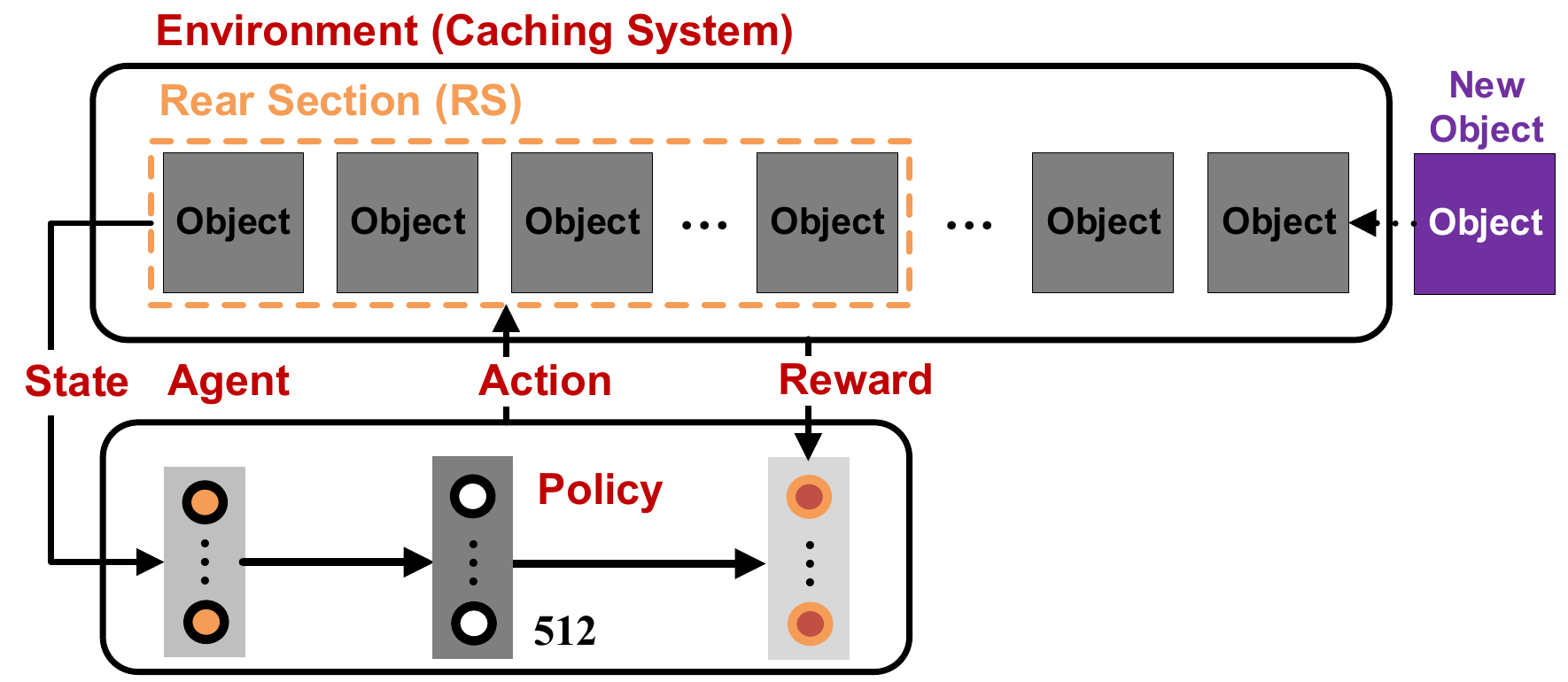}
	\end{center}
	\vspace{-0.2cm}
	\caption{The architecture of LRU-BaSE.}
	\vspace{-0.3cm}
	\label{fig:dqncache}
\end{figure}

We employ DQN~\cite{mnih2013playing} as the backbone of the architecture since object ID is a discrete value. We have also tried to compute fine-grained probabilities for eviction by the A3C~\cite{mnih2016asynchronous} or DDPG~\cite{lillicrap2015continuous} method without obtaining any performance improvement while increasing the parameter footprint.
DQN uses neural networks to learn the Q-score~\cite{kaelbling1996reinforcement} that is produced by continuous decisions. To ensure the efficiency of decision-making, we configure a lightweight network with only one hidden layer consisting of 512 neurons. The action space in LRU-BaSE is the rear section of the cache queue described in $\S$~\ref{sec:rear}, and an agent module is configured to learn the model and make decisions on this rear section. The architecture is shown in Figure~\ref{fig:dqncache}.

\begin{table}[t]
	\centering
 	\scalebox{0.78}{
		\begin{tabular}{ccccccc}
			\toprule
			\multicolumn{3}{c}{Hyper-Parameter} & \multicolumn{2}{c}{Trace-\textit{T}} & \multicolumn{2}{c}{Wikipedia} \\
			\cmidrule(lr){1-3} \cmidrule(lr){4-5} \cmidrule(lr){6-7}
			$\alpha$ & $\beta$ & $\gamma$ & OMR (\%) & BMR (\%) & OMR (\%) & BMR (\%) \\
			\midrule
			 0.4   & 0.4   & 0.2 & 31.38 & 32.77 & 1.02  & 3.17\\
             0.3   & 0.5   & 0.2 & +0.51 & +0.42 & +0.03 & +0.08\\
             0.5   & 0.3   & 0.2 & +0.23 & +0.34 & +0.05 & +0.04\\
             0.35  & 0.35  & 0.3 & +0.07 & +0.05 & +0.01 & +0.00\\
             0.45  & 0.45  & 0.1 & +0.02 & +0.05 & -0.01 & +0.01\\
			\bottomrule
		\end{tabular}
 	}
	\caption{ The changes of OMR and BMR under different hyper-parameters of the key-step reward function. Each result is the mean across 50 runs.}
	\label{table:rfp}
\end{table}

\subsubsection{Modeling}
\label{sec:model}
The LRU-BaSE architecture incorporates DQN to reflect six key elements, \ie, \textit{environment}, \textit{state}, \textit{reward}, \textit{policy}, \textit{action}, and \textit{agent}, as illustrated in Figure~\ref{fig:dqncache} and described below.

\noindent\textbf{Environment}: \textit{Environment} is the concrete caching system. It is not only the entity of the cache queue but also generates the \textit{state} and the performance after each decision.

\noindent\textbf{State}: \textit{State}, denoted by $\bm{s}$, is represented by the feature vector of the rear section in LRU-BaSE, which is an input to \textit{agent}.

\noindent \textbf{Reward}: \textit{Reward} is a scalar denoted as $r$. It is calculated by a function that defines the expectation for performance changes. Generally, if the current performance is better than the last performance, the \textit{Reward} will be positive, otherwise, it will be negative (See $\S$~\ref{sec:reward}).

\noindent \textbf{Policy}: \textit{Policy} is a function, \ie, a network with parameters, that transforms a \textit{state} into \textit{action}. In LRU-BaSE, the \textit{policy} is learned by the network, \textit{state} and \textit{reward}.

\noindent \textbf{Action}: \textit{Action} is a scalar denoted as $a$, which represents the index of the victim object in the queue array. We denote the \textit{action} as $\bm{a}_t$ corresponding to the \textit{state} $\bm{s}_t$.

\noindent\textbf{Agent}: \textit{Agent} is a black box consisting of \textit{policy}. Like a brain, it inputs the \textit{state} to the network, adjusts the parameters by the \textit{reward}, and outputs the \textit{action} to the \textit{environment}. Note that during online decision at runtime, the \textit{agent} merely executes the process of input and output.

\subsubsection{Key-step Reward Function}
\label{sec:reward}

The reward function is crucial in RL. We desire that the function can encourage the behavior that maintains OMR and minimizes BMR. In addition, to minimize the feedback delay, RL in the cache can only make a decision in one step. Thus, we design the \textit{key-step reward function} focusing on the local benefits rather than the global ones. Assume that $B_i$ and $O_i$ represent BMR and OMR respectively at the $i$-th step, $\Delta B_i=\frac{-B_i+B_{i-1}}{B_{i-1}}$, $\Delta O_i=\frac{-O_i+O_{i-1}}{O_{i-1}}$, $\Delta B^{0}_i=\frac{-B_i+B_{0}}{B_0}$, $\Delta O^{0}_i=\frac{-O_i+O_{0}}{O_0}$, and $\Delta D_i= \Delta B_i-\Delta O_i$. The reward function is described below.
\begin{equation}
r=\alpha\frac{\Delta B^{0}_i}{1-\Delta B_i}+\beta\frac{\Delta O^{0}_i}{1-\Delta O_i}+\gamma\Delta D_i ,
\end{equation}
where $\alpha+\beta+\gamma=1$. In theory, we require $\alpha=\beta>\gamma$ to give priority to the consistent reduction of OMR and BMR, and then to encourage a faster decline for BMR. When $r<0$, DQN will terminate the learning process and discard this transition. If $r\geq0$ and $\Delta D_i>0$, DQN will terminate the learning process and store this transition, which is our \textit{key step}. Otherwise, DQN will continue the learning process in the current episode.

With a cache size of 1024GB, we test OMR and BMR at different settings of $\alpha$, $\beta$, and $\gamma$ on Trace-\textit{T} and Wikipedia. According to the results shown in Table~\ref{table:rfp}, we determine that $\alpha = \beta = 0.4$ and $\gamma = 0.2$ is a desired setting.

\subsubsection{Model Usage}
Assume that the set of objects in the rear section is $O=\{o_1, ..., o_k, ..., o_N\}$, where $o_k$ denotes the $k$-th object and the size of the rear section is equal to \textit{N}.
 
\noindent\textbf{Training:} Assume that the cache queue is full and replacement by LRU has occurred at least 1000 times. When a new object is loaded into the cache and the object $o_{N+1}$ is loaded into the rear section, the \textit{environment} selects $\{o_1,o_2,...,o_{N}\}$ to participate in training. With features from $\S$~\ref{sec:ll}, \textit{state} $\bm{s}_t$ is input to the \textit{agent}. The \textit{agent} learns the \textit{policy} by the style of DQN, outputs \textit{action} by the \textit{policy}, and informs the \textit{environment} which object should be evicted. The \textit{environment} will generate a new \textit{state} and \textit{reward} after eviction. The \textit{state} and \textit{reward} will be input into the \textit{agent} to train a new \textit{policy}. A new \textit{action} will be generated by the new \textit{policy} and a new  \textit{state} will be produced from the \textit{environment}. This cycle will continue until the termination requirement is satisfied. The training result (\ie, \textit{policy}) of each model will be saved as a 512-dimensional floating-point vector.

\noindent\textbf{Decision:} With a trained \textit{policy}, the current \textit{state} gathered from the rear section is input into the \textit{agent}, and the \textit{agent} outputs an \textit{action}. According to this \textit{action}, the system selects an object to evict from the cache queue. Specifically, the \textit{agent} only outputs the index corresponding to the \textit{action} that acquires the maximum score. The index is equal to 0 when the object is at the end of the queue. \textit{Environment} locates the object via the index and evicts it. Since the traversing using a key is omitted, the decision is efficient.

\noindent\textbf{Update:} To maintain the model's reliability by data locality, we must update the model by training the data from 1 day ago. To maintain decision-making consistency, the training process must be continuous, and in parallel with another model that is making decisions. Since each model's training time does not exceed the time span, CPU resources outside of cache running and decision making are sufficient to maintain the update process. When it comes to another time region, the corresponding \textit{policy} transfers to the current model. We describe the detailed process in $\S$~\ref{sec:workflow}.

\subsection{Workflow}
\label{sec:workflow}

\begin{figure}[t]
	\begin{center}
		\includegraphics[width=0.9\linewidth]{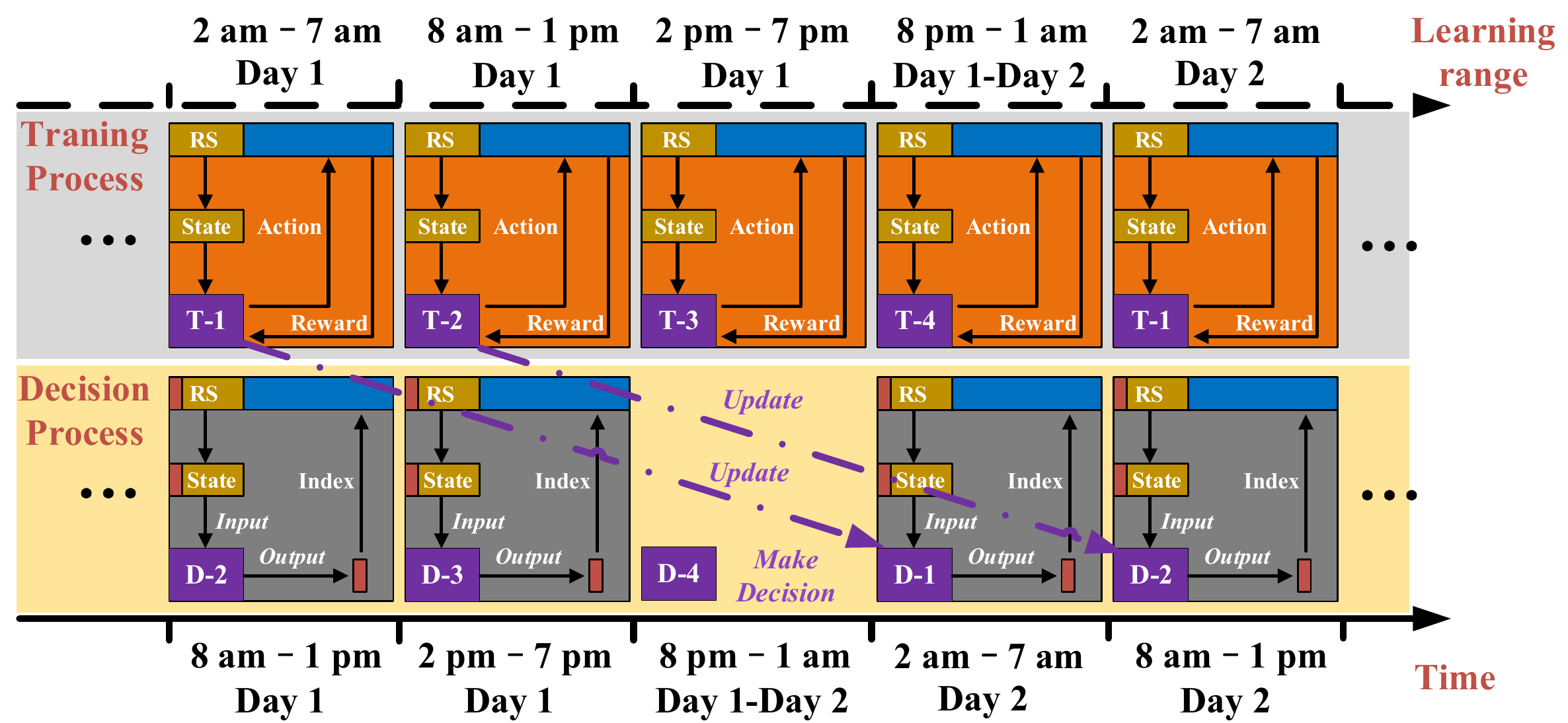}
	\end{center}
	\vspace{-0.3cm}
	\caption{The flow diagram of LRU-BaSE.}
	\vspace{-0.3cm}
	\label{fig:work_flow}
\end{figure}

Based on the above design, we describe the workflow of LRU-BaSE as shown in Figure~\ref{fig:work_flow}. The training process is depicted in parallel to the decision process along the timeline. T-1 represents the LRU-BaSE model used for training the data in the 1$^{st}$ time region (\ie, 2 a.m. to 7 a.m.) while D-1 represents another LRU-BaSE model used for decision on the 1$^{st}$ time region. Based on the segment in Figure~\ref{fig:work_flow} and assume that D-1 to D-4 have deployed the \textit{policy} on Day 1, T-1 begins training upon entering the 2$^{nd}$ time region (\ie, 8 a.m. to 1 p.m.) because the requests in the 1$^{st}$ time region have already been completed and collected. D-2 makes replacement decisions and produces the index of eviction candidates at the same time. Next, T-2 begins training upon entering the 3$^{nd}$ time region (\ie, 2 p.m. to 7 p.m.) with the initial parameters,~\ie, the \textit{policy} of T-1. Meanwhile, D-3 starts the decision. In this order, when the 1$^{st}$ time region of Day 2 arrives, D-1 makes decisions after updating the \textit{policy} to that of T-1, which was trained one day ago. The update time is within one second, and LRU takes over the replacement during this time. Next, T-1 re-trains its \textit{policy} at the 2$^{nd}$ time region. D-2 updates its \textit{policy} from T-2 and then makes decisions. The whole process will repeat. Note that there is actually only one training model and one decision model. Training and decision in different time regions simply necessitate the transformation of inputs and \textit{policies}.

\subsection{Parallel Decision}
As shown in Figure~\ref{fig:work_flow}, the training and decision processes are performed on a single CPU core continuously, resulting in a high CPU resource demand. In a real system implementation supporting compatibility with multi-threaded data access and avoiding consistency problems, we propose a parallel decision scheme for LRU-BaSE, with a multi-threaded execution approach working on multiple CPU cores for the replacement policy to ensure ample computational resources and online efficiency. For example, the OC layer of XXPhoto is deployed on a CPU with 56 cores and is working with multiple threads, where one thread is used for data gathering and cleaning, another thread is used for model training, and the rest of the threads are used for data access from users.

We update the \textit{action} to a vector that consists of the index sequence of the recommended objects for eviction and is sorted in descending order of the most deserving of eviction. Note that the extent of deserving of eviction stems from the Q-scores of object ID in the rear section of the cache queue. Generally, we will set the dimension of the vector to be equal to the maximum number of threads. When a batch of threads conducts eviction operations, LRU-BaSE will delineate the top-$T$ indexes from the vector for eviction, where $T$ is equal to the number of threads. By parallelizing eviction operations and avoiding multiple waits for DQN's recommendations, this scheme helps to ensure efficiency in real systems.

\section{Deployment and Performance of LRU-BaSE on Real System TDC}

In this section, we will detail the deployment of LRU-BaSE on a real system and show the performance improvements and overhead measured from the monitor system.

\subsection{Deployment}
\label{section:pro}
T Disk Cache (TDC) is a SSD-based disk cache system of company \textit{T}. The system already has 3000+ physical machines online that provide large-capacity and high-performance distributed key-value cache service to 2500+ businesses including XXPhoto. TDC consists primarily of two modules: \textit{Cell Master} and \textit{Cache Server}. The \textit{Cell Master}, as a resource management module, is responsible for managing the routing information of the whole system, periodically detecting the running status of each server, and receiving the statistical information reported by each server. The \textit{Cache Server} is the data storage and processing engine and its prototype is a storage node based on MCP++, multi-ccd/multi-smcd process model, raw disk, inode, and asynchronous disk I/O technologies(\eg, libaio/SPDK). It is used for storing and processing business data, \ie, objects, where data processing can be achieved by the deployed GPUs. On the top layer of the disk, a memory cache stores the keys and indexes of objects in addition to the learning optimization processes described in $\S$~\ref{sec:learnforcdn}, where each process works on an exclusive memory space and the index in the memory cache is in perfect sync with the objects on the disk. We replace LRU with LRU-BaSE on the memory cache rather than on the disk, since it helps the endurance of SSDs by reducing write operations due to frequent updates to the index.


\subsection{Performance}
\label{sec:improve}

\begin{figure}[t]
	\centering
	\includegraphics[width=0.34\linewidth]{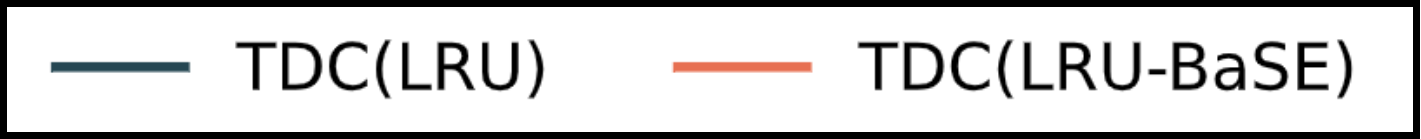}
	\vspace{-0.2cm}
	\\
	\subfigure[Overall system: bandwidth]{
	    \includegraphics[width=0.46\linewidth]{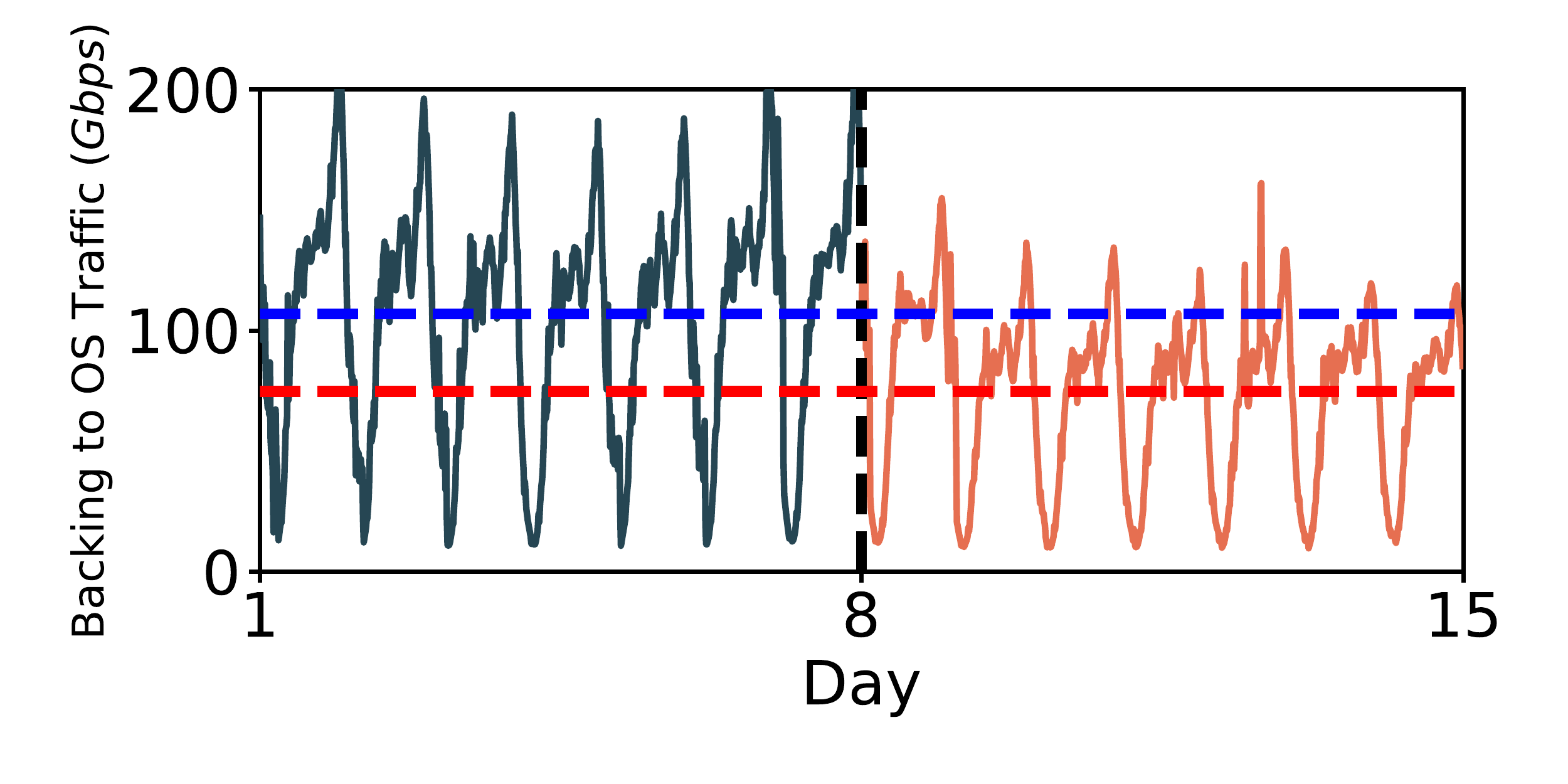}
	    \label{fig:real_traffic}
	}
	\subfigure[Overall system: latency]{
	    \includegraphics[width=0.46\linewidth]{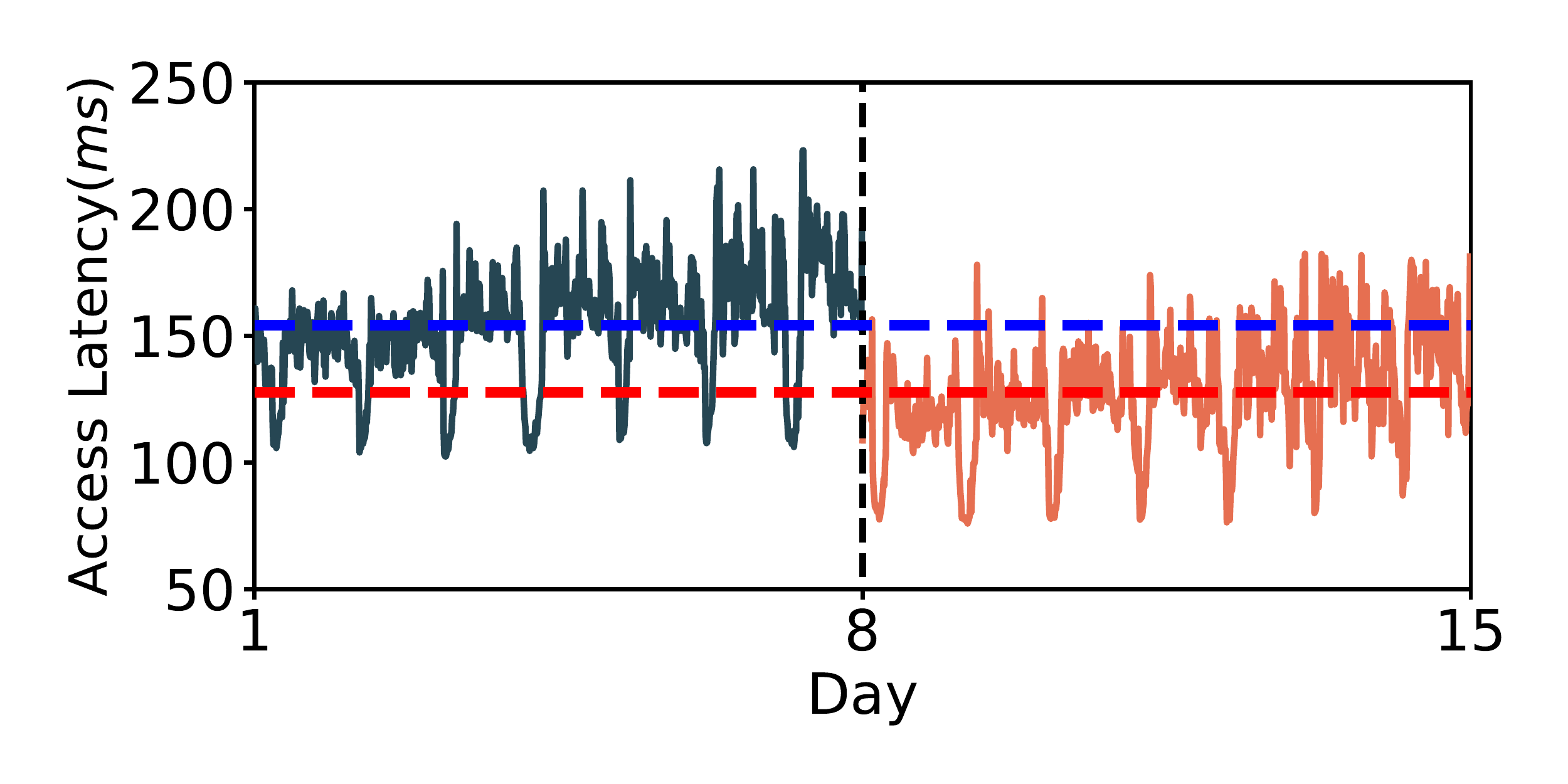}
	    \label{fig:real_latency}
	}
	\subfigure[One machine: CPU utilization]{
	    \includegraphics[width=0.46\linewidth]{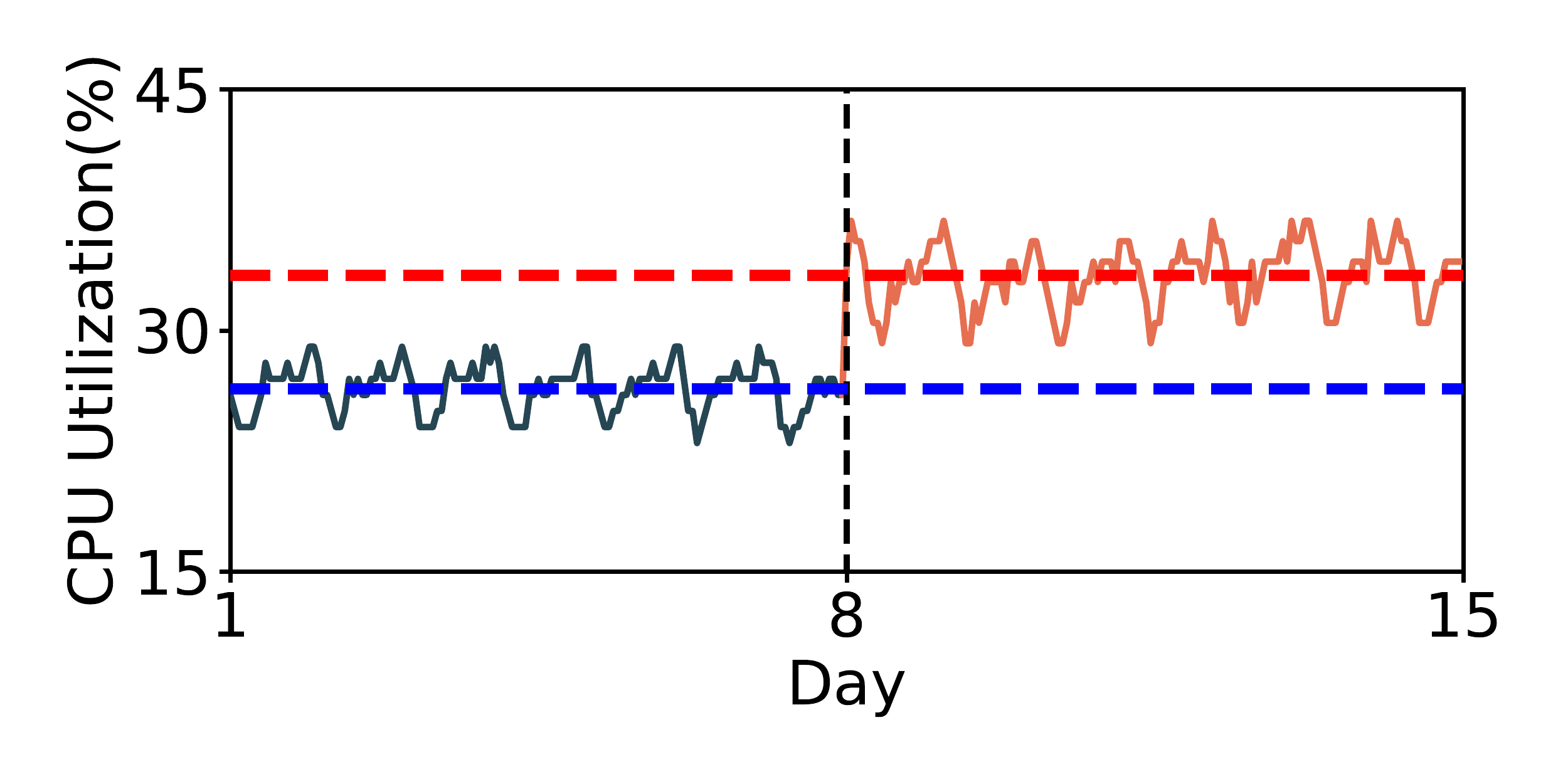}
	    \label{fig:real_cpu}
	}
	\subfigure[One machine: throughput]{
	    \includegraphics[width=0.46\linewidth]{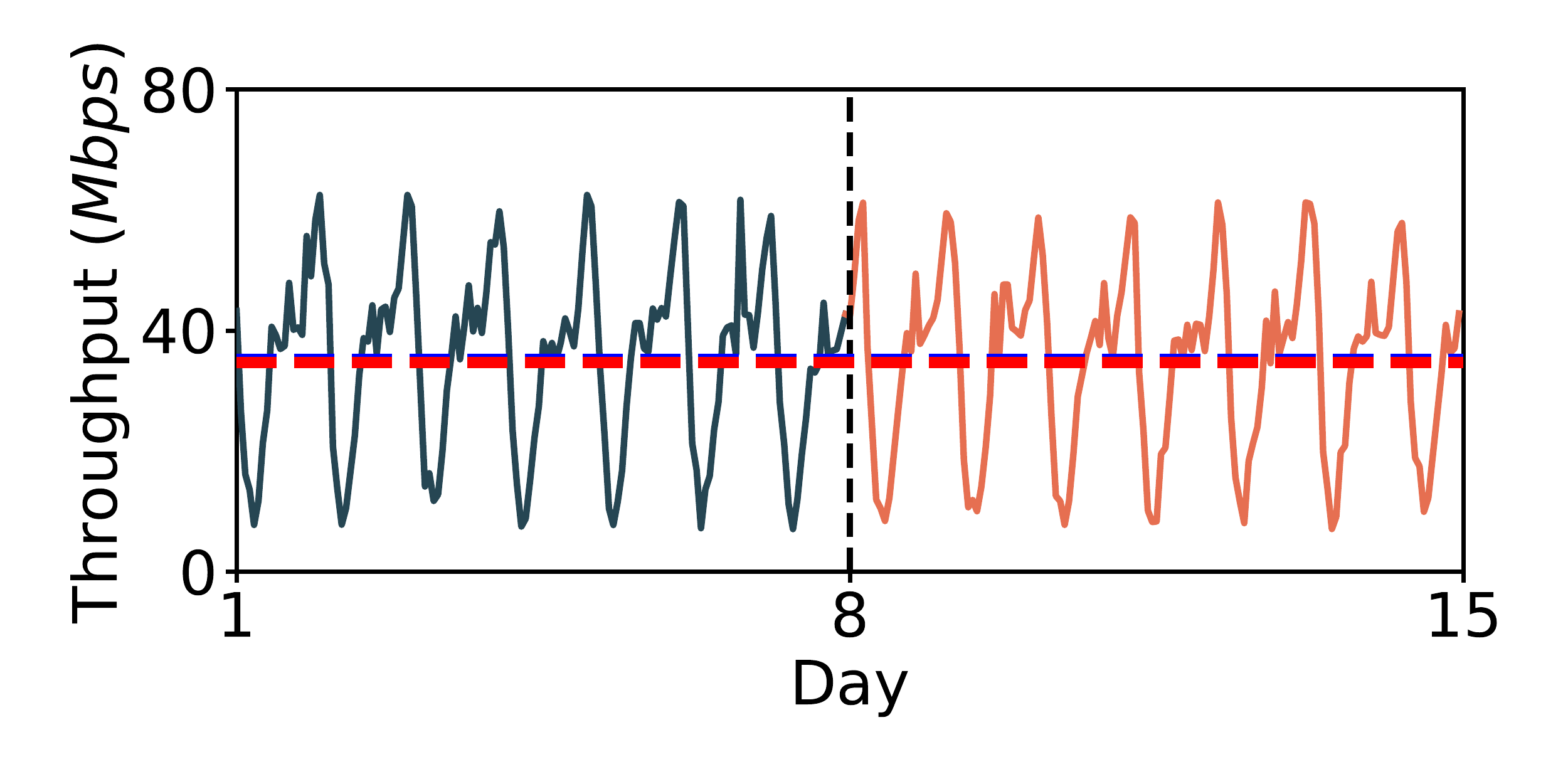}
	    \label{fig:real_throughput}
	}
	\caption{Changes in performance and overhead of TDC after the online deployment of LRU-BaSE. LRU was replaced by LRU-BaSE at 24:00 on Day 7. The purple and red dashed lines represent average values of LRU's results in 7 days and LRU-BaSE's results in 7 days, respectively. (a) and (b) represent performance measured from the overall system, while (c) and (d) represent overhead and performance measured from a physical machine randomly selected from the system.}
	\label{fig:real}
\end{figure}

\begin{table}[t]
\centering
\scalebox{0.95}{
\begin{tabular}{lrr}
\toprule
\textbf{Metric}           & \textbf{TDC(LRU)} & \textbf{TDC(LRU-BaSE)} \\ \hline
Average OMR               & 52.5\%            & 28.2\%                 \\ 
Average BMR               & 65.9\%            & 33.6\%                 \\ 
$99.9^{th}$ Latency       & 1.45s             & 0.48s                  \\ 
Peak CPU                  & 29.01\%           & 36.84\%                \\ 
Peak Memory               & 14.3GB            & 15.08GB                \\ 
Max Throughput            & 62.5Gbps          & 61.25Gbps              \\ 
\bottomrule
\end{tabular}
}
\caption{Important metrics for TDC(LRU) and TDC(LRU-BaSE).}
\label{tabel:resource}
\end{table}




As shown in Figure~\ref{fig:real_traffic} and Figure~\ref{fig:real_latency}, we measured the performance changes in "backing to OS" traffic and average user access latency from the monitor system. After deploying LRU-BaSE, the average "backing to OS" traffic dropped by 31.96Gbps (30.05\%), and the average user access latency dropped by 28.75ms (17.07\%). Furthermore, the request tail latency was reduced from 1.45s to 0.48s at the $99.9^{th}$ percentile, or 66.90\%. We attribute this result to the judicious eviction of large files by LRU-BaSE. Note that there is a hot event emerging on Day 12 at 10 a.m. Although the "backing to OS" traffic spiked for a moment, it quickly fell back with LRU-BaSE's online learning and decision making. In addition, since Day 6, Day 7, Day 13 and Day 14 are weekends, there were more requests for the same data, resulting in system congestion and fluctuations in the user access latency during these periods. 

In the absence of LRU-BaSE, an administrator would likely resort to resource overprovisioning by expansion of equipment to cope with the surge of traffic and removing the overprovisioned equipment when the traffic ebbed, resulting in human labor costs and equipment costs. Moreover, as the cache space of TDC is about 3000TB, adding about 100TB of SSD can only achieve a 1\% reduction in OMR. As shown in Table~\ref{tabel:resource}, LRU-BaSE can reduce OMR and BMR by 24.3\% and 32.3\%, respectively. Although LRU-BaSE will consume more of TDC's CPU resources that are relatively plentiful, it can save roughly \$795,000 per year in bandwidth costs.

\subsection{Overhead}


To comprehensively evaluate the utility of LRU-BaSE, we measured the overhead of using LRU-BaSE in the same time period as $\S$~\ref{sec:improve}.

\noindent\textbf{CPU Utilization.} As shown in Figure~\ref{fig:real_cpu}, the average CPU utilization of all devices was 29.41\% before the LRU-BaSE was deployed and increased to 36.42\% afterward, where the peek CPU utilization observed is 38.77\%. Since LRU-BaSE keeps the CPU utilization within 50\%, it does not affect the stability of the system.

\noindent\textbf{Throughput.} As shown in Figure~\ref{fig:real_throughput}, the change in average throughput is from 35.41Mbps to 34.62Mbps. 

\noindent\textbf{Memory.} LRU-BaSE uses an extra 751MB RAM for storing model-related metadata such as network structure parameters and policy parameters. In addition, the extra memory is used to store the learning data comprised of the timestamp (long long int, 8 Bytes), object key (string, 16 Bytes), and size (long int, 4 Bytes). Given the 18 billion inodes in the system, the required capacity is 48MB. Based on this, we conclude that the total additional memory overhead for LRU-BaSE is 799MB or 5.94\% of the inherent consumption. This additional capacity can only bring less than a 0.001\% drop in OMR and BMR when used directly as a memory replenishment cache for LRU. 


\section{Evaluation}
In this section, we will compare LRU-BaSE to state-of-the-art caching algorithms on the simulator.


\begin{table}[t]
\centering
\scalebox{0.77}{
\begin{tabular}{|l|l|r|r|r|}
\hline
\multicolumn{2}{|l|}{}                              & Trace-\textit{T} & Wikipedia  & CDN-Q                         \\ \hline
\multicolumn{2}{|l|}{Total Requests (Millions)}          & 78.75            & 268.57     & 39.71                       \\ \hline
\multicolumn{2}{|l|}{Unique Object Requested (Millions)} & 24.71            & 3.75       & 15.83                                    \\ \hline
\multicolumn{2}{|l|}{Total Bytes Requested (GB)}         & 3346.72          & 9219.61    & 1136.54                     \\ \hline
\multicolumn{2}{|l|}{Unique Bytes Requested (GB)}        & 1097.63          & 568.92     & 349.91                               \\ \hline
\multicolumn{2}{|l|}{Warm-up Requests (Millions)}        & 40               & 130        &    20                       \\ \hline 
\multirow{3}{*}{Request Object Size}     & Mean (KB)     & 46.58            & 158.97     & 23.18                           \\ \cline{2-5} 
                                         & Max (MB)      & 19.97            & 674.38     & 7.99                              \\ \cline{2-5}
                                         & Min (B)       & 2                & 10         & 31                                         \\ \hline
\end{tabular}
}
\caption{Summary of the three production traces that are used throughout our evaluation.}
\label{table:trace}
\end{table}

\begin{figure*}[t]
	\centering
	\includegraphics[width=0.45\linewidth]{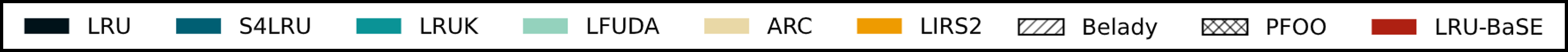}
	\\
	\subfigure[Trace-\textit{T}: BMR]{
	    \includegraphics[width=0.185\linewidth]{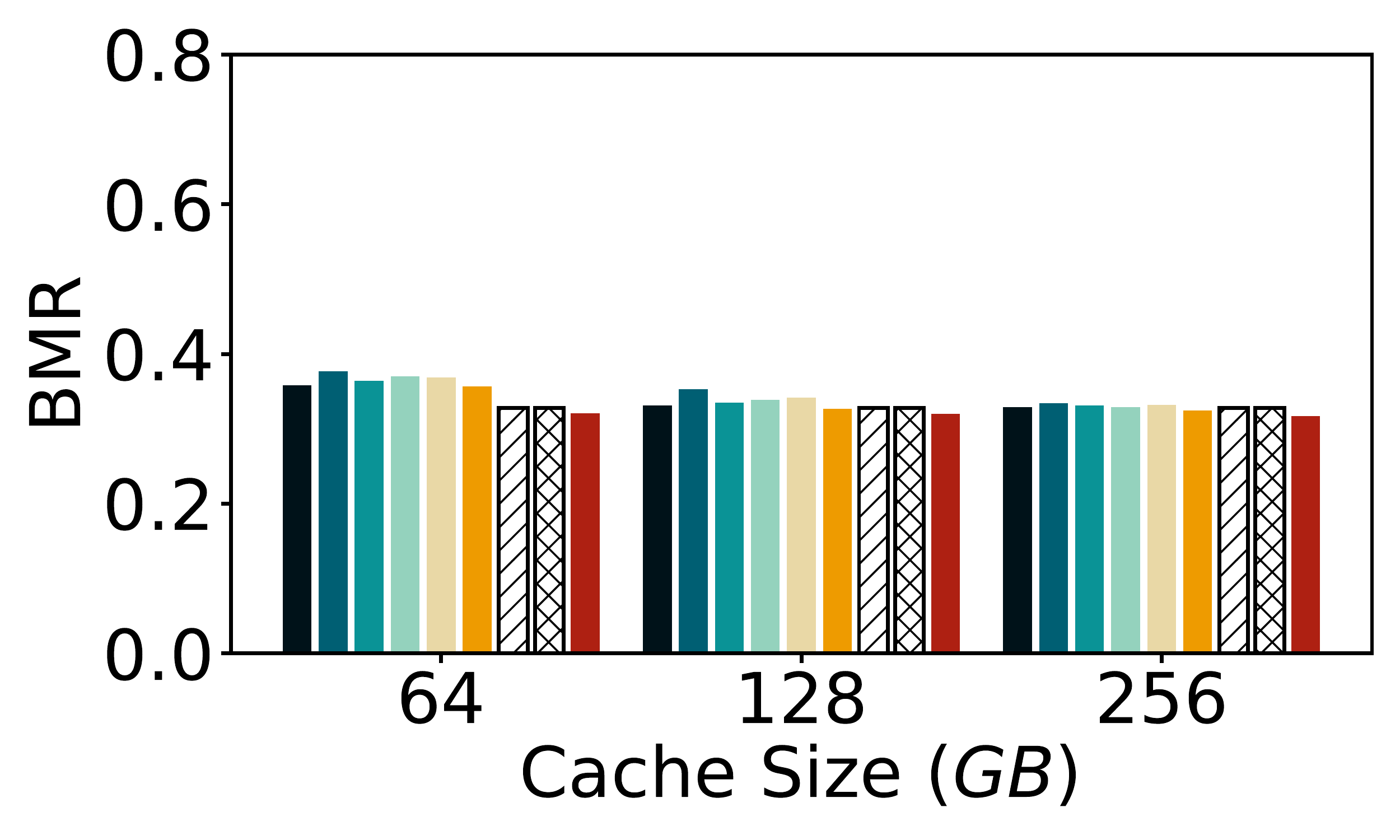}
	    \label{fig:trace_t_bmr}
	}
	\subfigure[Trace-\textit{T}: OMR]{
	    \includegraphics[width=0.185\linewidth]{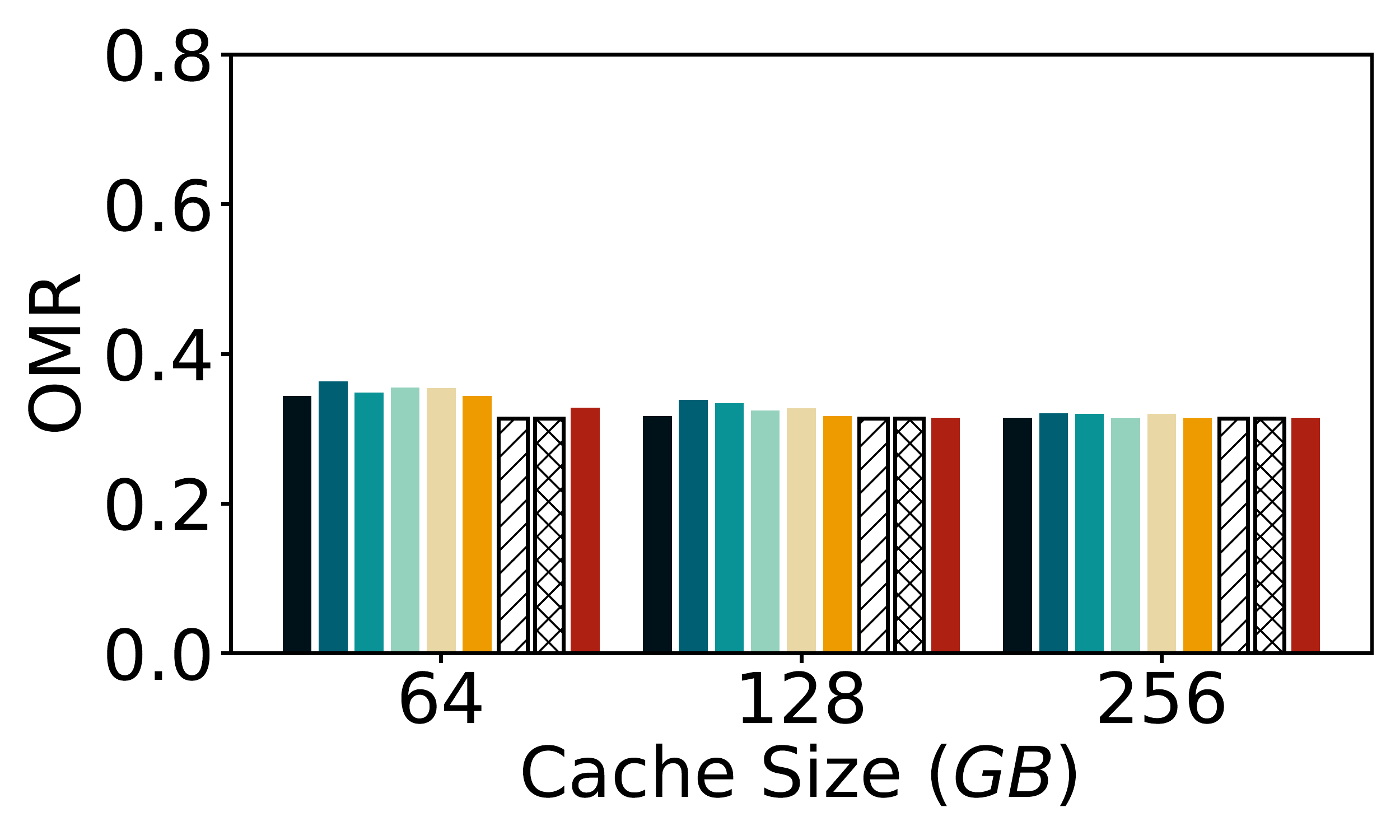}
	    \label{fig:trace_t_omr}
	}
	\subfigure[Trace-\textit{T}: Peak CPU]{
	    \includegraphics[width=0.185\linewidth]{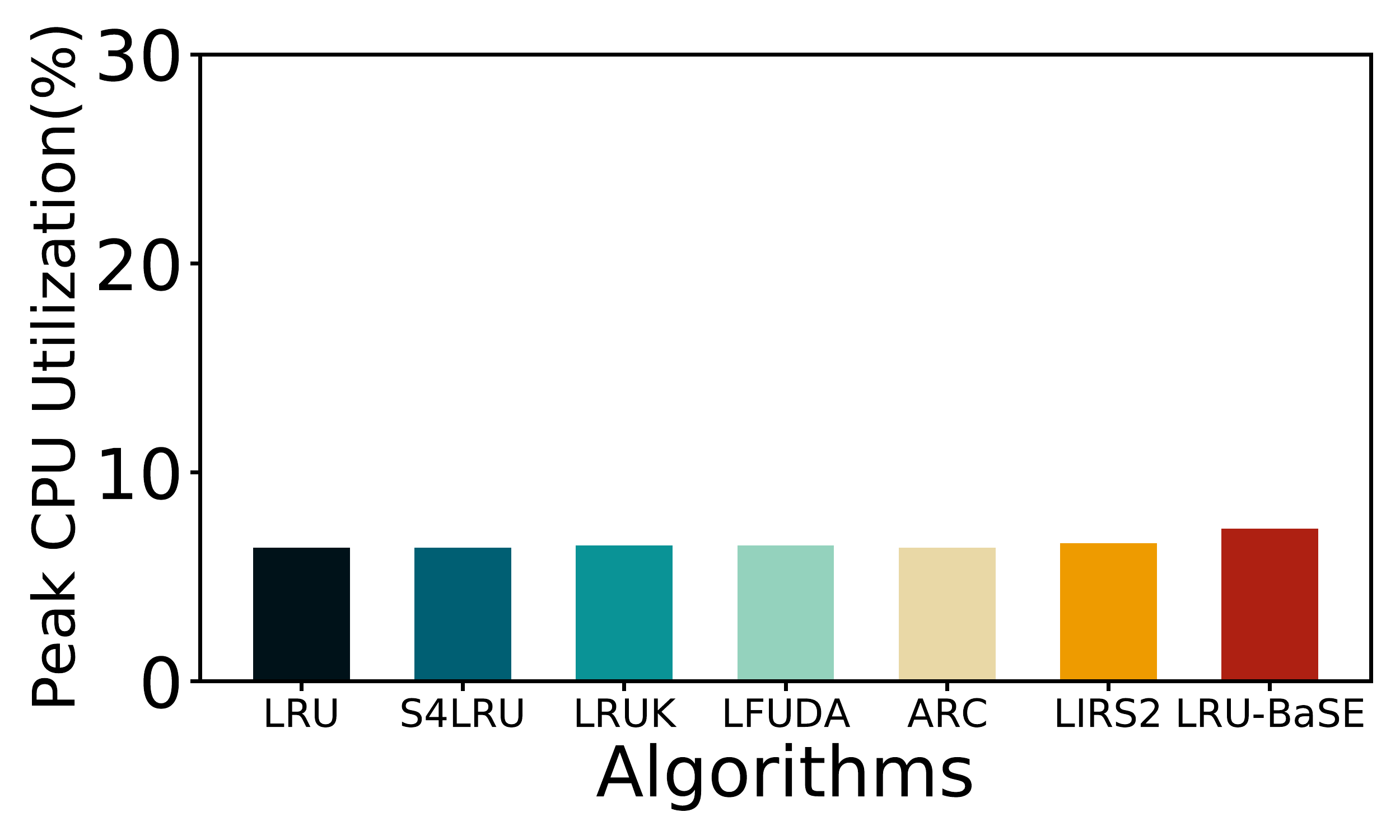}
	    \label{fig:trace_t_cpu}
	}
	\subfigure[Trace-\textit{T}: TPS]{
	    \includegraphics[width=0.185\linewidth]{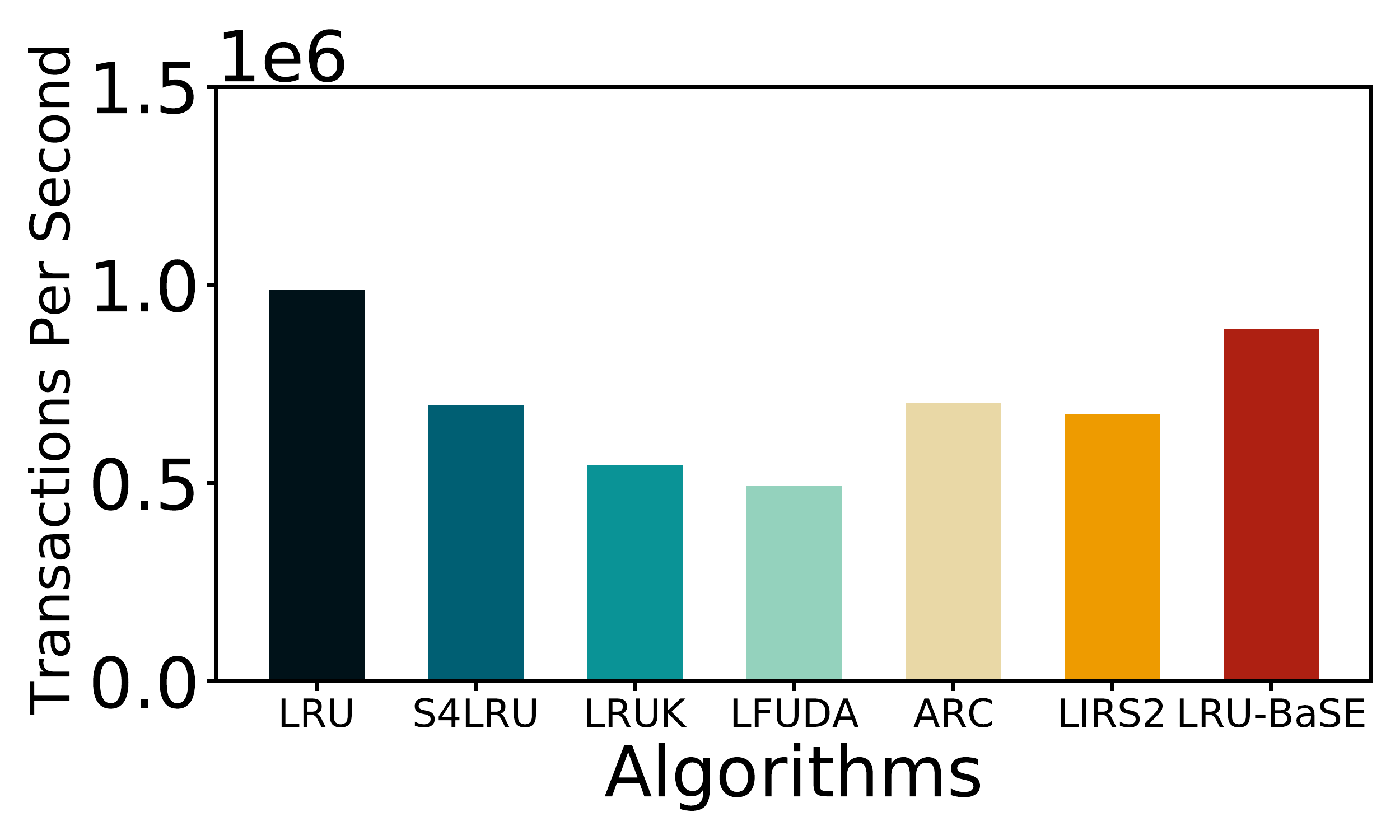}
	    \label{fig:trace_t_throughput}
	}
	\subfigure[Trace-\textit{T}: Peak Memory]{
	    \includegraphics[width=0.185\linewidth]{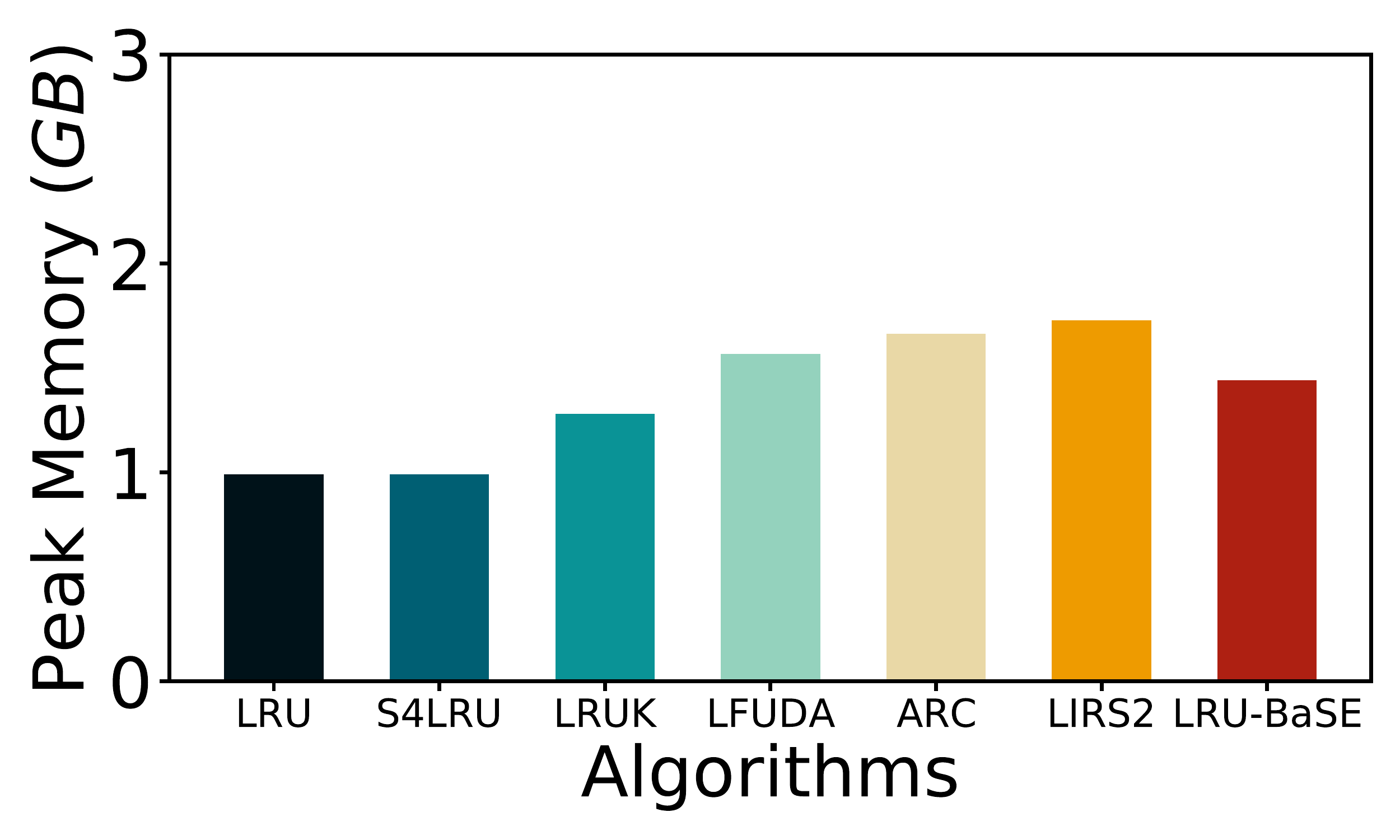}
	    \label{fig:trace_t_memory}
	}
	\\
	\subfigure[Wikipedia: BMR]{
	    \includegraphics[width=0.185\linewidth]{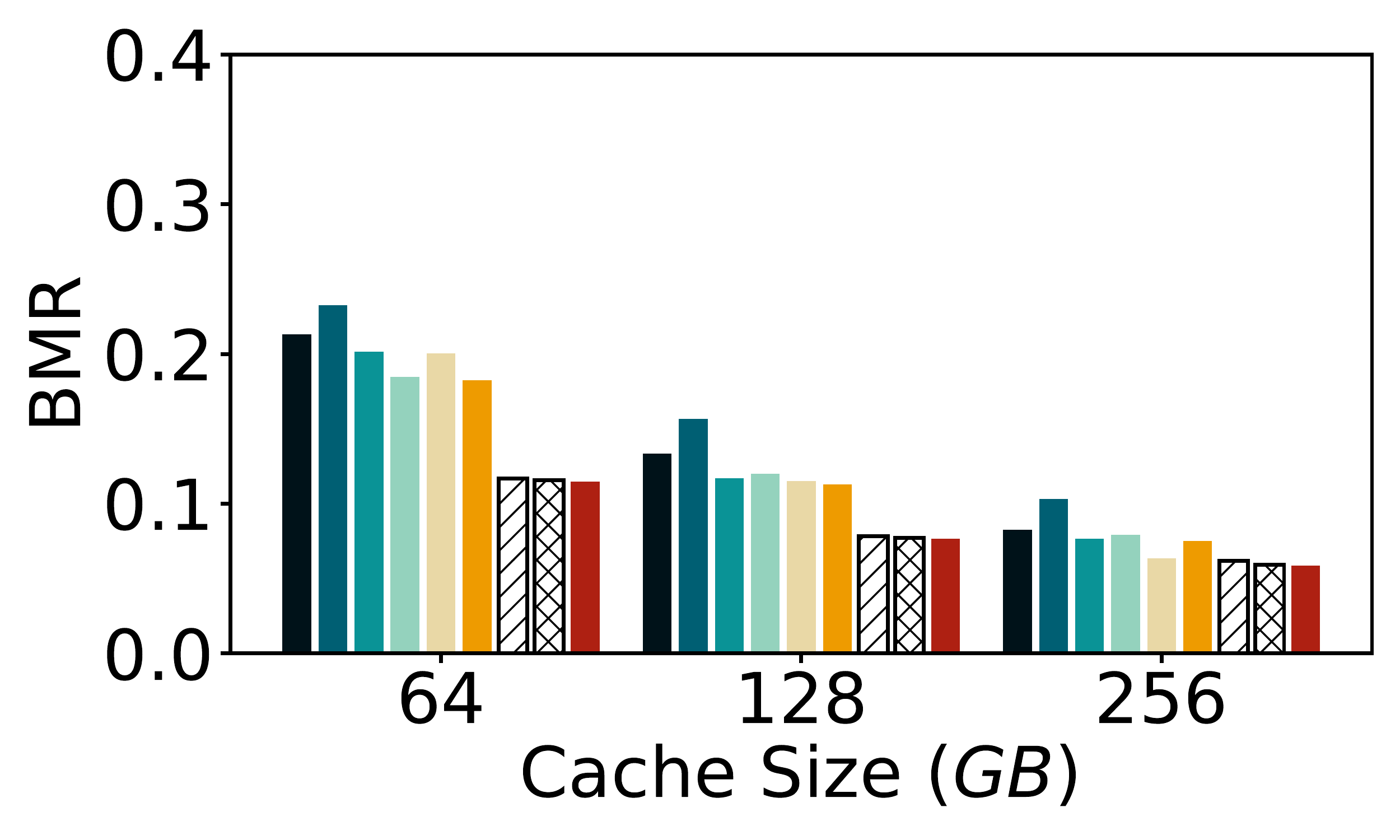}
	    \label{fig:wikipedia_bmr}
	}
	\subfigure[Wikipedia: OMR]{
	    \includegraphics[width=0.185\linewidth]{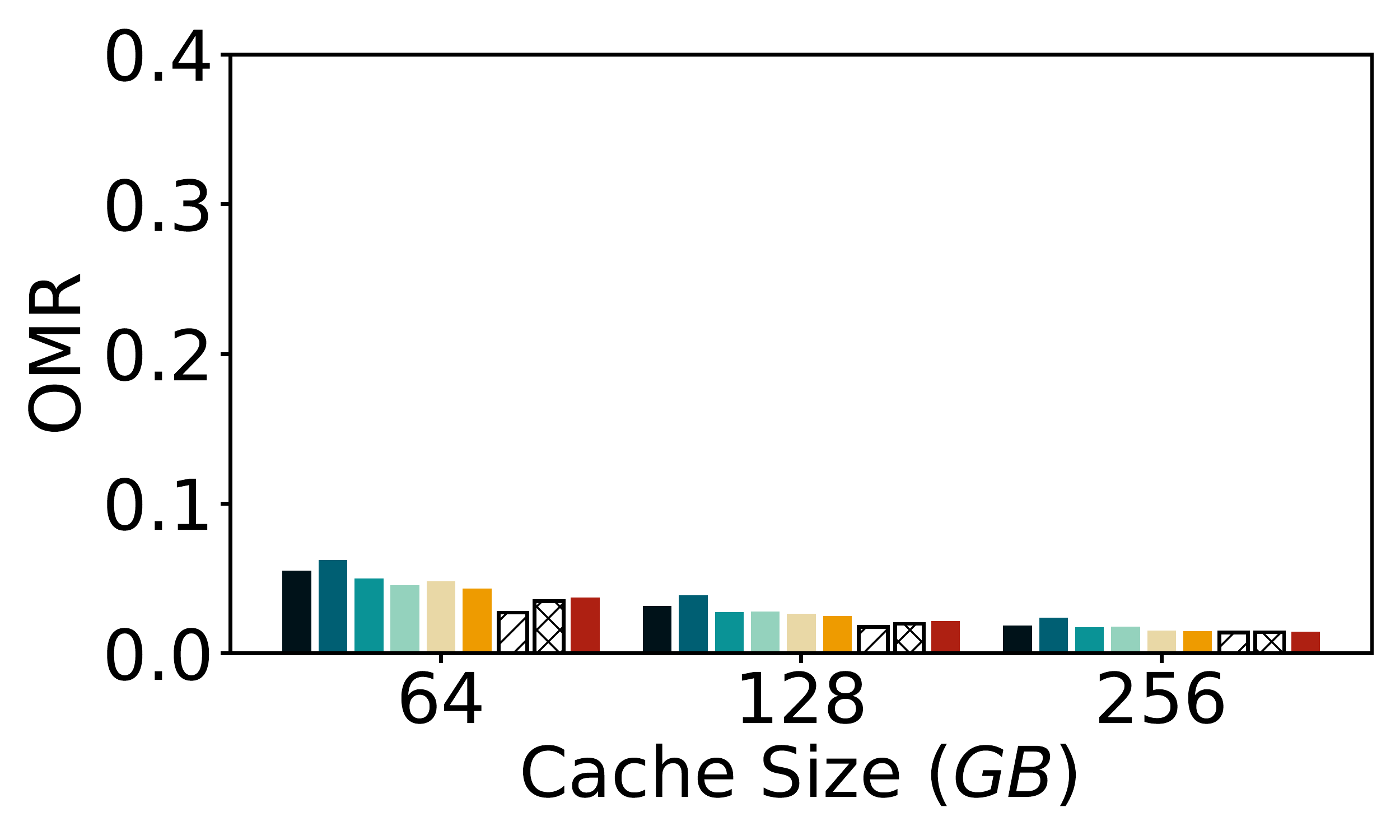}
	    \label{fig:wikipedia_omr}
	}
	\subfigure[Wikipedia: Peak CPU]{
	    \includegraphics[width=0.185\linewidth]{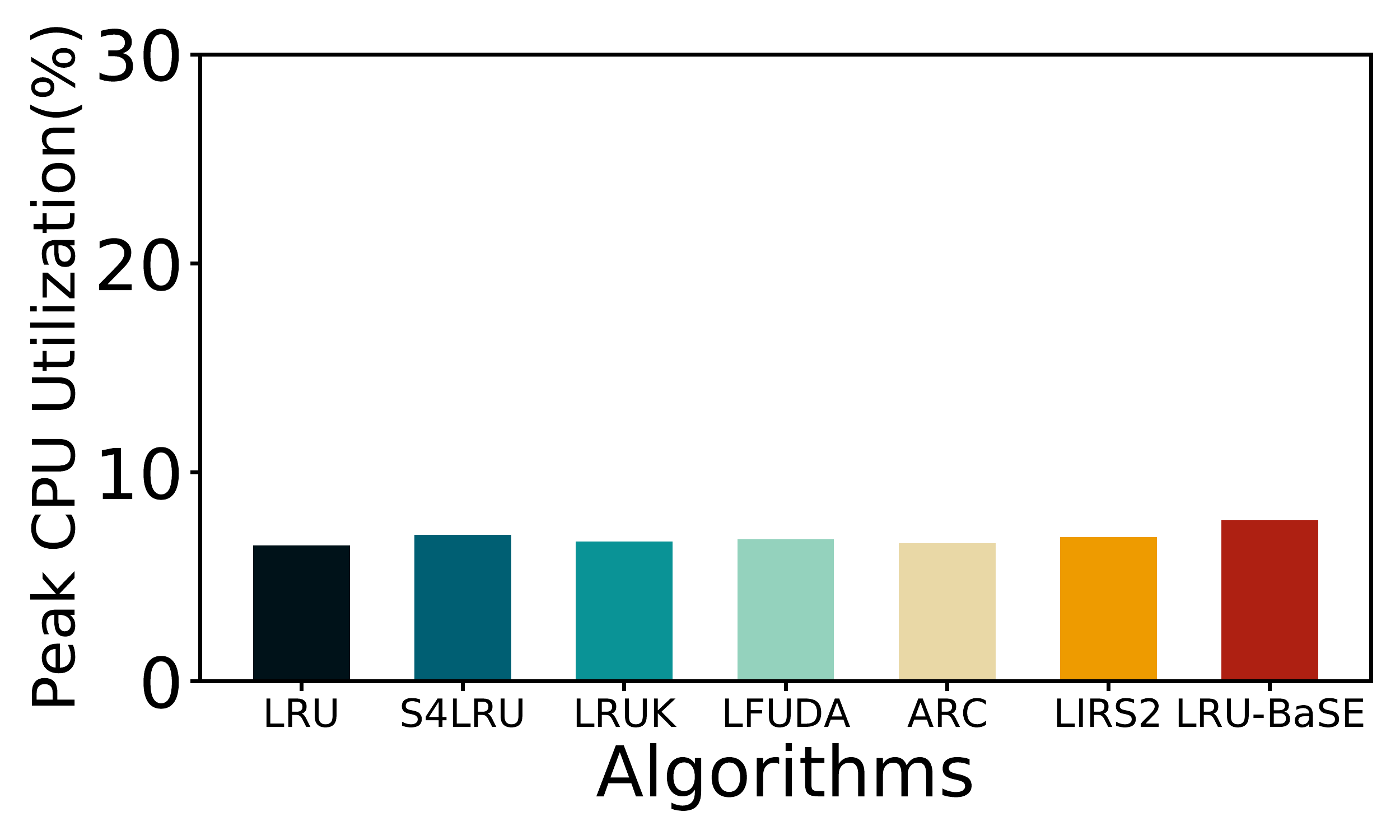}
	    \label{fig:wikipedia_cpu}
	}
	\subfigure[Wikipedia: TPS]{
	    \includegraphics[width=0.185\linewidth]{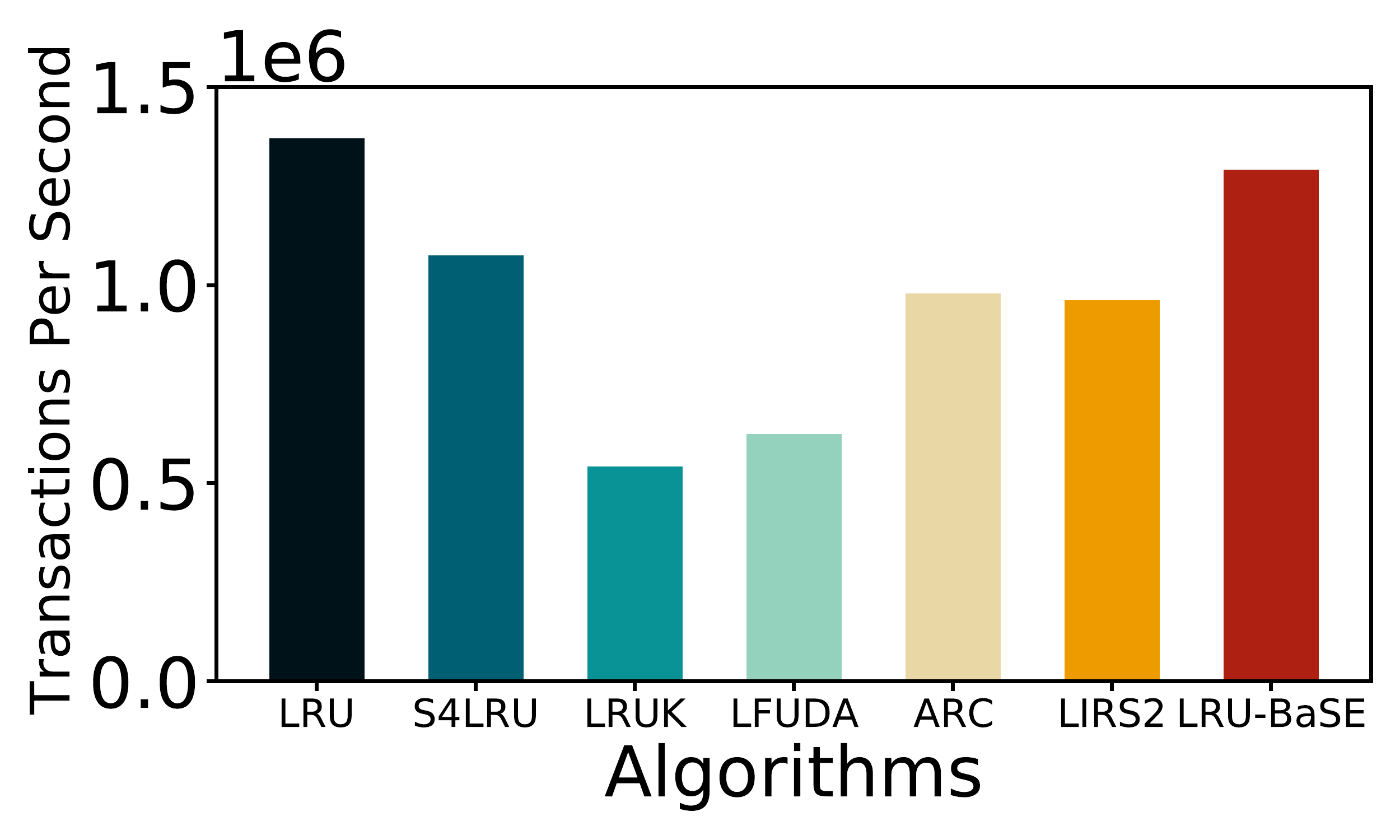}
	    \label{fig:wikipedia_throughput}
	}
	\subfigure[Wikipedia: Peak Memory]{
	    \includegraphics[width=0.185\linewidth]{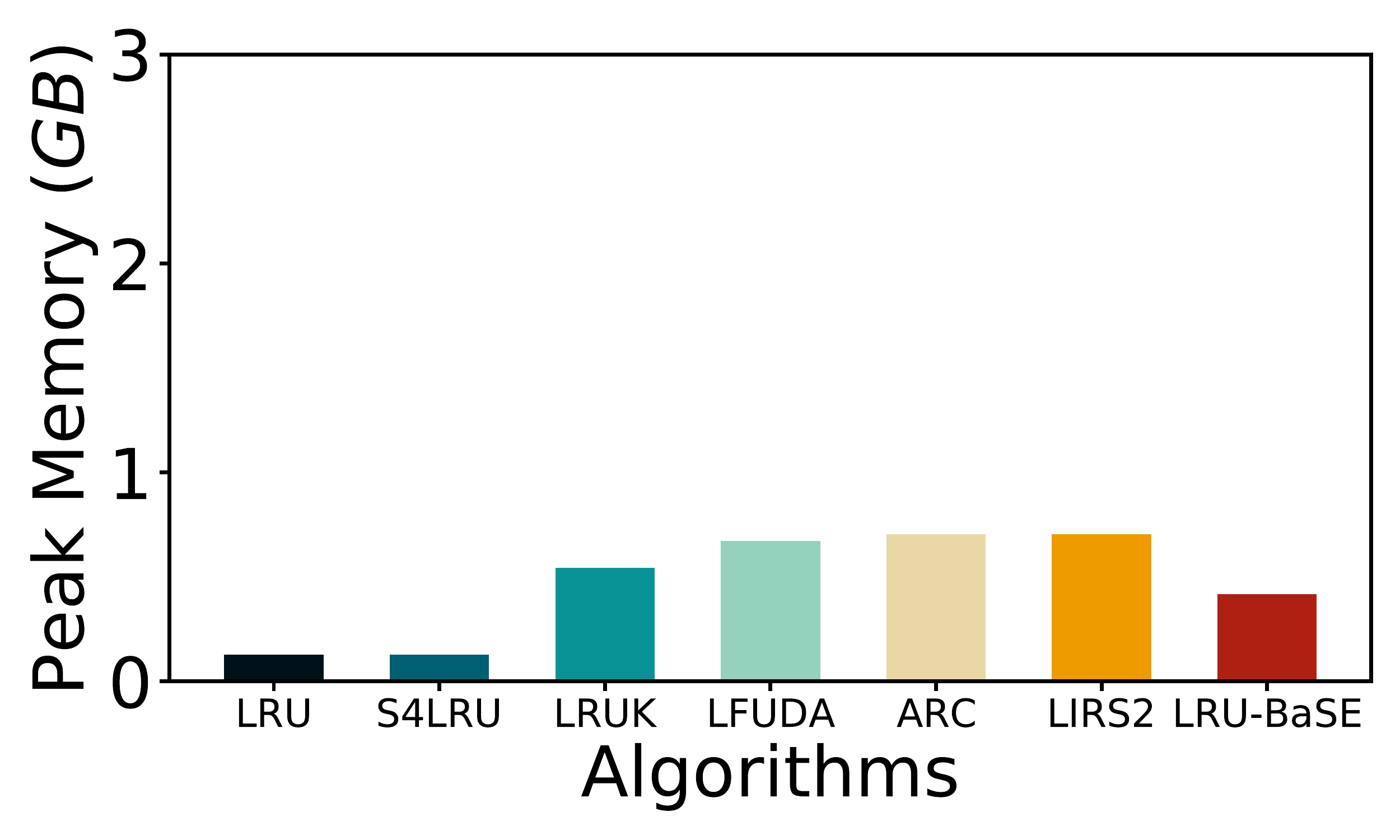}
	    \label{fig:wikipedia_memory}
	}
	\\
	\subfigure[CDN-Q: BMR]{
	    \includegraphics[width=0.185\linewidth]{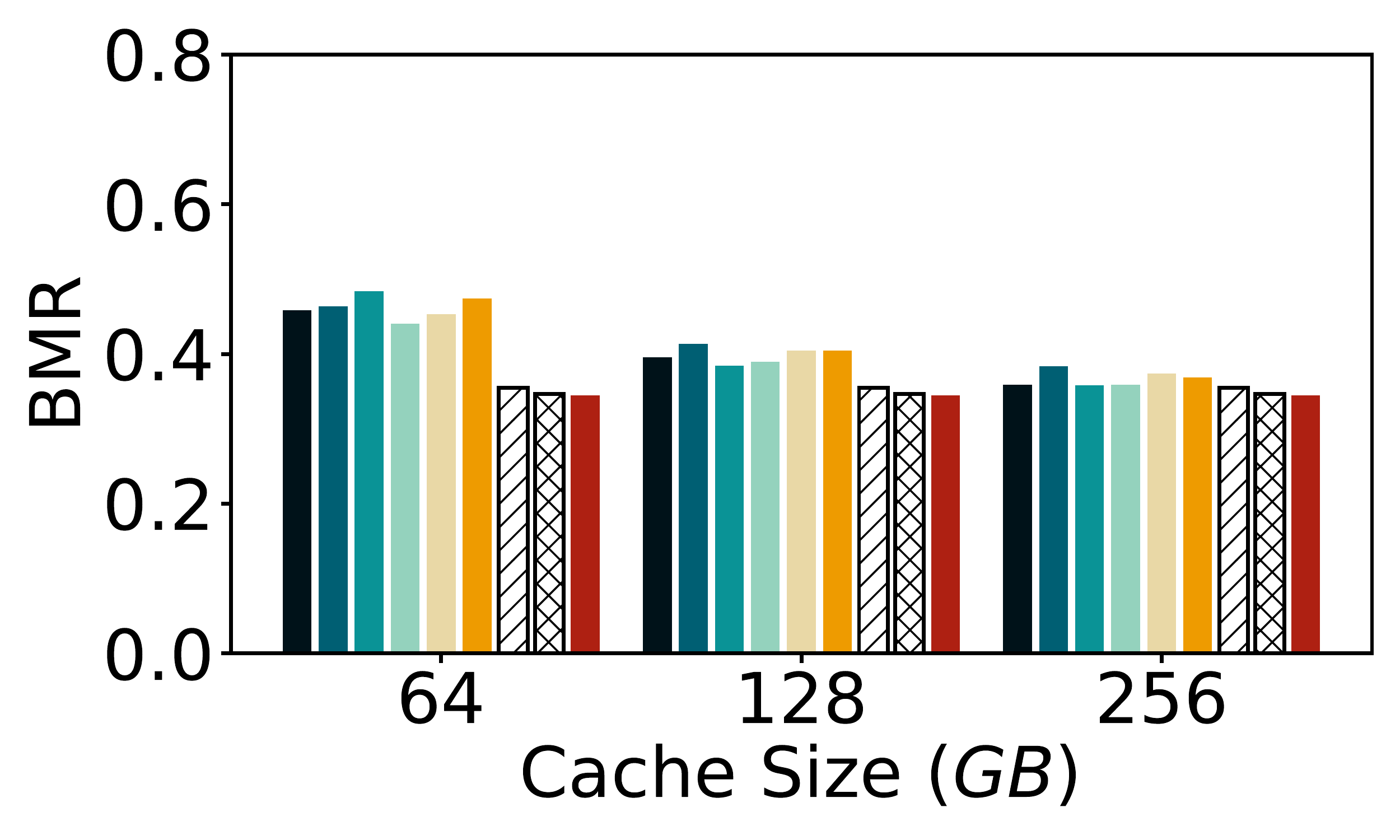}
	    \label{fig:cdn_q_bmr}
	}
	\subfigure[CDN-Q: OMR]{
	    \includegraphics[width=0.185\linewidth]{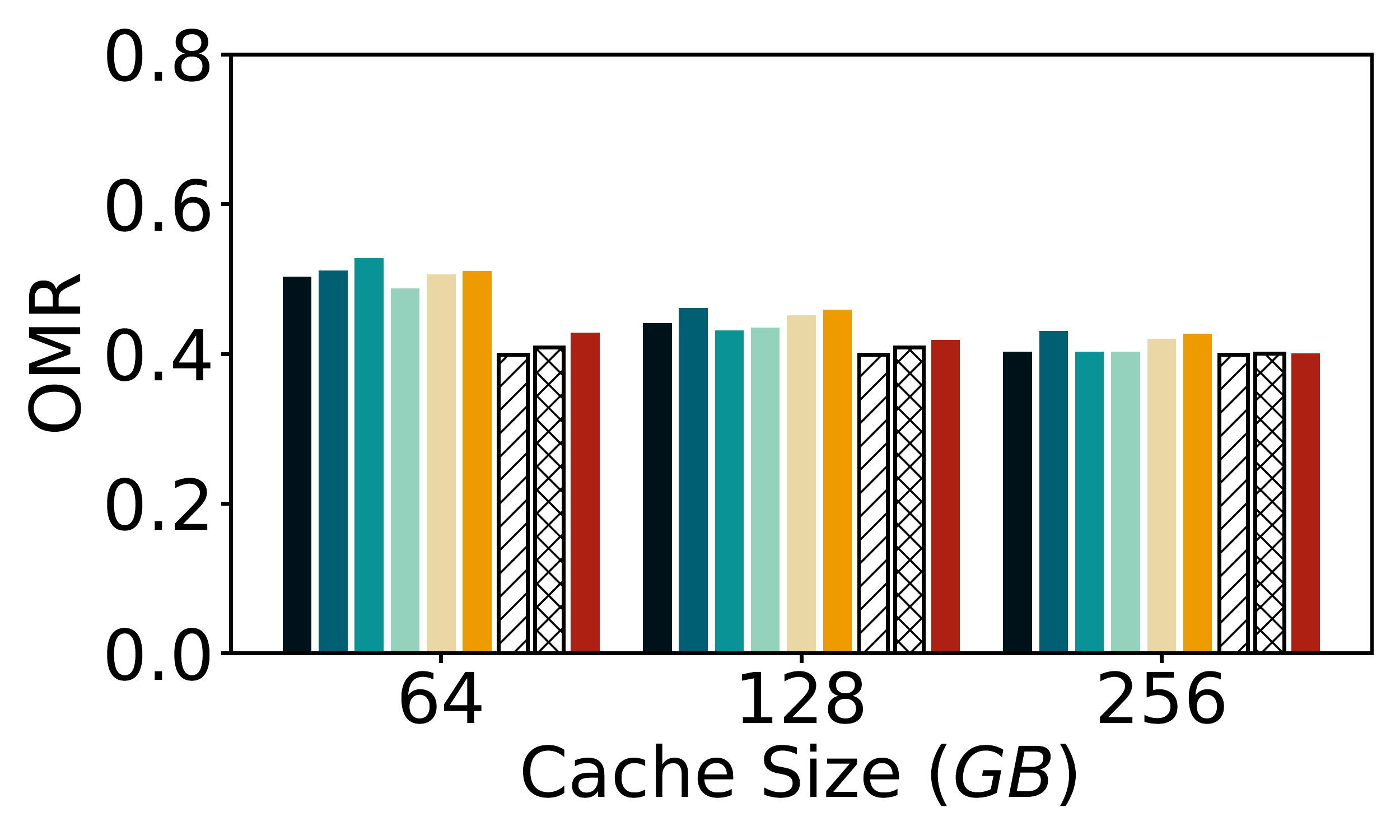}
	    \label{fig:cdn_q_omr}
	}
	\subfigure[CDN-Q: Peak CPU]{
	    \includegraphics[width=0.185\linewidth]{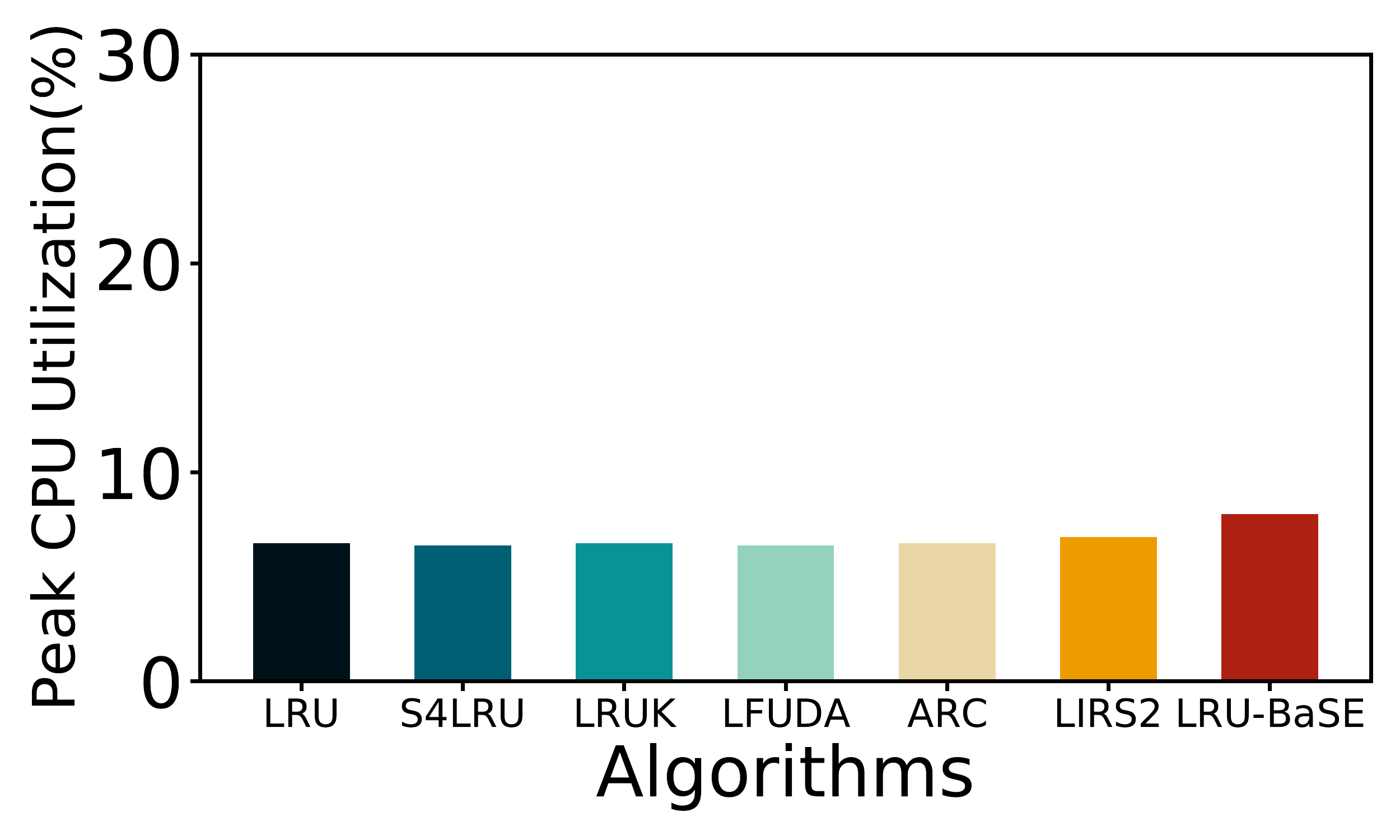}
	    \label{fig:cdn_q_cpu}
	}
	\subfigure[CDN-Q: TPS]{
	    \includegraphics[width=0.185\linewidth]{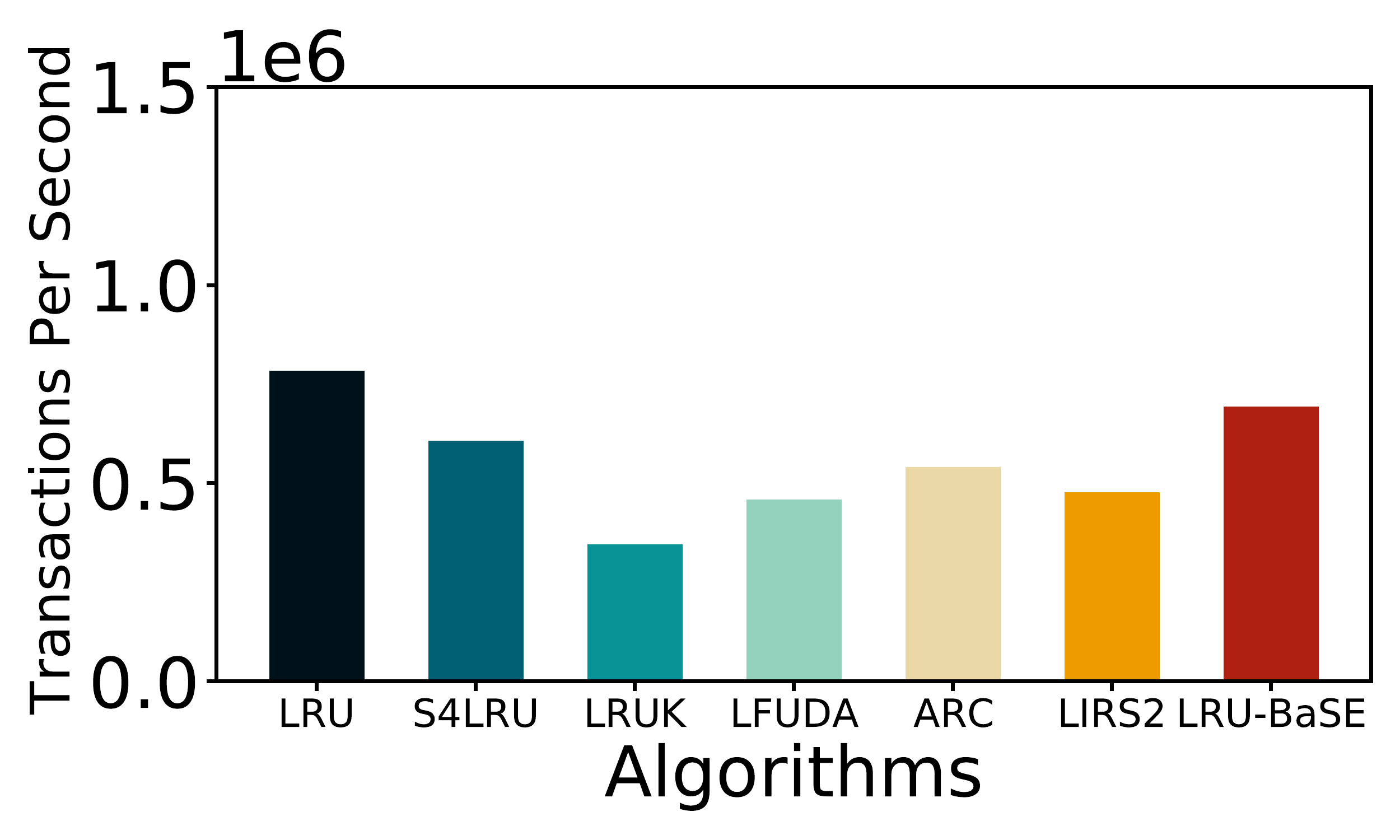}
	    \label{fig:cdn_q_throughput}
	}
	\subfigure[CDN-Q: Peak Memory]{
	    \includegraphics[width=0.185\linewidth]{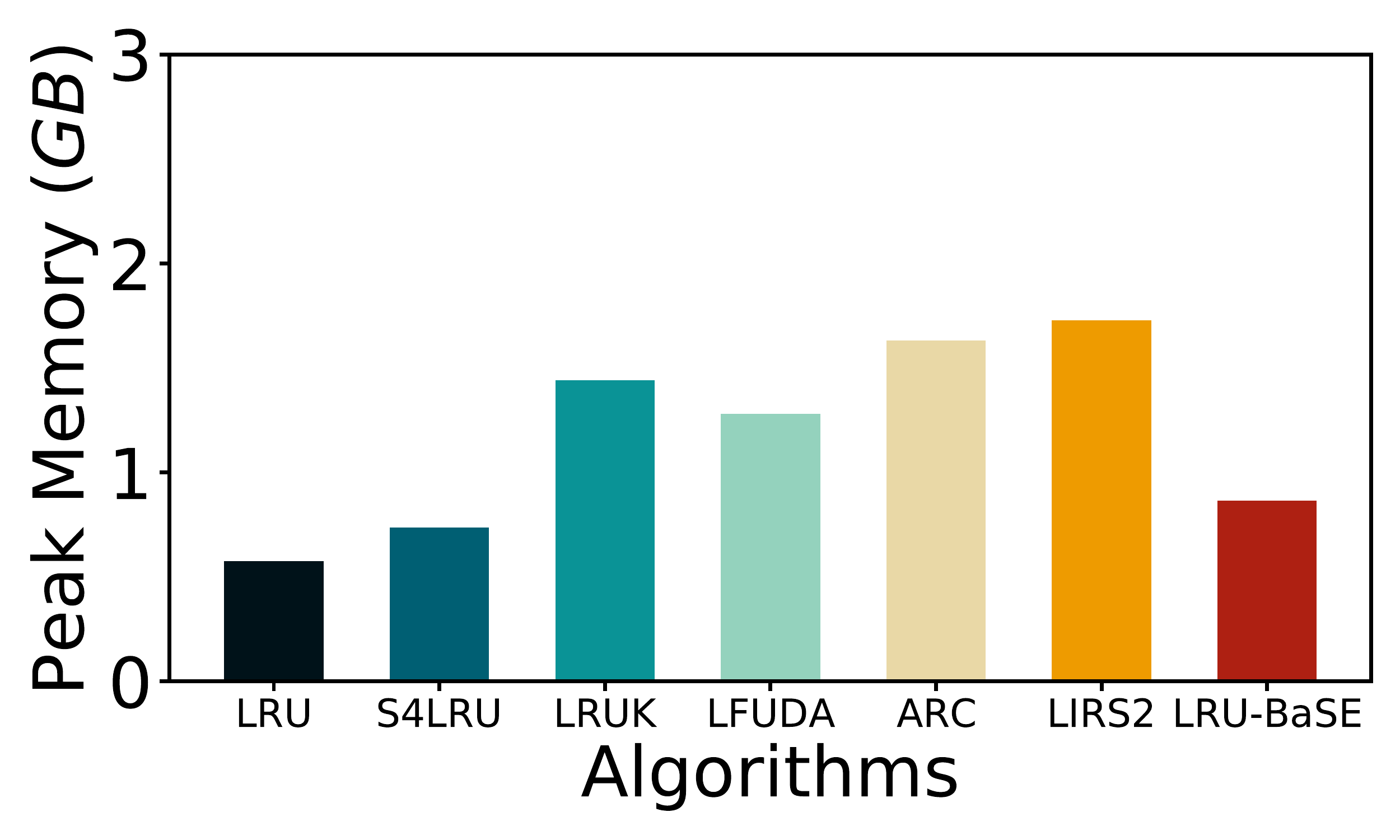}
	    \label{fig:cdn_q_memory}
	}
	\caption{Comparisons of Belady, PFOO, LRU-BaSE and six classical cache algorithms in terms of BMR, OMR, peak CPU utilization, transactions per second (TPS), and peak memory at different cache sizes on three traces.}
	\vspace{-0.2cm}
	\label{fig:performance_tro}
\end{figure*}

\subsection{Evaluation Methodology}
\label{sec:setttings}


\noindent \textbf{Traces and warmup traces}. Our evaluation uses CDN traces from three CDNs, two of which, Wikipedia and CDN-Q, are open source and the third, Trace-\textit{T} is captured from a real-world production system, as shown in Table~\ref{table:trace}. Each experiment below allows for a warmup period during which no metrics are recorded. The end of the warmup period is deﬁned by the time when BMR is stable. We list the number of warmup requests for different traces in Table~\ref{table:trace}.

\noindent- \textbf{Wikipedia}: A trace collected on a node, serving photos and other media content for Wikipedia pages~\cite{song2020learning}.

\noindent- \textbf{CDN-Q}: A open source trace collected from an instant messaging application~\cite{zhou2018demystifying}.

\noindent- \textbf{Trace-\textit{T}}: A trace collected from a real-world production system, serving a different mixture of photos, texts, and videos from many different content providers.


\noindent \textbf{State-of-the-art algorithms}. We compare LRU-BaSE with 16 state-of-the-art algorithms, including the classic, size-aware, and machine-learning-based cache algorithms.

\noindent- \textbf{Classical cache algorithms}: The classical cache algorithms include LRU, S4LRU~\cite{huang2013analysis}, ARC~\cite{megiddo2003arc}, LRUK~\cite{o1993lru}, LFUDA~\cite{arlitt2000evaluating}, and LIRS2~\cite{zhong2021lirs2}.

\noindent- \textbf{Size-aware cache algorithms}: The size-aware cache algorithms include ThLRU~\cite{einziger2017tinylfu}, ThS4LRU~\cite{einziger2017tinylfu}, GDSF~\cite{cherkasova2001role}, GDWheel~\cite{li2015gd}, LHD~\cite{beckmann2018lhd}, and AdaptSize~\cite{berger2017adaptsize}.

\noindent- \textbf{Machine learning (ML) based algorithms}: The cache algorithms based on machine learning include LRB~\cite{song2020learning}, LeCaR~\cite{vietri2018driving}, CACHEUS~\cite{rodriguez2021learning}, and UCB~\cite{costa2017mlcache}.






We use the \emph{results of Belady~\cite{belady1966study} and PFOO~\cite{berger2018practical} as the lower bound of OMR and BMR, respectively}. Although PFOO requires all information about traces in advance and does not always achieve optimal BMR, PFOO has been shown to achieve lower BMR than Belady and Belady-Size.





\noindent\textbf{Simulator and Testbed}. We use the DQN model~\cite{dqn_code} to implement LRU-BaSE on the LRB simulator~\cite{lrb_code}. The execution of LRU-BaSE is divided into three parts. We first create a cache environment with a process consisting of parameter initialization, attribute replacement, and reward function feedback. Second, the DQN model deploys the network structure, the learning function, and the replacement interface. Finally, data processing (\eg, the rear section and time regions) and training are implemented. The simulator is running on a device with a 56-core 2.40GHz CPU, 32GB of RAM, and a 6TB SSD. In addition, we use the simulator to integrate the aforementioned state-of-the-art algorithms for fair comparison.

\begin{figure*}[t]
	\centering
	\includegraphics[width=0.5\linewidth]{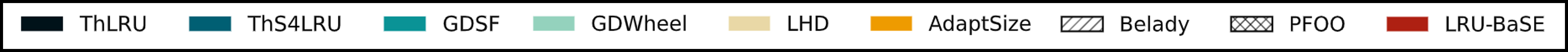}
	\\
	\subfigure[Trace-\textit{T}: BMR]{
	    \includegraphics[width=0.185\linewidth]{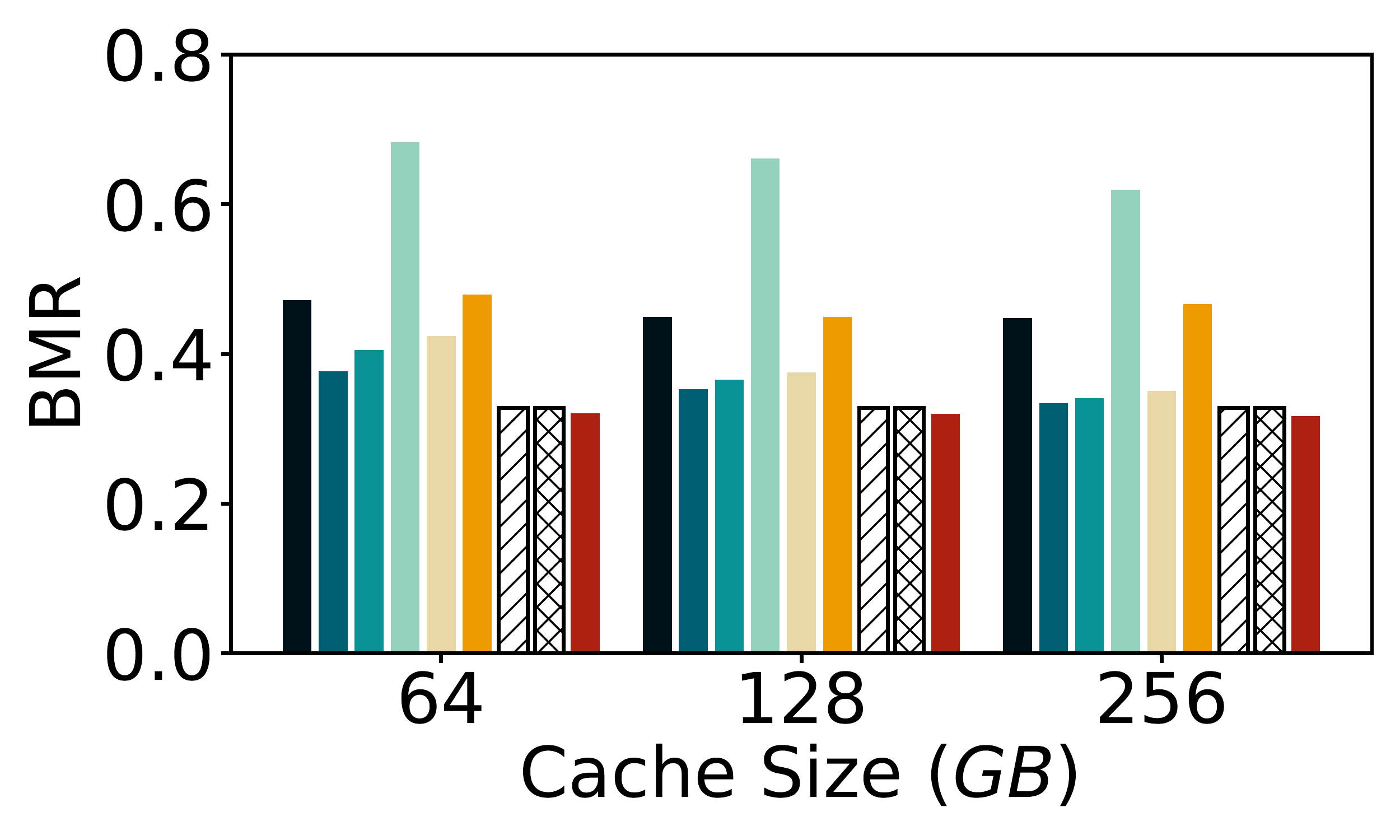}
	    \label{fig:trace_t_bmr2}
	}
	\subfigure[Trace-\textit{T}: OMR]{
	    \includegraphics[width=0.185\linewidth]{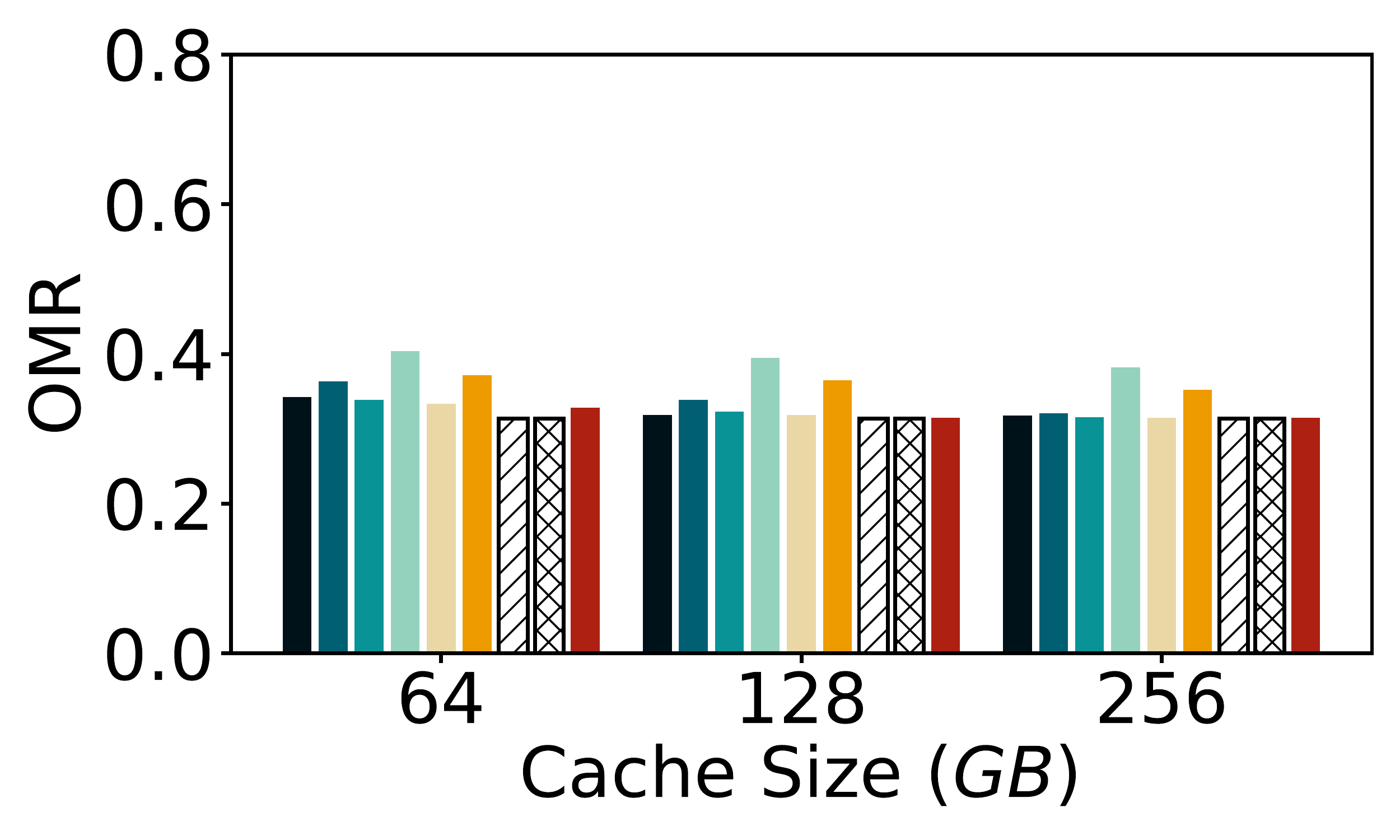}
	    \label{fig:trace_t_omr2}
	}
	\subfigure[Trace-\textit{T}: CPU]{
	    \includegraphics[width=0.185\linewidth]{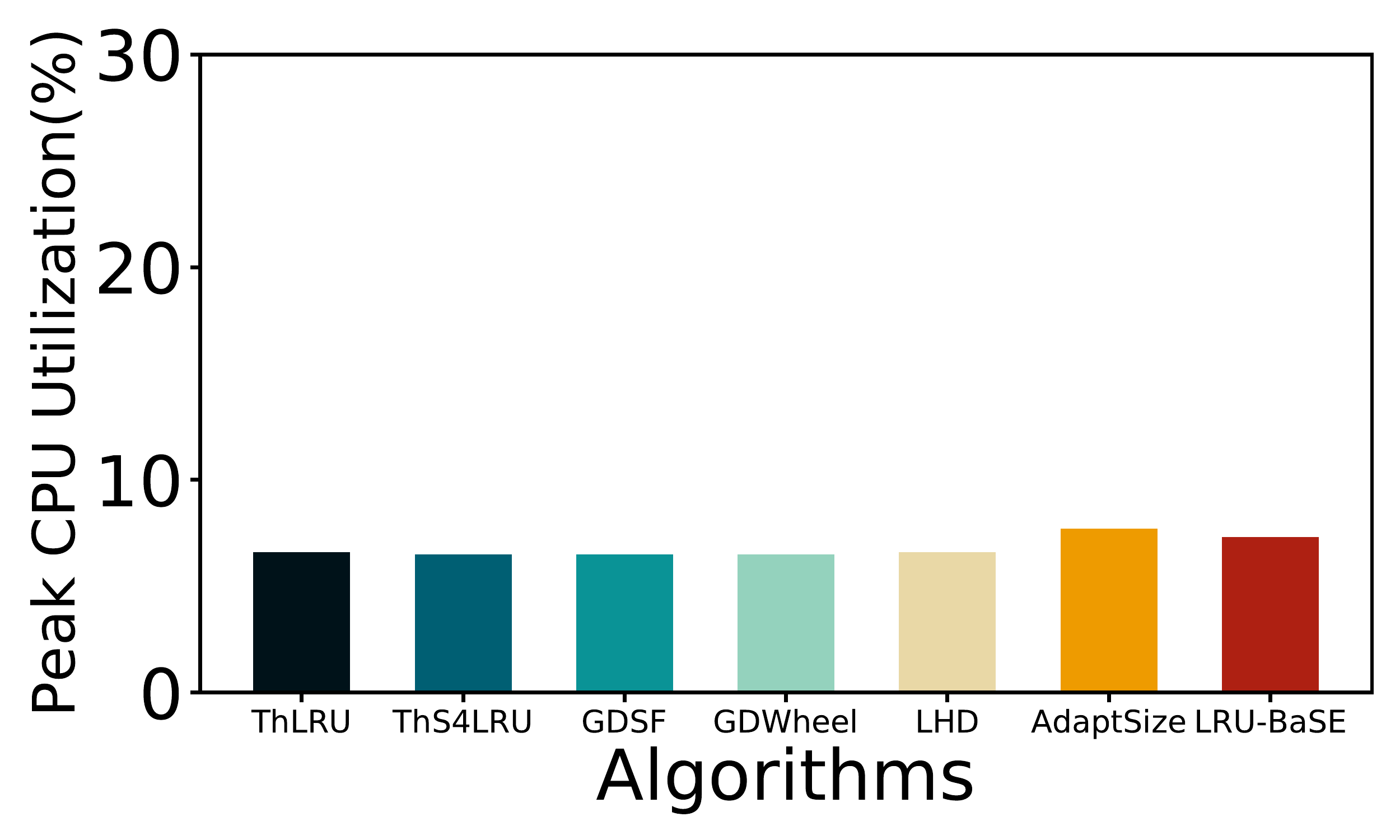}
	    \label{fig:trace_t_cpu2}
	}
	\subfigure[Trace-\textit{T}: TPS]{
	    \includegraphics[width=0.185\linewidth]{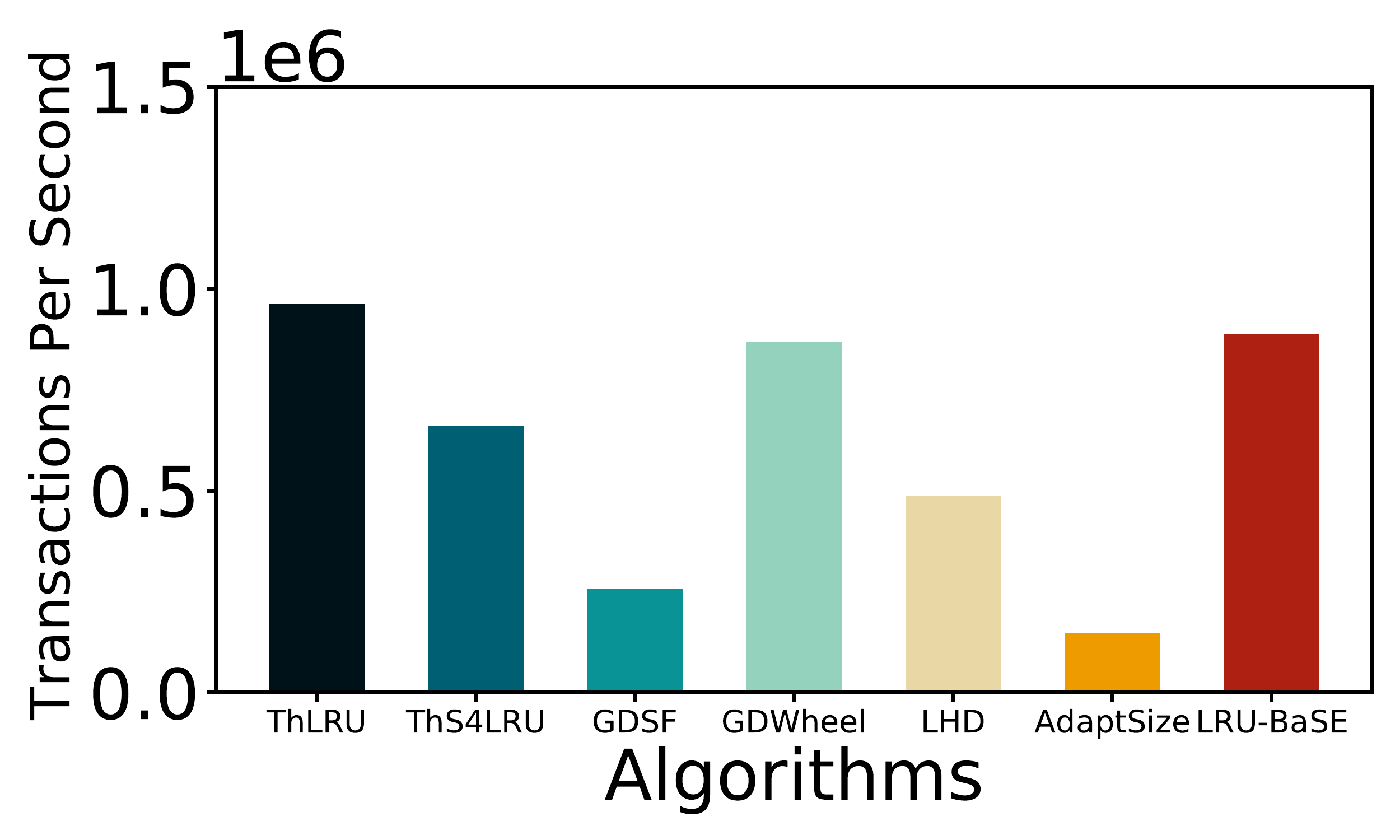}
	    \label{fig:trace_t_throughput2}
	}
	\subfigure[Trace-\textit{T}: Memory]{
	    \includegraphics[width=0.185\linewidth]{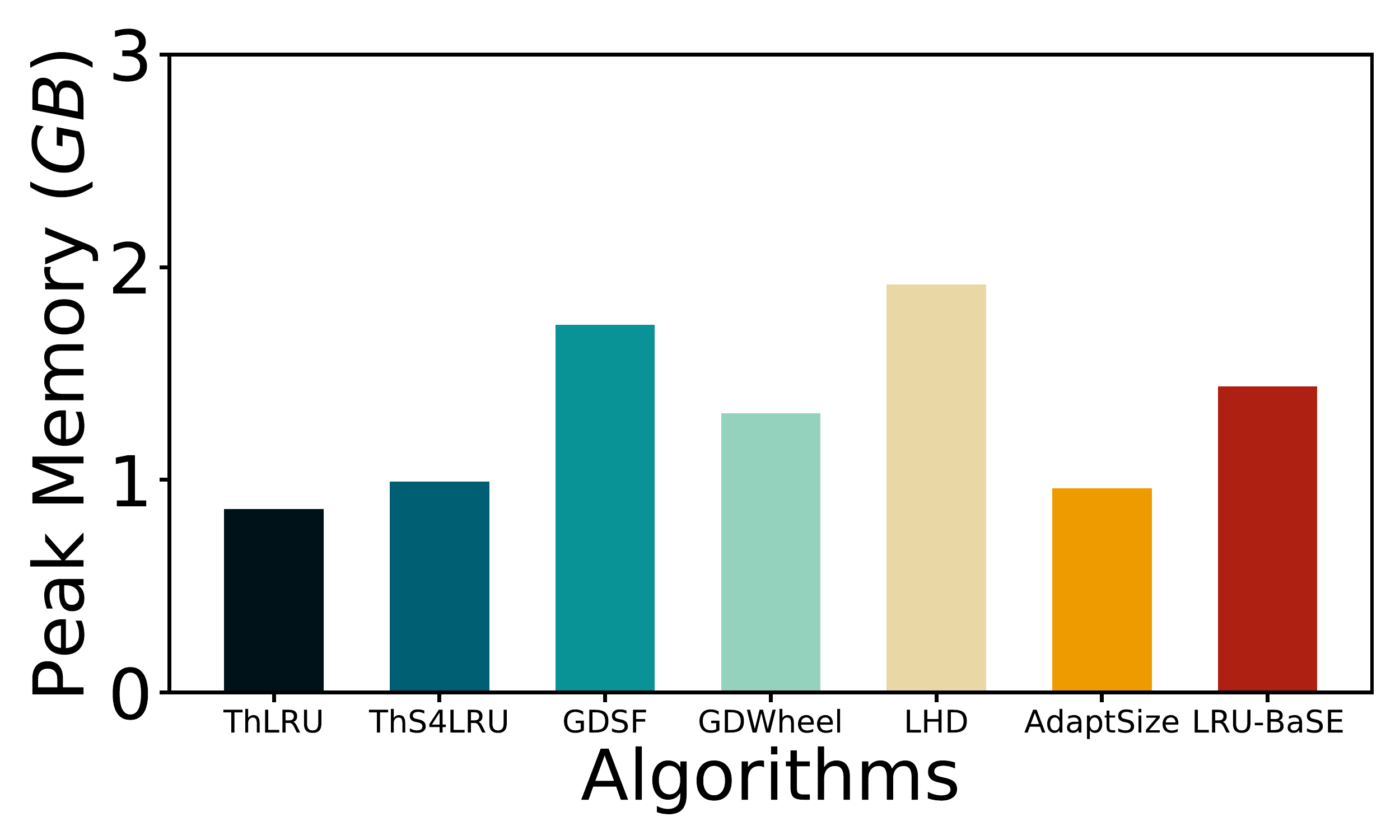}
	    \label{fig:trace_t_memory2}
	}
	\\
	\subfigure[Wikipedia: BMR]{
	    \includegraphics[width=0.185\linewidth]{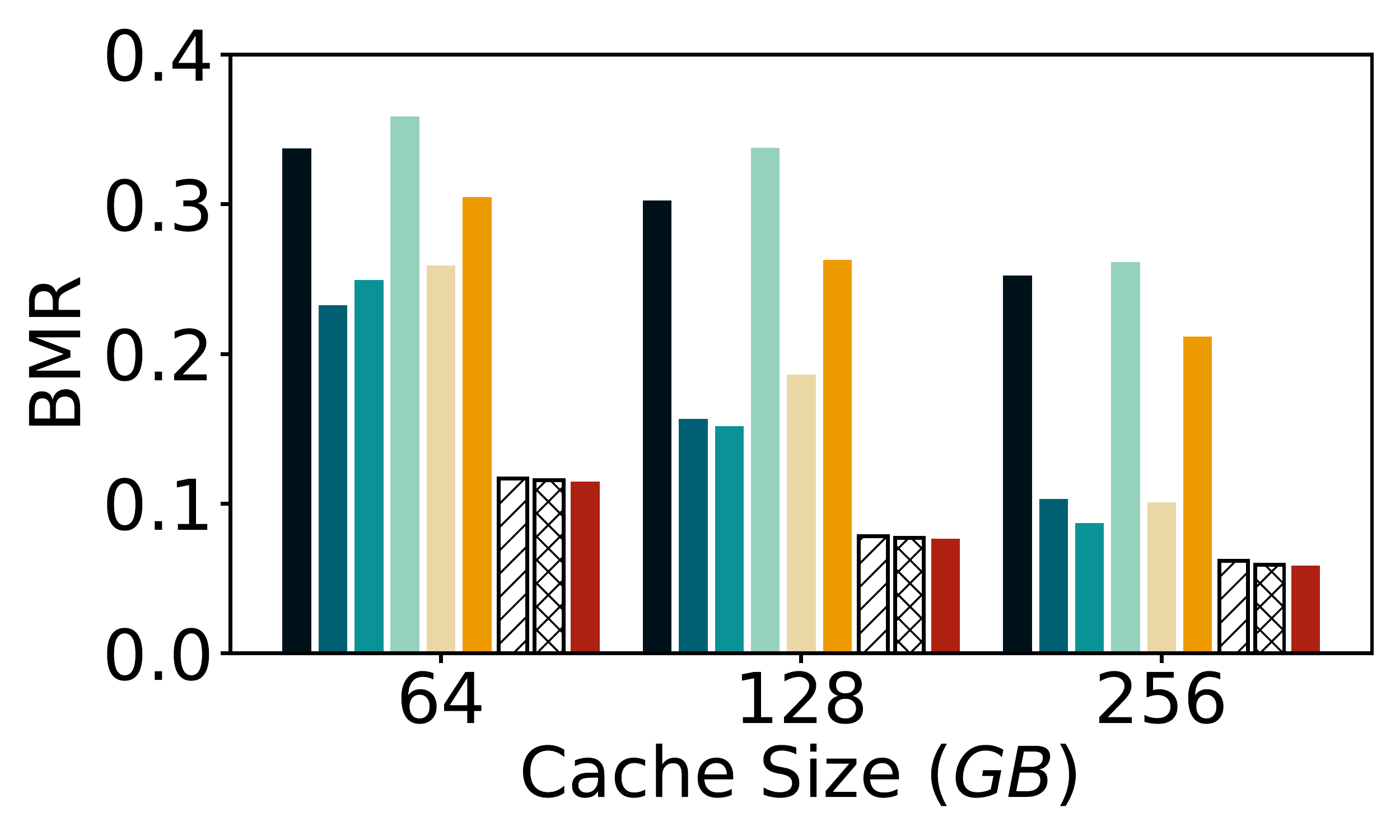}
	    \label{fig:wikipedia_bmr2}
	}
	\subfigure[Wikipedia: OMR]{
	    \includegraphics[width=0.185\linewidth]{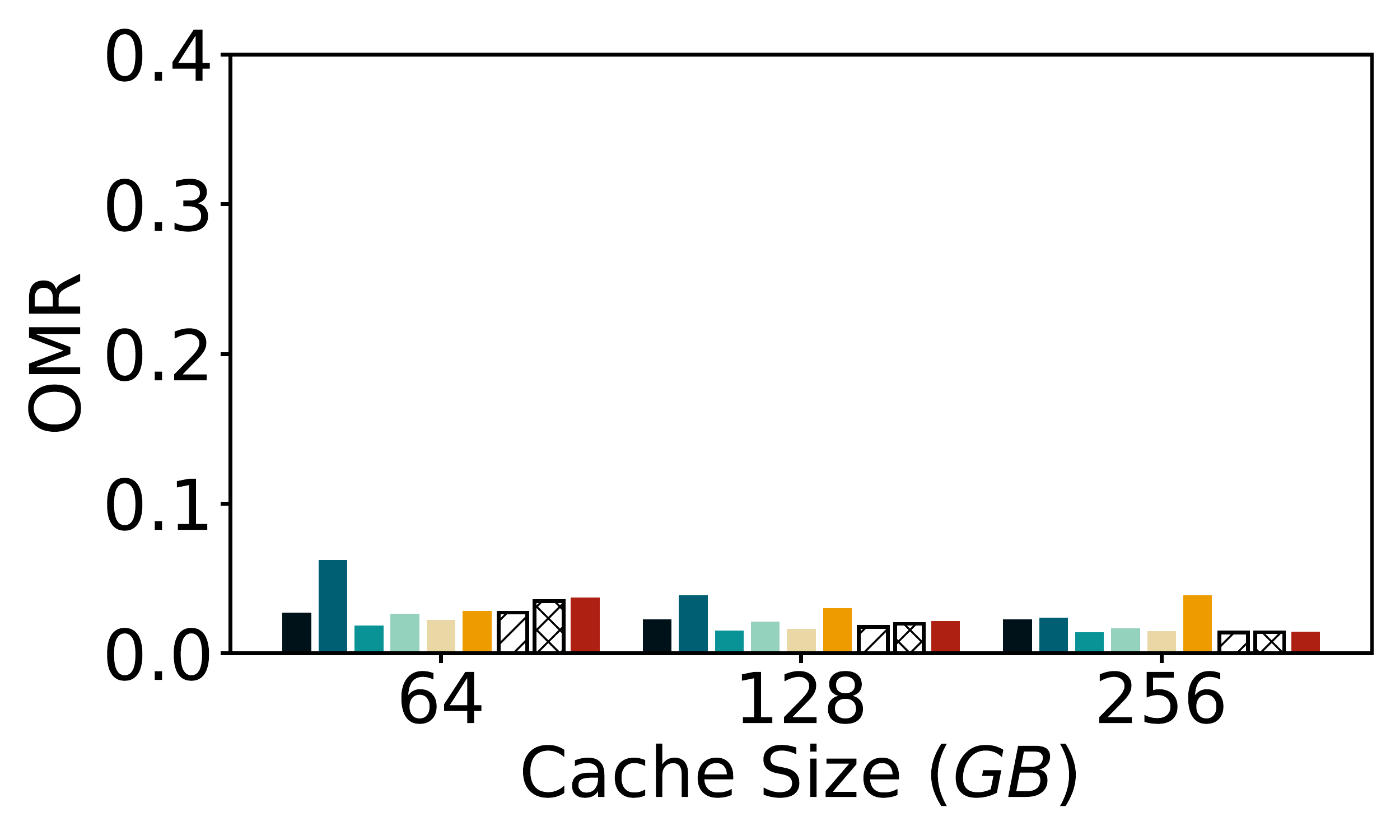}
	    \label{fig:wikipedia_omr2}
	}
	\subfigure[Wikipedia: CPU]{
	    \includegraphics[width=0.185\linewidth]{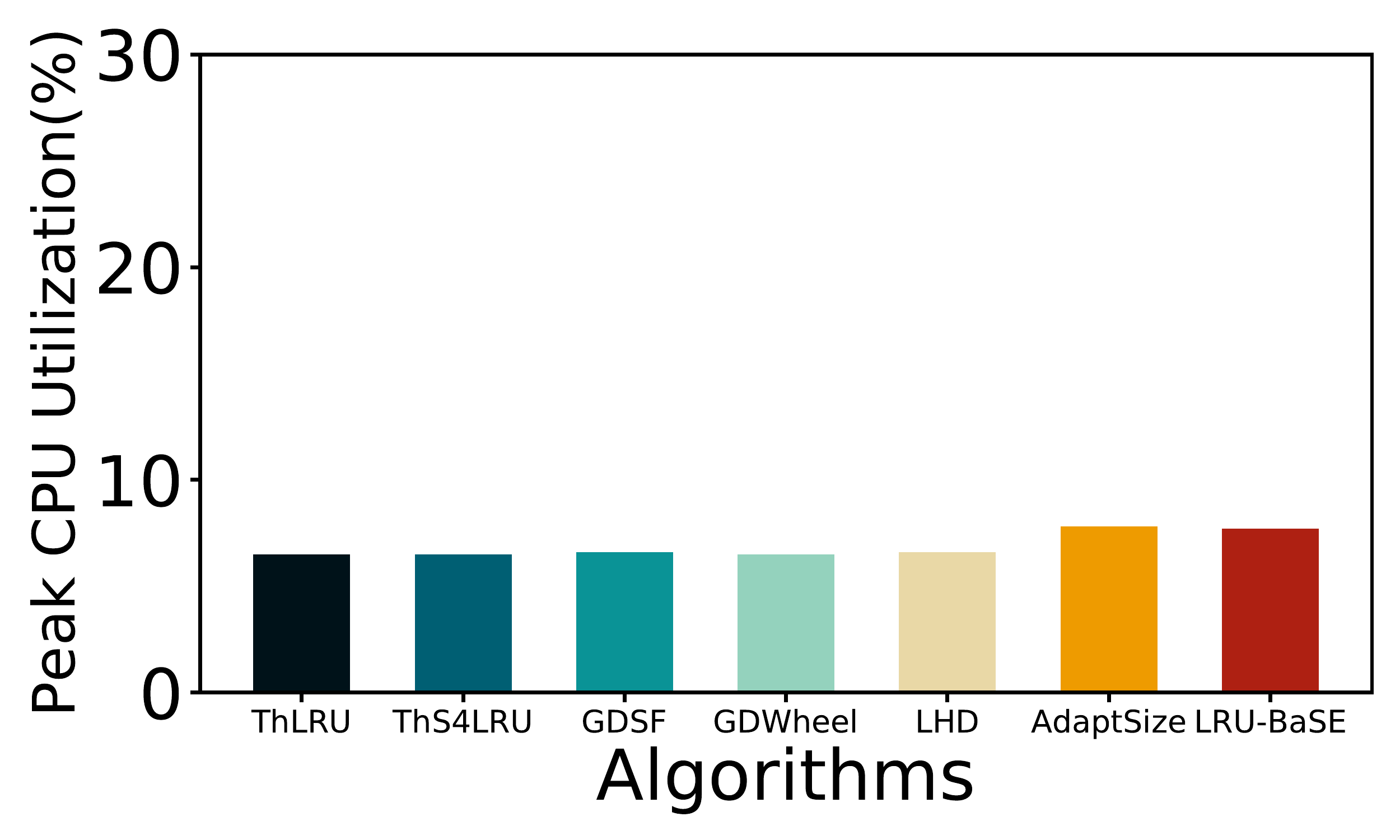}
	    \label{fig:wikipedia_cpu2}
	}
	\subfigure[Wikipedia: TPS]{
	    \includegraphics[width=0.185\linewidth]{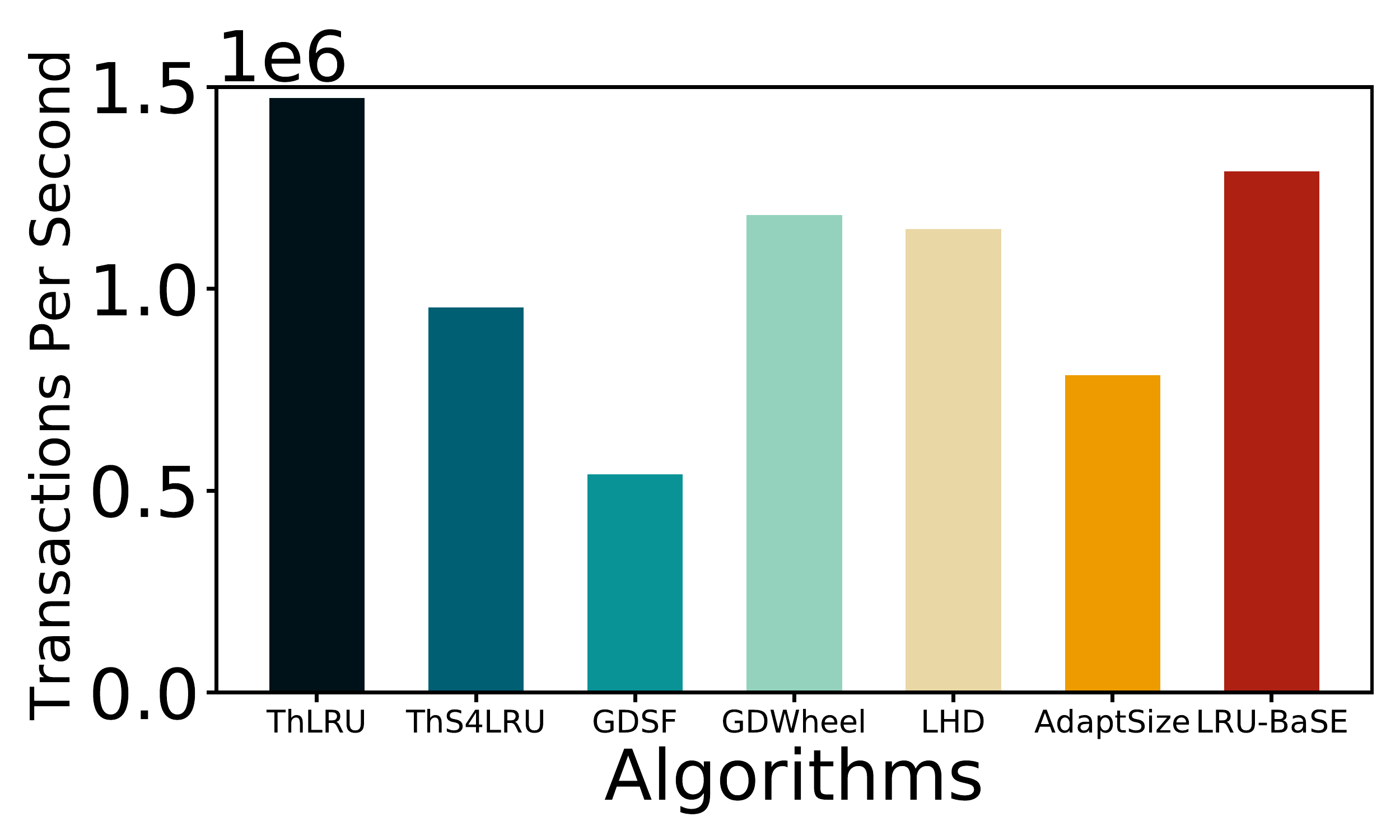}
	    \label{fig:wikipedia_throughput2}
	}
	\subfigure[Wikipedia: Memory]{
	    \includegraphics[width=0.185\linewidth]{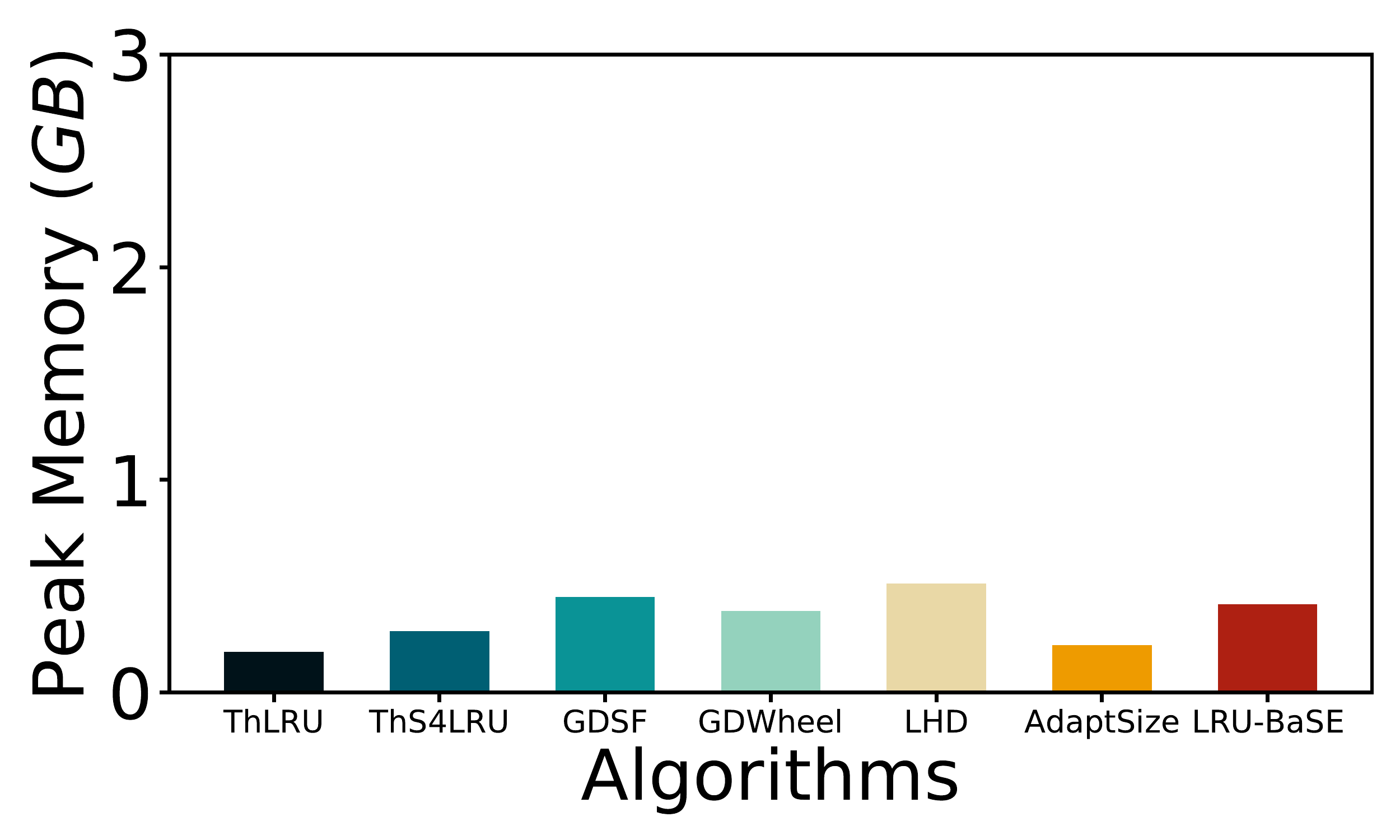}
	    \label{fig:wikipedia_memory2}
	}
	\\
	\subfigure[CDN-Q: BMR]{
	    \includegraphics[width=0.185\linewidth]{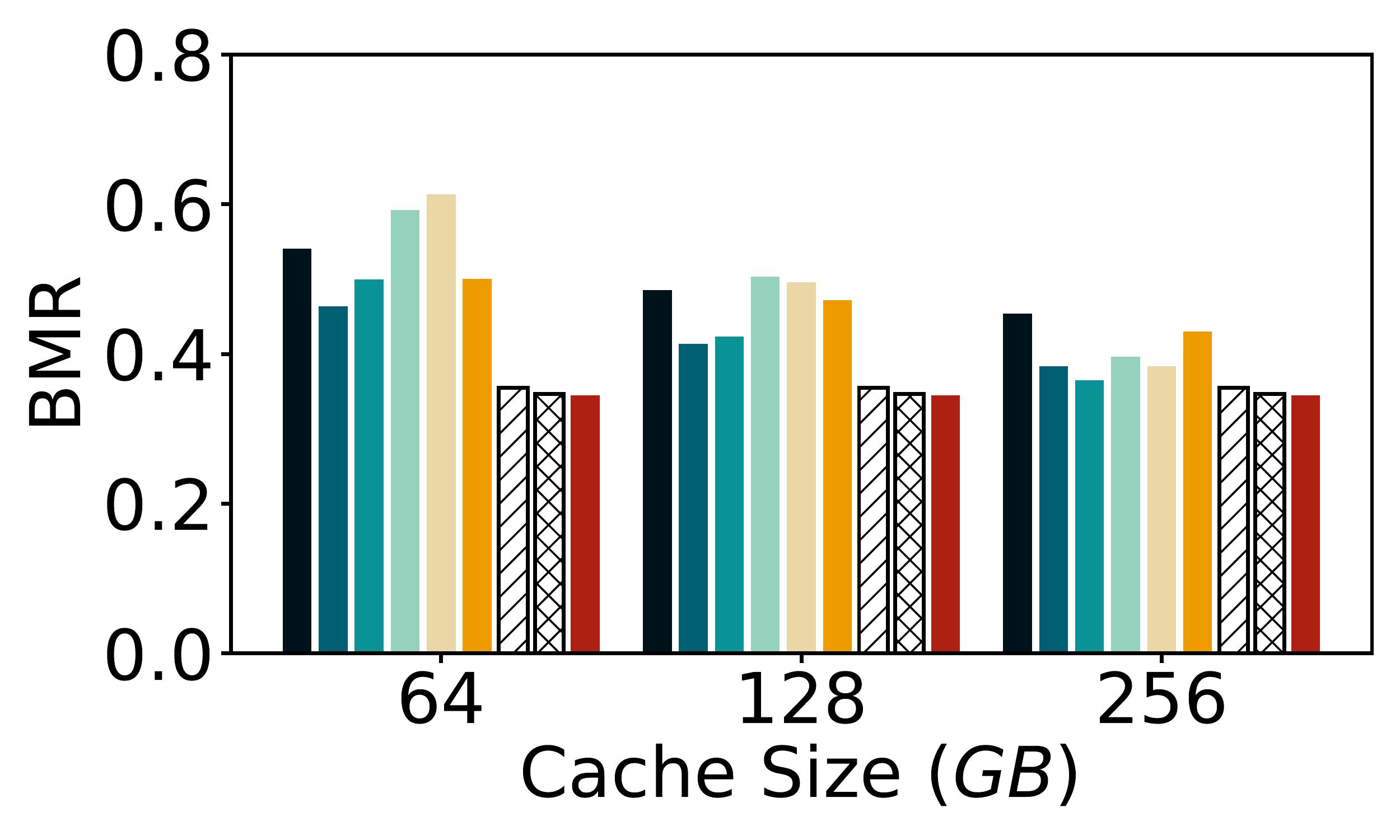}
	    \label{fig:cdn_q_bmr2}
	}
	\subfigure[CDN-Q: OMR]{
	    \includegraphics[width=0.185\linewidth]{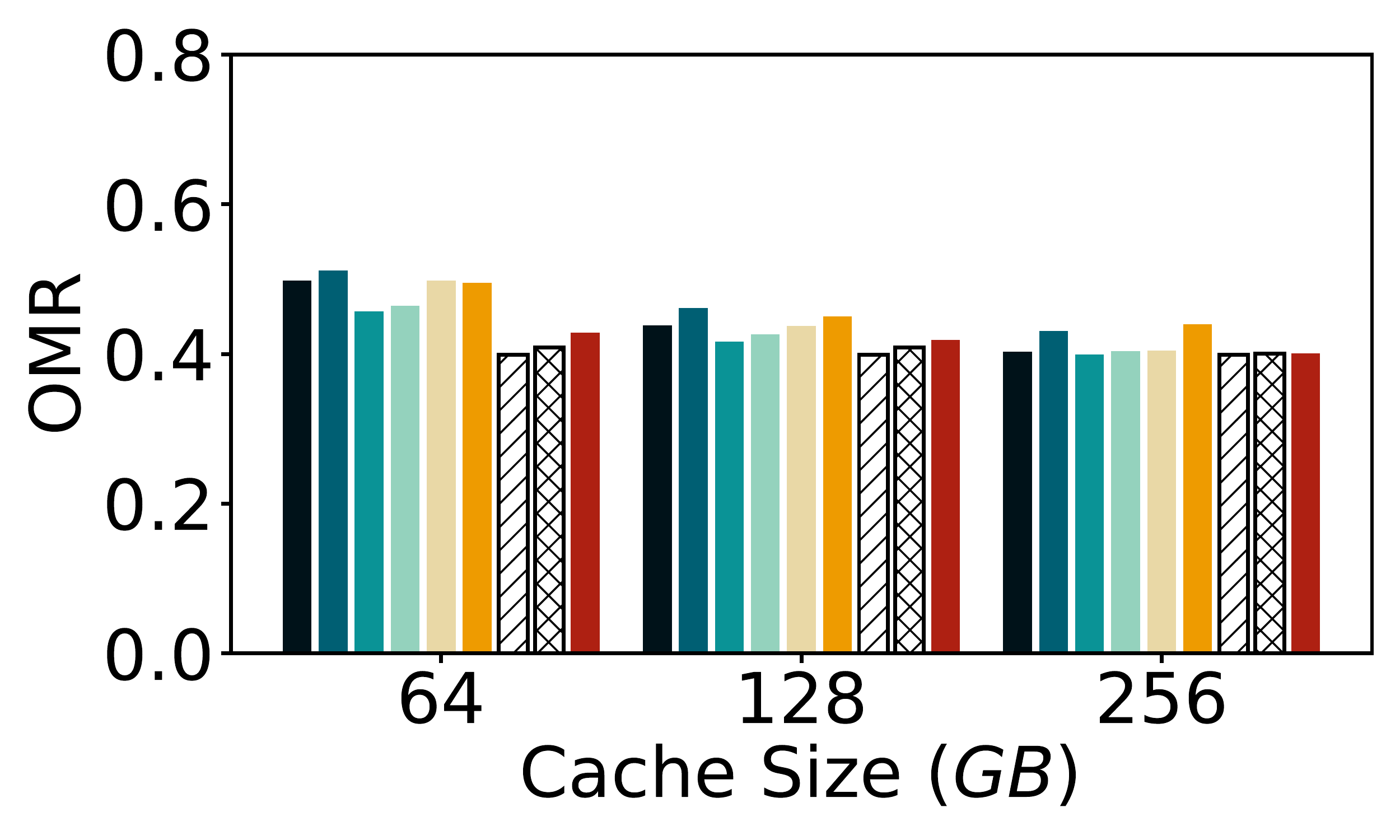}
	    \label{fig:cdn_q_omr2}
	}
	\subfigure[CDN-Q: CPU]{
	    \includegraphics[width=0.185\linewidth]{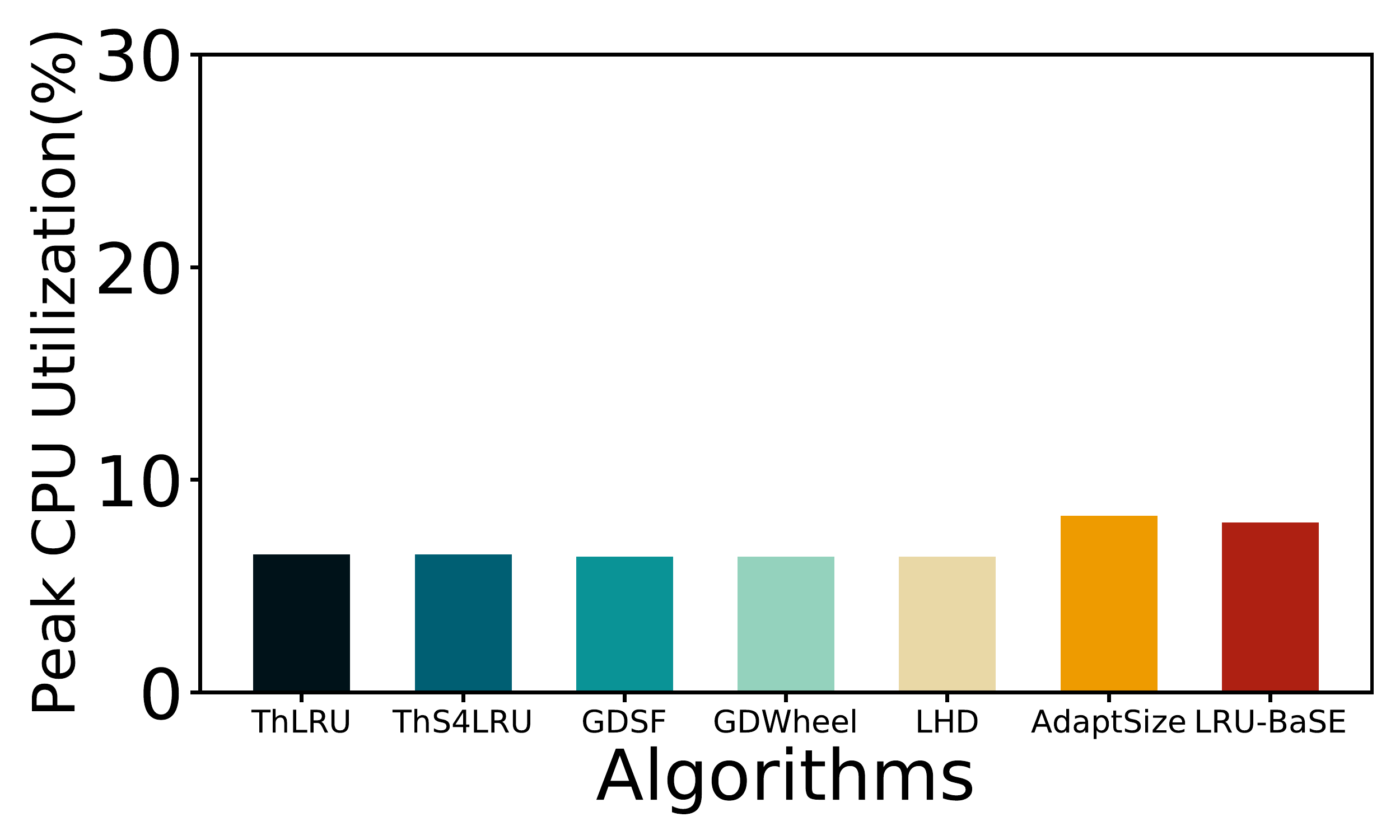}
	    \label{fig:cdn_q_cpu2}
	}
	\subfigure[CDN-Q: TPS]{
	    \includegraphics[width=0.185\linewidth]{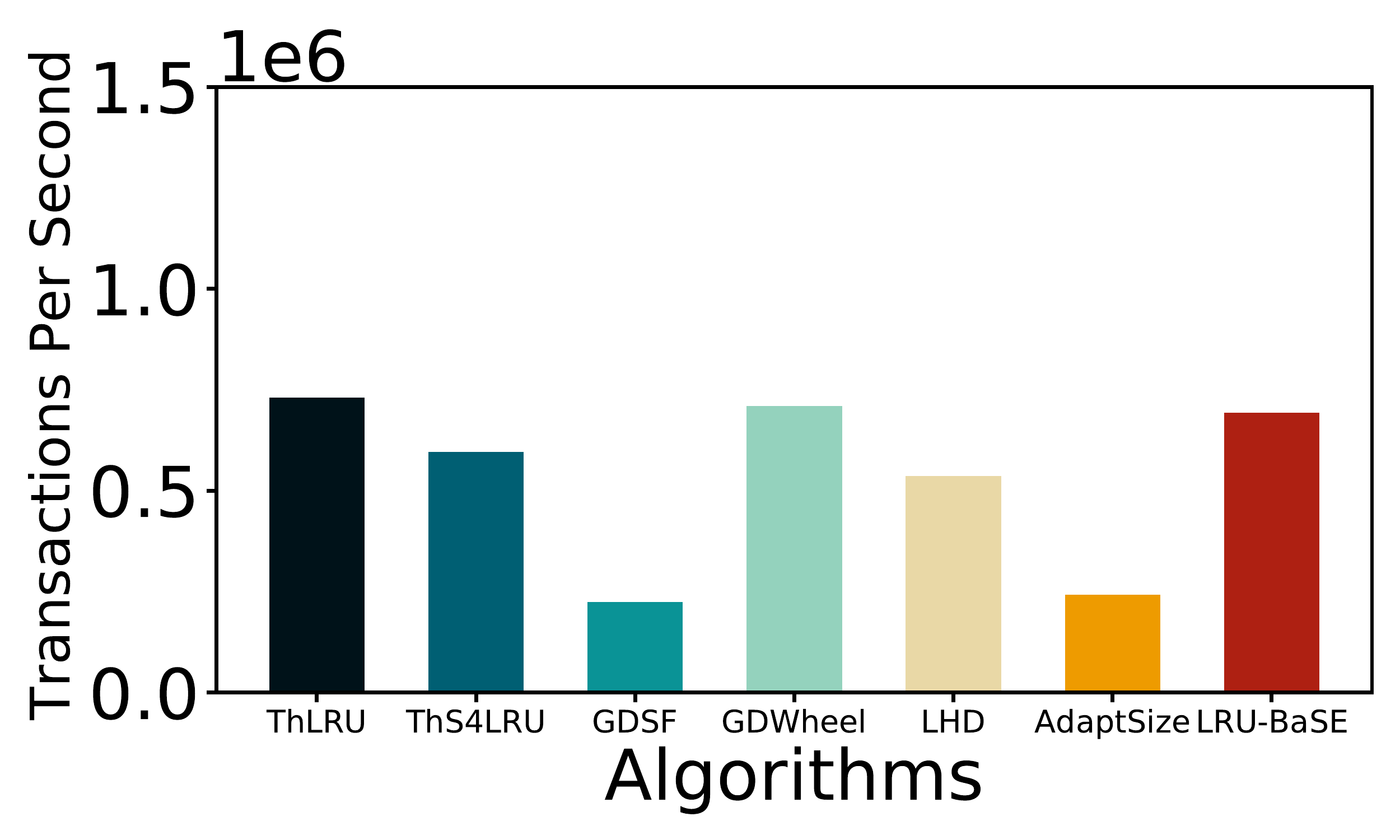}
	    \label{fig:cdn_q_throughput2}
	}
	\subfigure[CDN-Q: Memory]{
	    \includegraphics[width=0.185\linewidth]{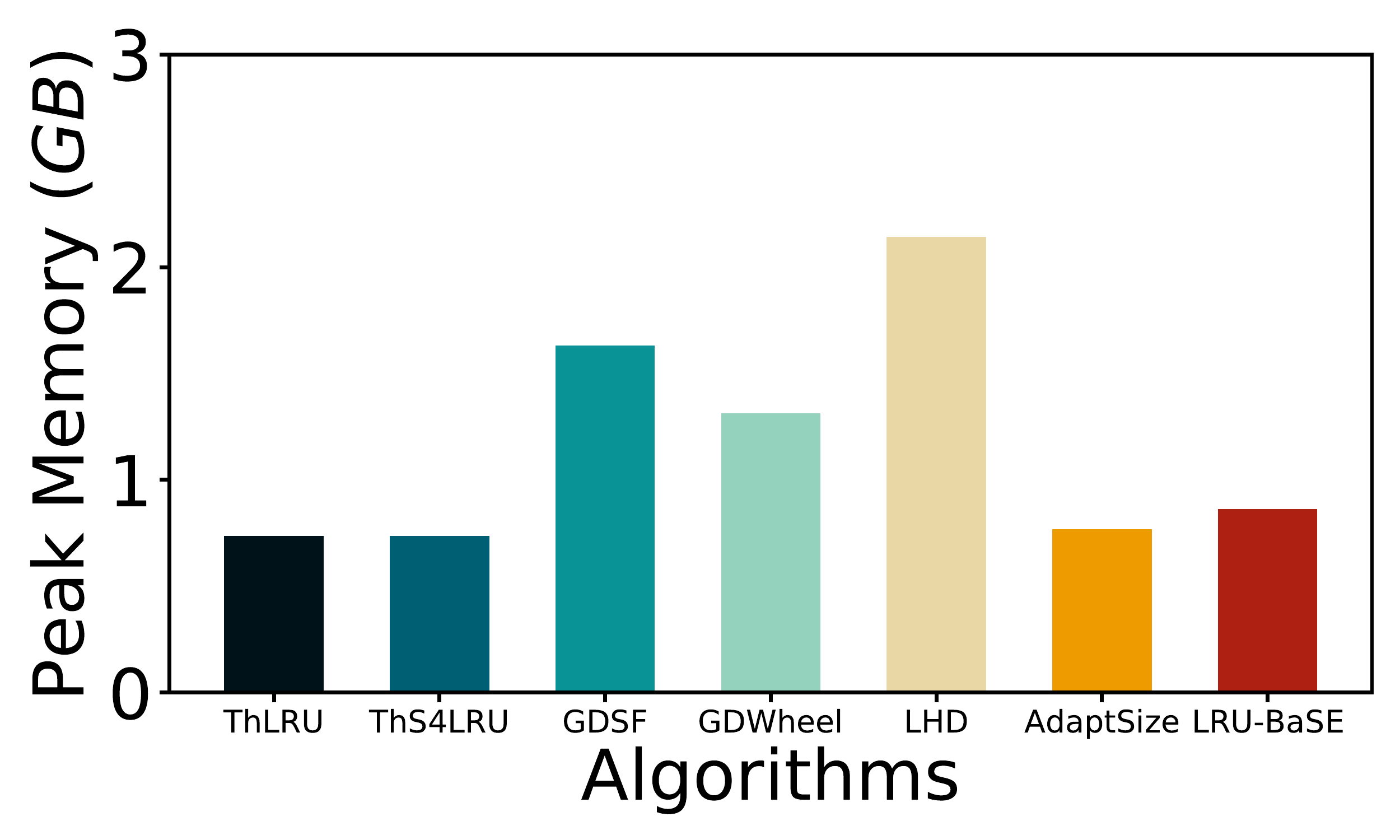}
	    \label{fig:cdn_q_memory2}
	}
	\caption{Comparisons of Belady, PFOO, LRU-BaSE and six size-aware cache algorithms in terms of BMR, OMR, peak CPU utilization, transactions per second (TPS), and peak memory at different cache sizes on three traces.}
	\vspace{-0.2cm}
	\label{fig:performance_pro2}
\end{figure*}

\subsection{LRU-BaSE vs. state-of-the-art algorithms}

\begin{figure*}[t]
	\centering
	\includegraphics[width=0.4\linewidth]{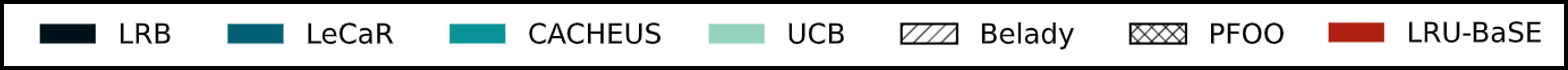}
	\\
	\subfigure[Trace-\textit{T}: BMR]{
	    \includegraphics[width=0.15\linewidth]{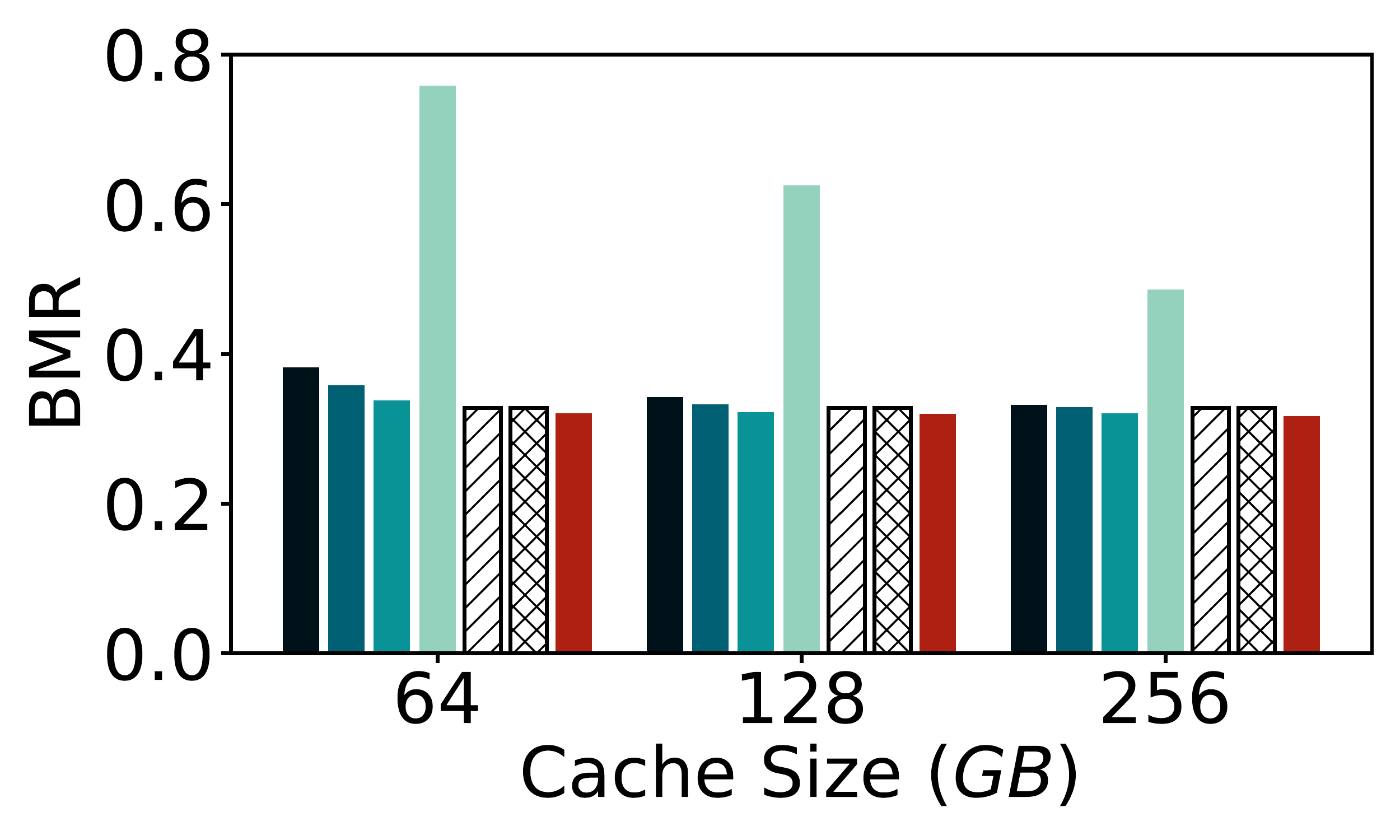}
	    \label{fig:trace_t_bmr3}
	}
	\subfigure[Trace-\textit{T}: OMR]{
	    \includegraphics[width=0.15\linewidth]{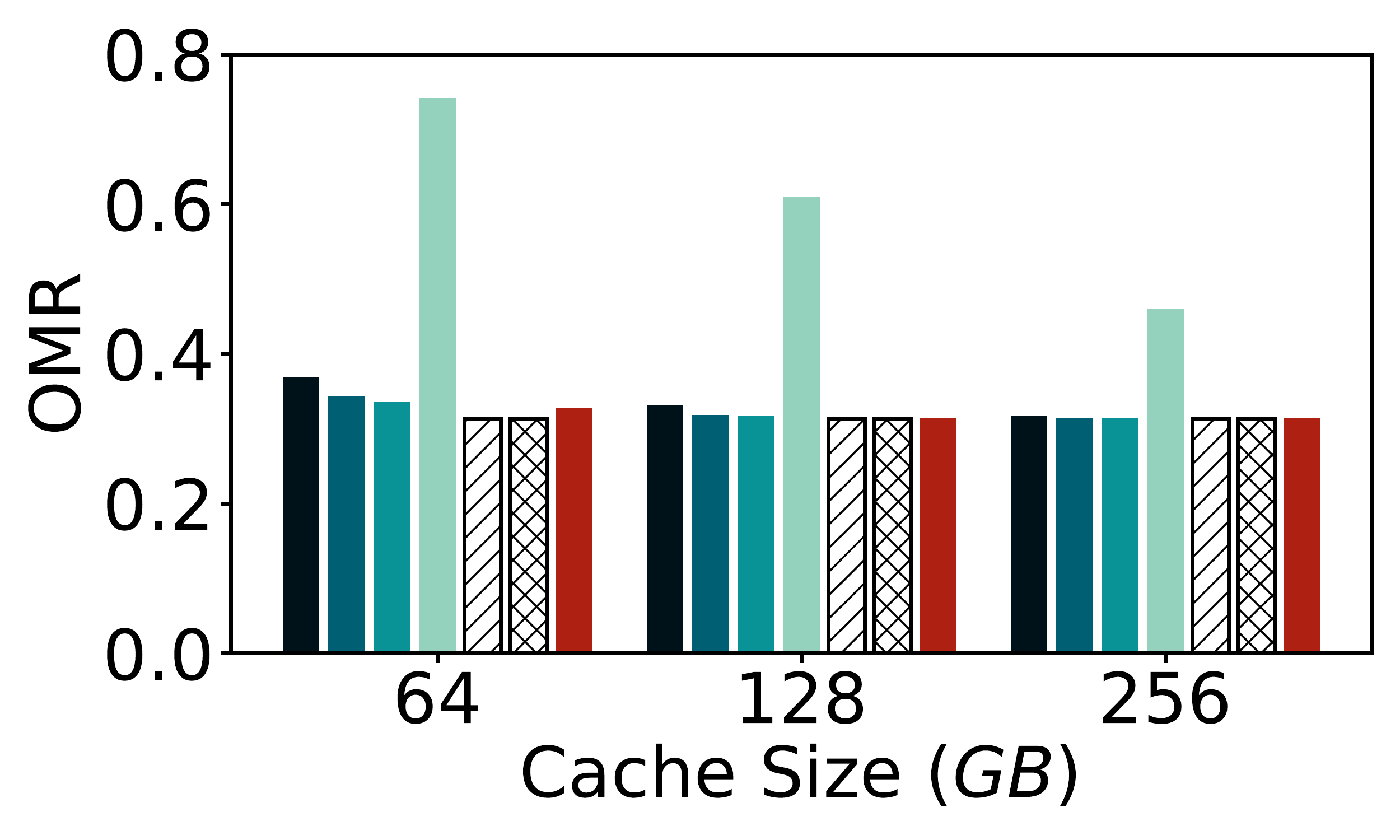}
	    \label{fig:trace_t_omr3}
	}
	\subfigure[Trace-\textit{T}: CPU]{
	    \includegraphics[width=0.15\linewidth]{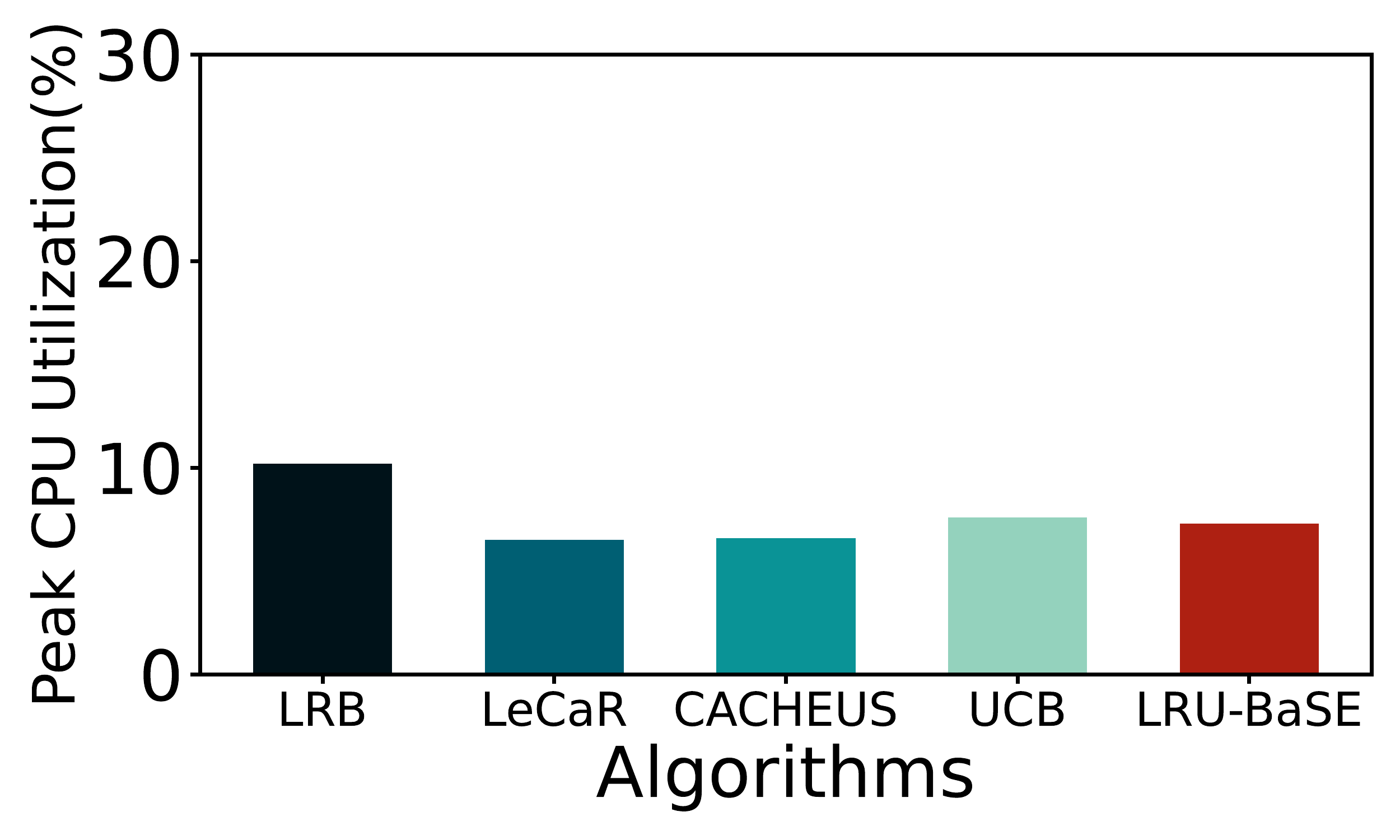}
	    \label{fig:trace_t_cpu3}
	}
	\subfigure[Trace-\textit{T}: TPS]{
	    \includegraphics[width=0.15\linewidth]{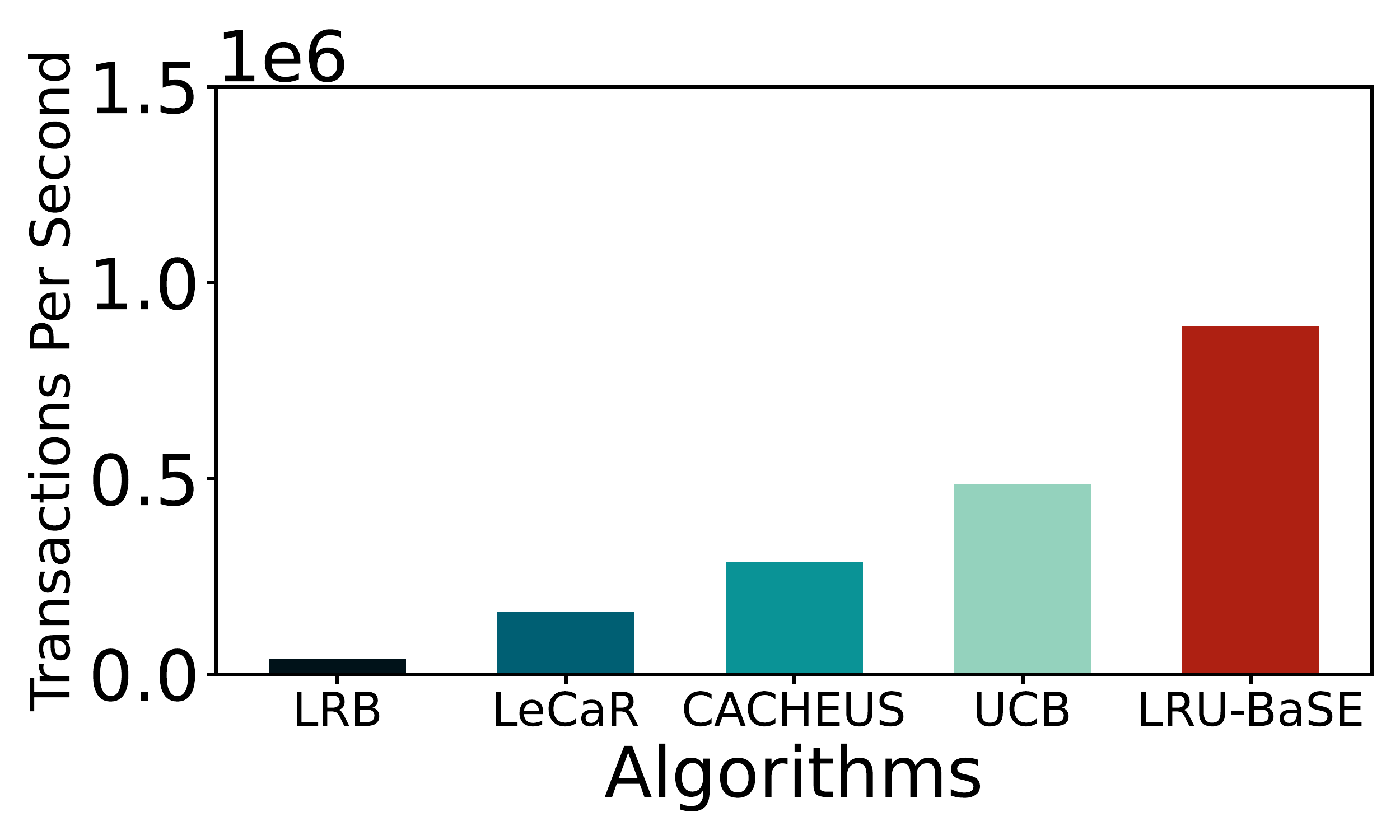}
	    \label{fig:trace_throughput3}
	}
	\subfigure[Trace-\textit{T}: Memory]{
	    \includegraphics[width=0.15\linewidth]{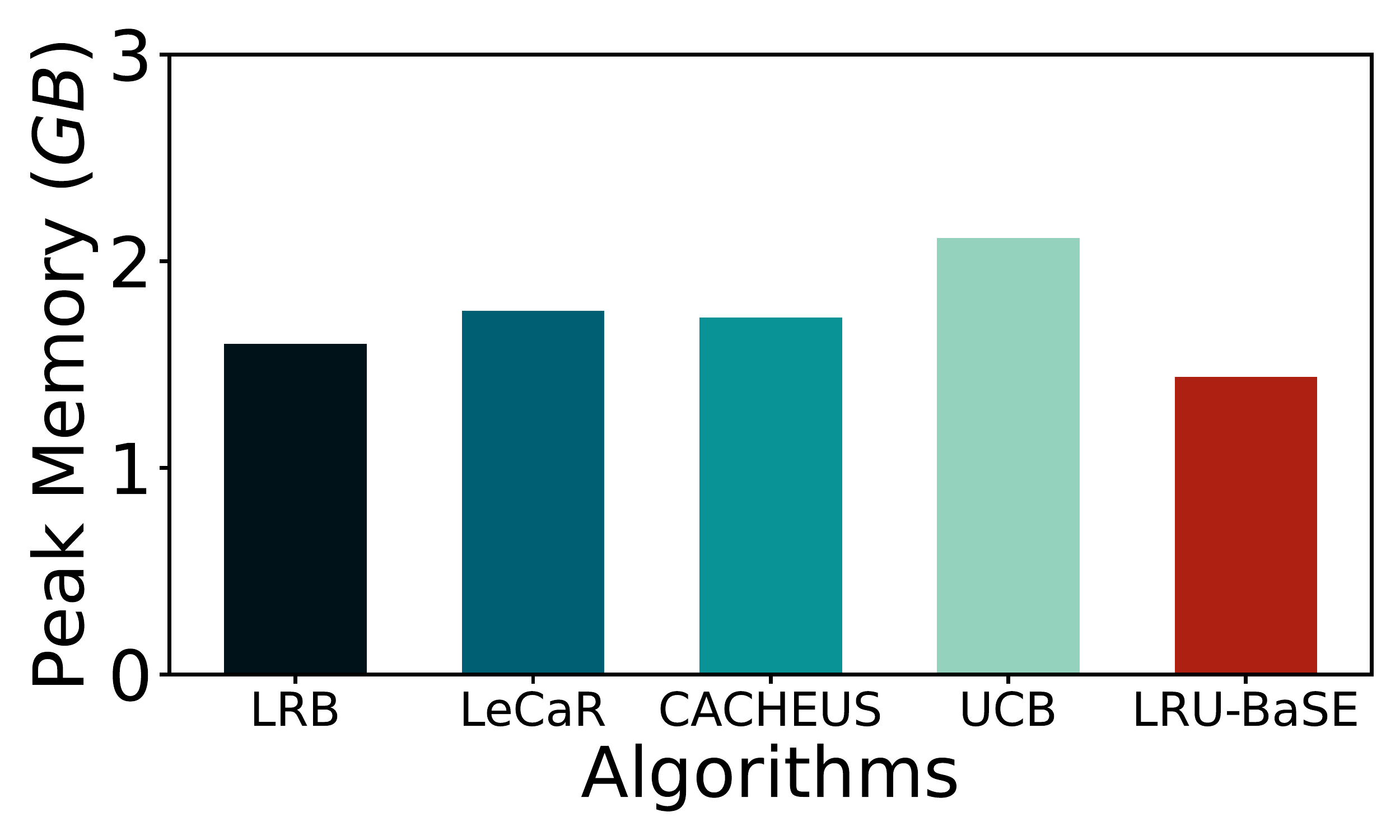}
	    \label{fig:trace_memory3}
	}
	\subfigure[Trace-\textit{T}: D-Time]{
	    \includegraphics[width=0.15\linewidth]{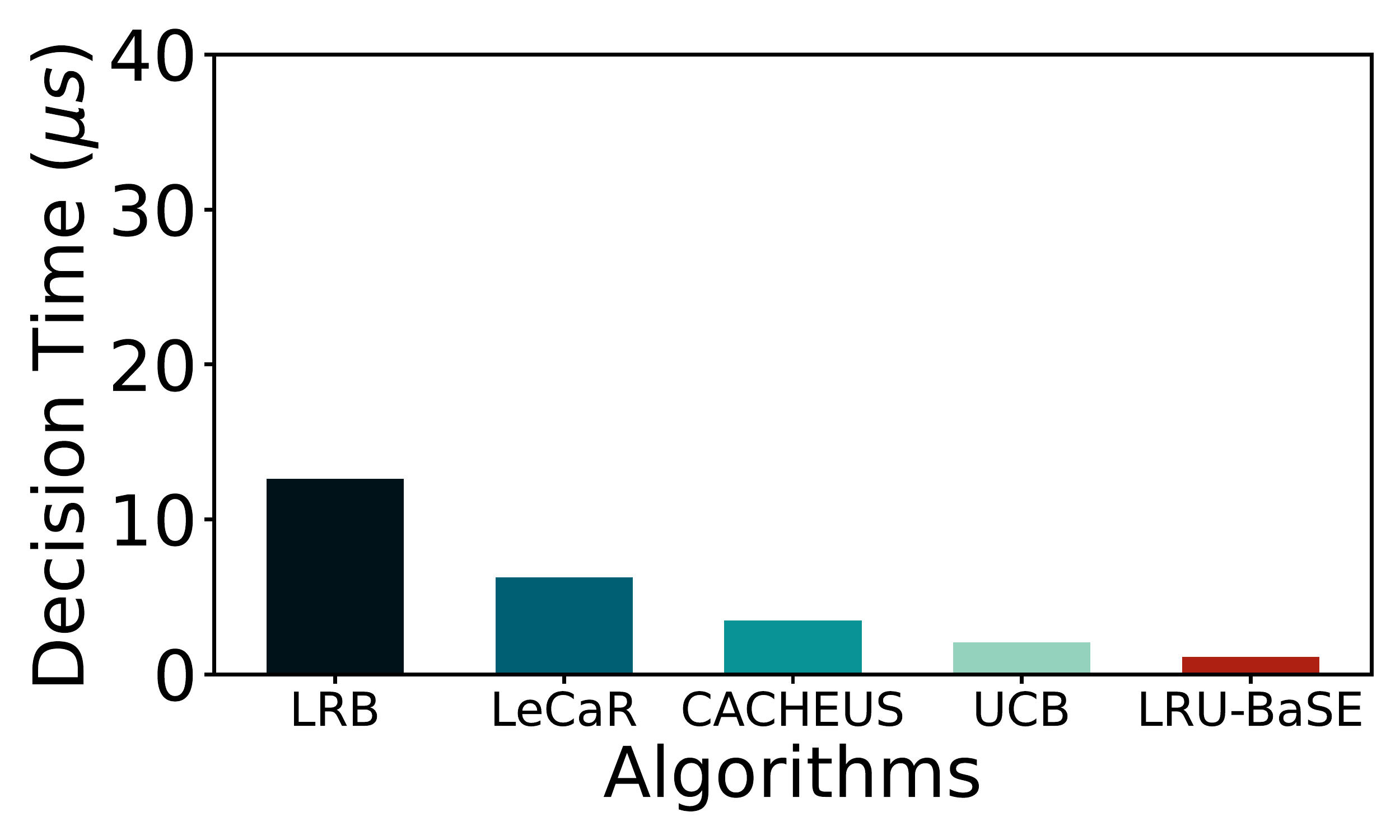}
	    \label{fig:trace_dt}
	}
	\\
	\subfigure[Wikipedia: BMR]{
	    \includegraphics[width=0.15\linewidth]{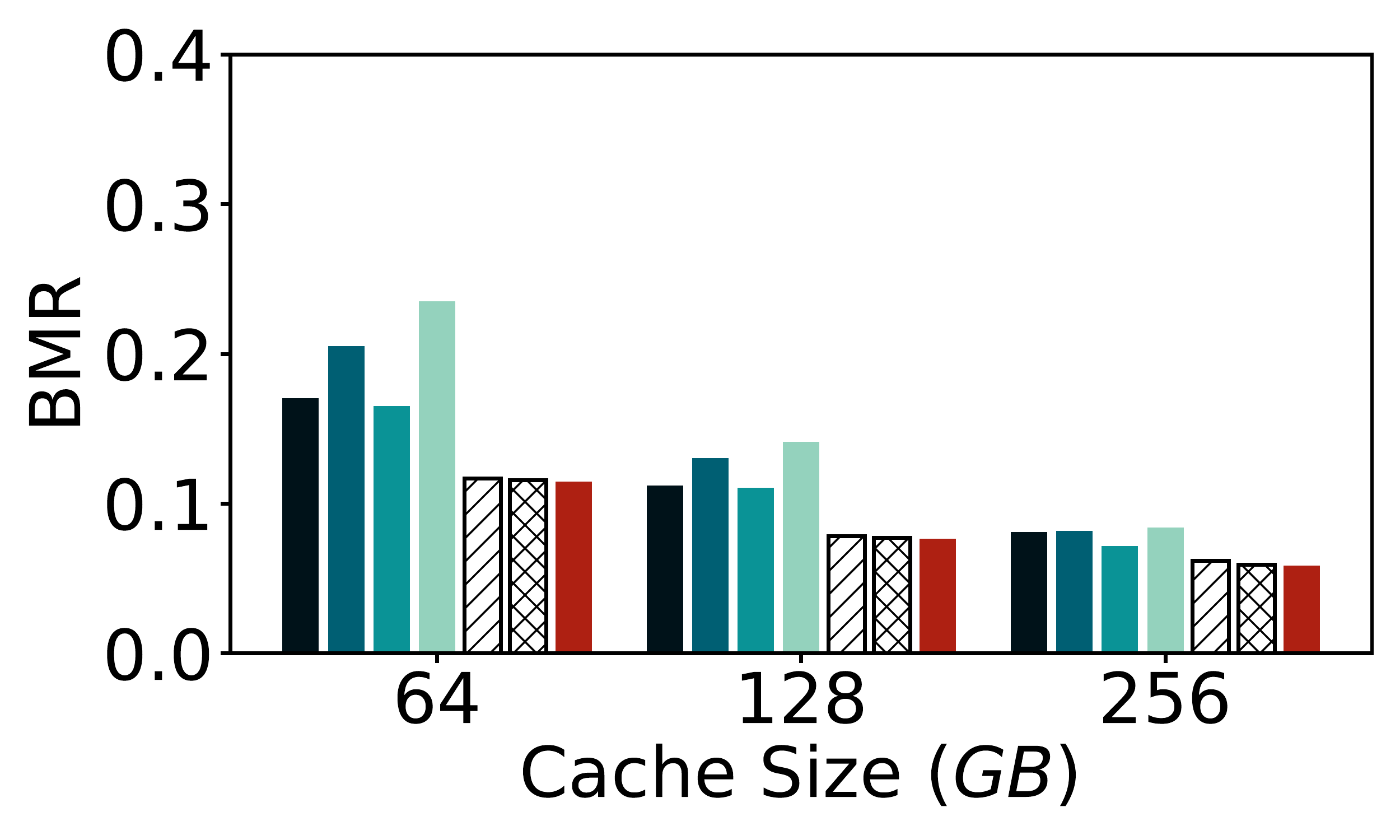}
	    \label{fig:wikipedia_bmr3}
	}
	\subfigure[Wikipedia: OMR]{
	    \includegraphics[width=0.15\linewidth]{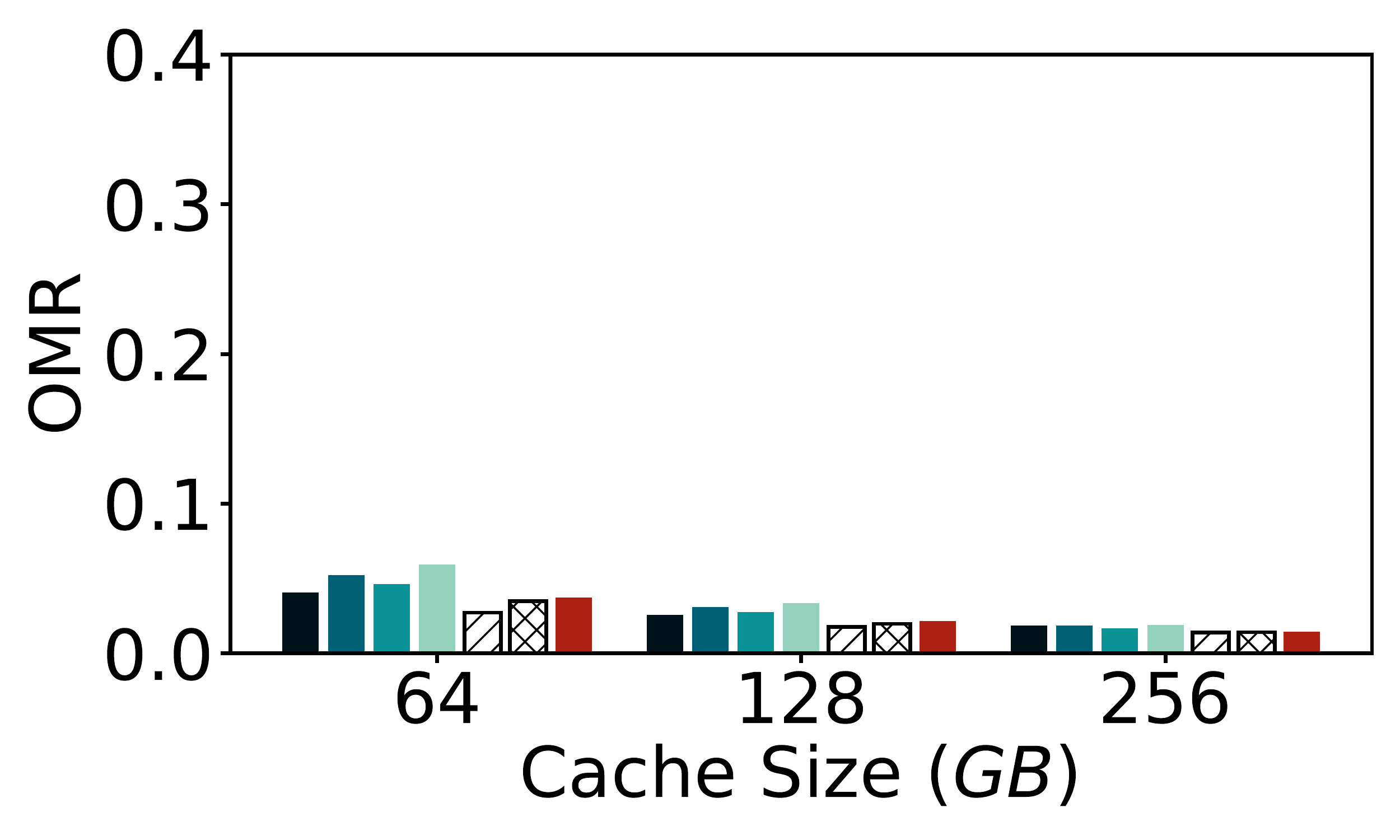}
	    \label{fig:wikipedia_omr3}
	}
	\subfigure[Wikipedia: CPU]{
	    \includegraphics[width=0.15\linewidth]{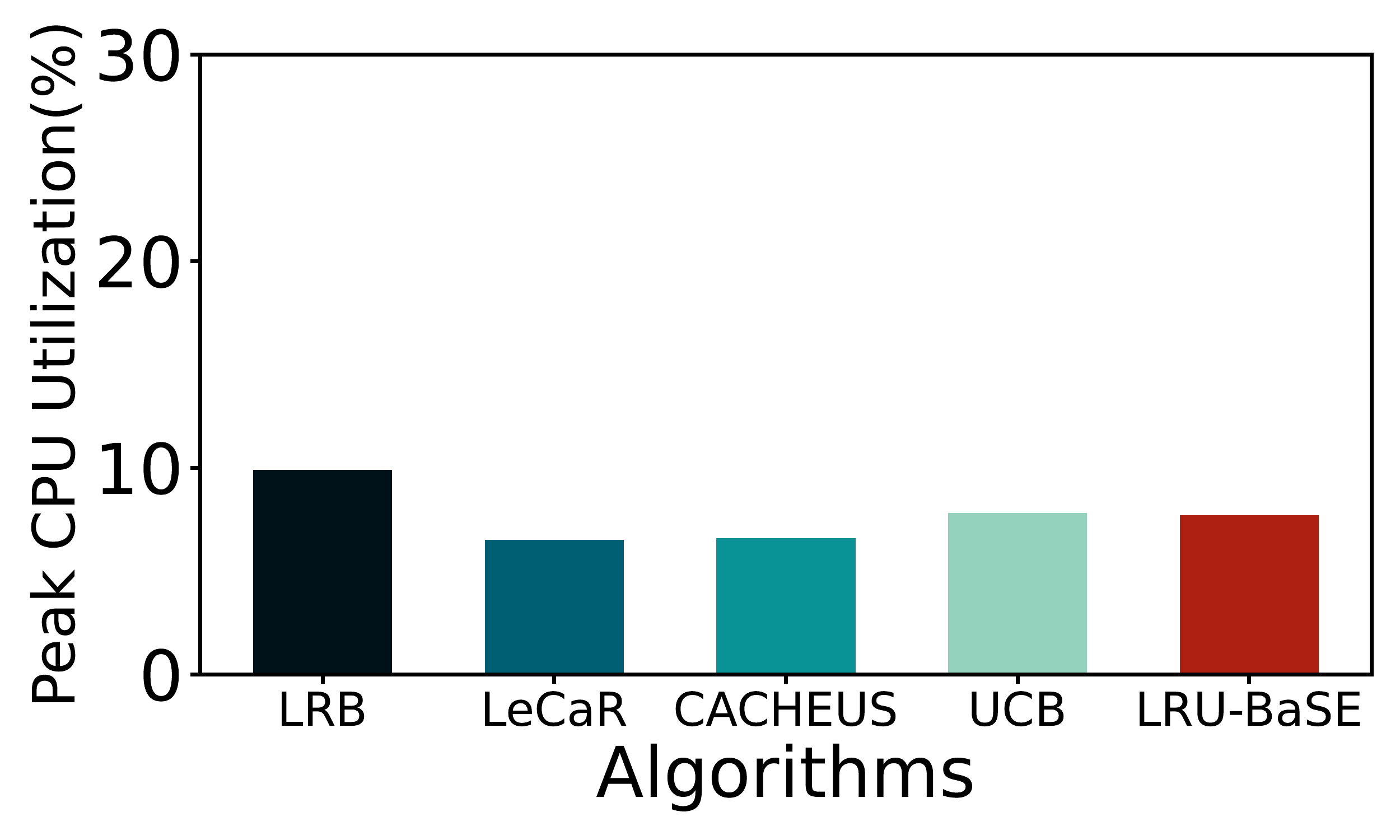}
	    \label{fig:wikipedia_cpu3}
	}
	\subfigure[Wikipedia: TPS]{
	    \includegraphics[width=0.15\linewidth]{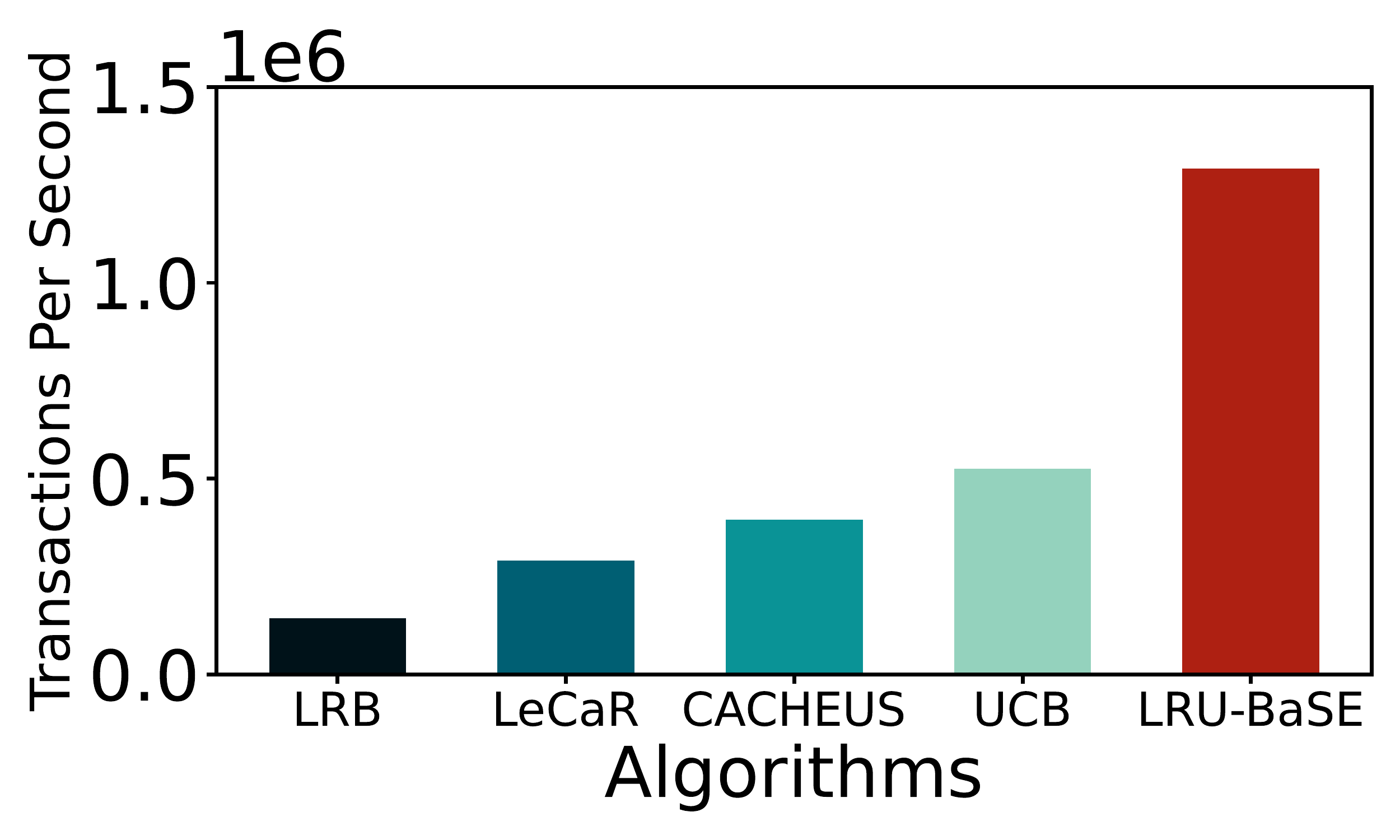}
	    \label{fig:wikipedia_throughput3}
	}
	\subfigure[Wikipedia: Memory]{
	    \includegraphics[width=0.15\linewidth]{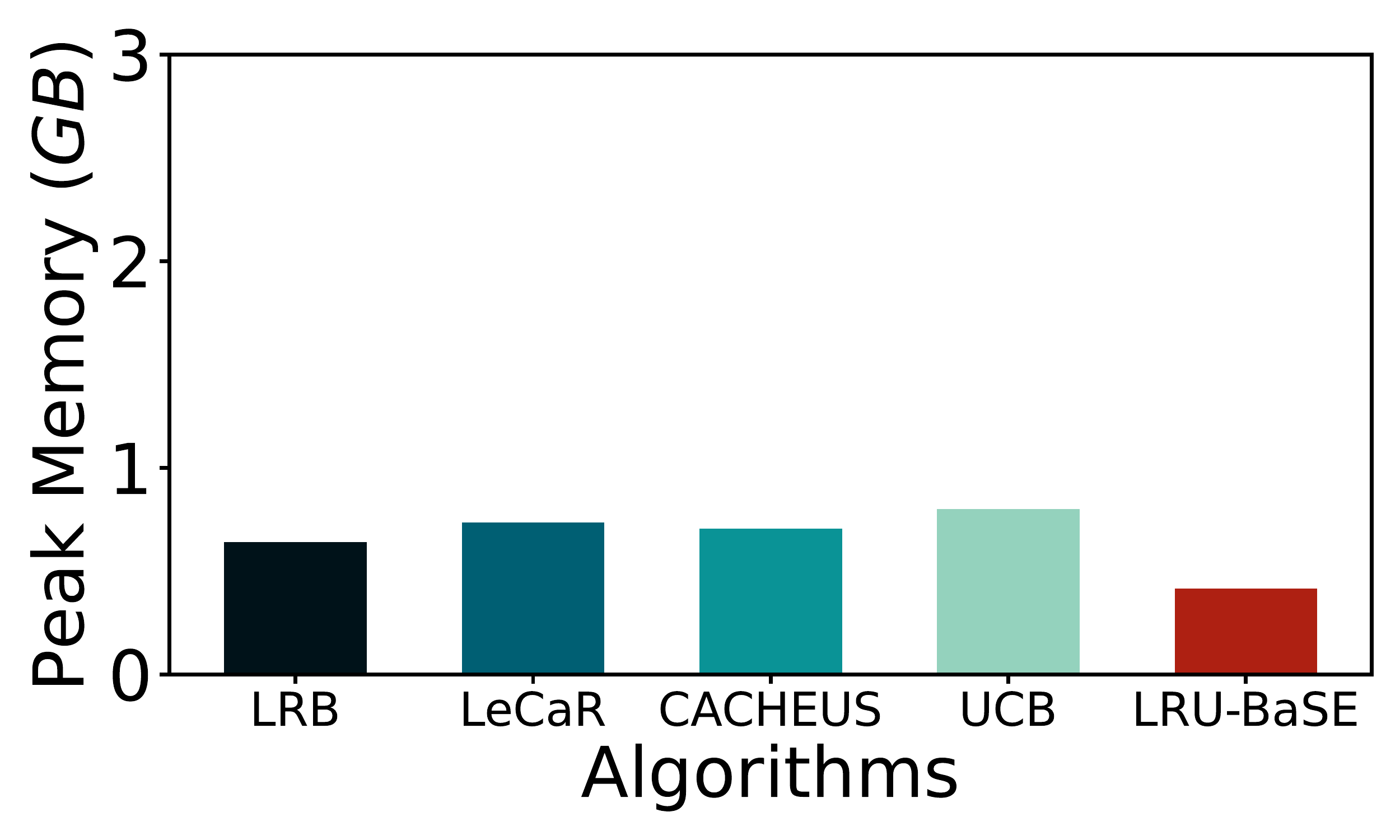}
	    \label{fig:wikipedia_memory3}
	}
	\subfigure[Wikipedia: D-Time]{
	    \includegraphics[width=0.15\linewidth]{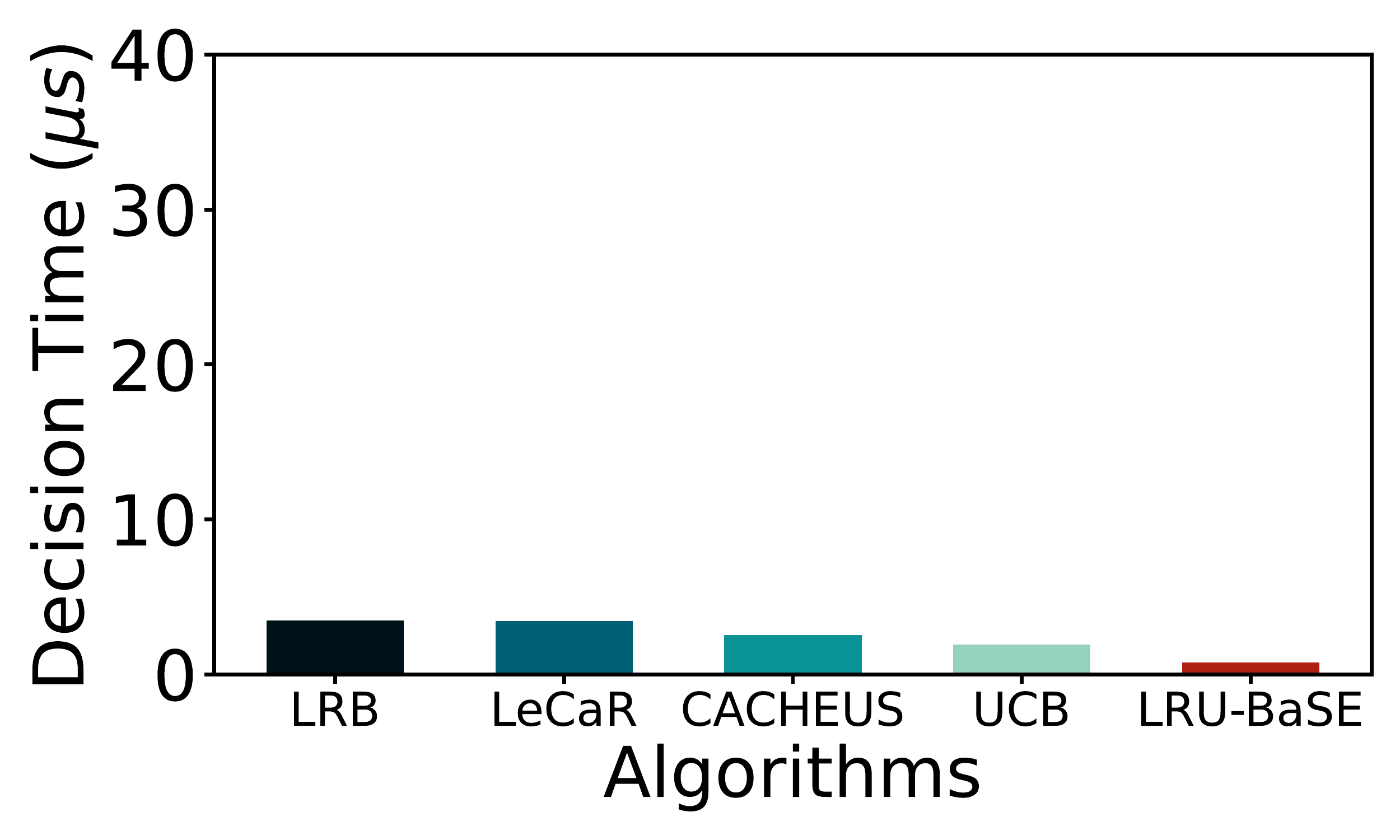}
	    \label{fig:wikipedia_dt}
	}
	\\
	\subfigure[CDN-Q: BMR]{
	    \includegraphics[width=0.15\linewidth]{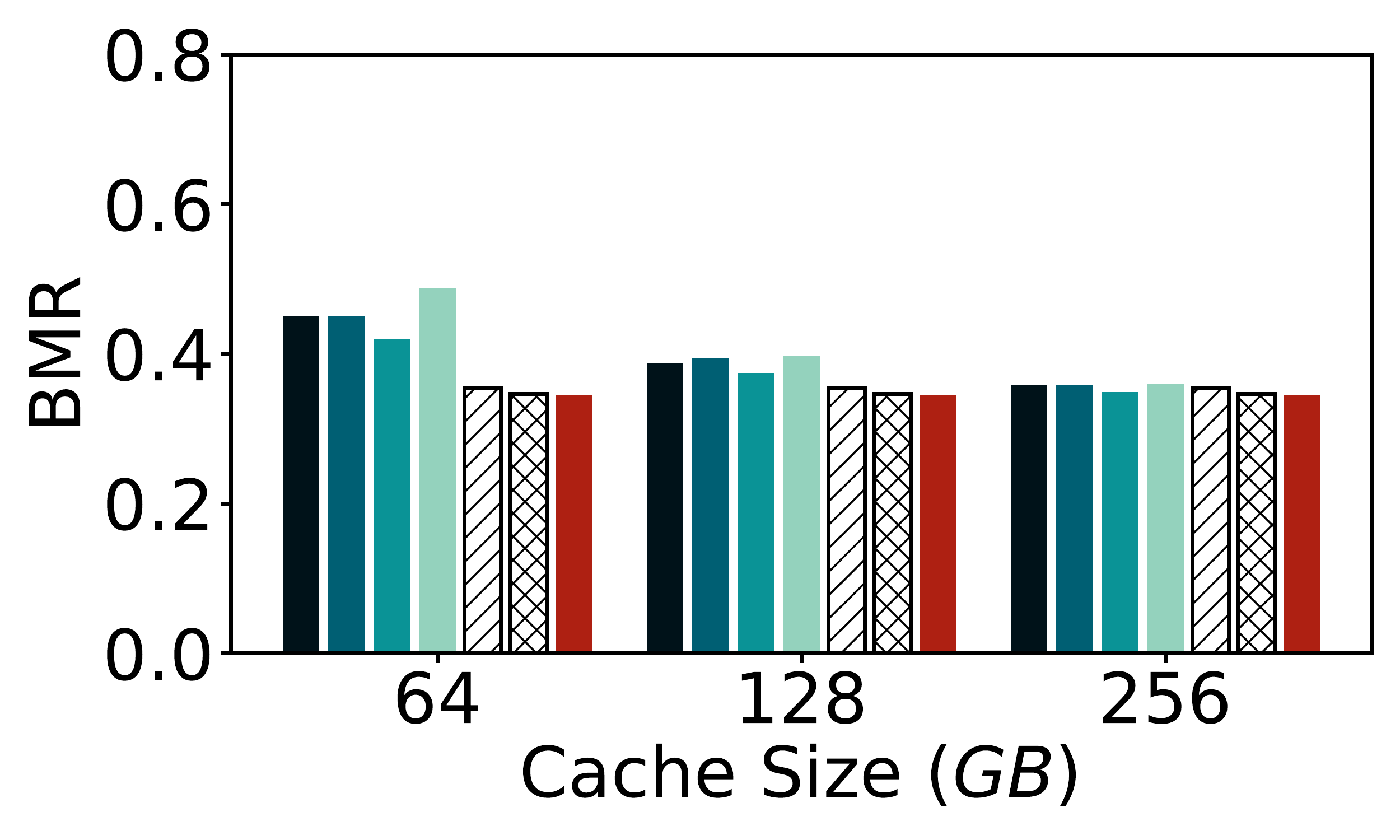}
	    \label{fig:cdn_q_bmr3}
	}
	\subfigure[CDN-Q: OMR]{
	    \includegraphics[width=0.15\linewidth]{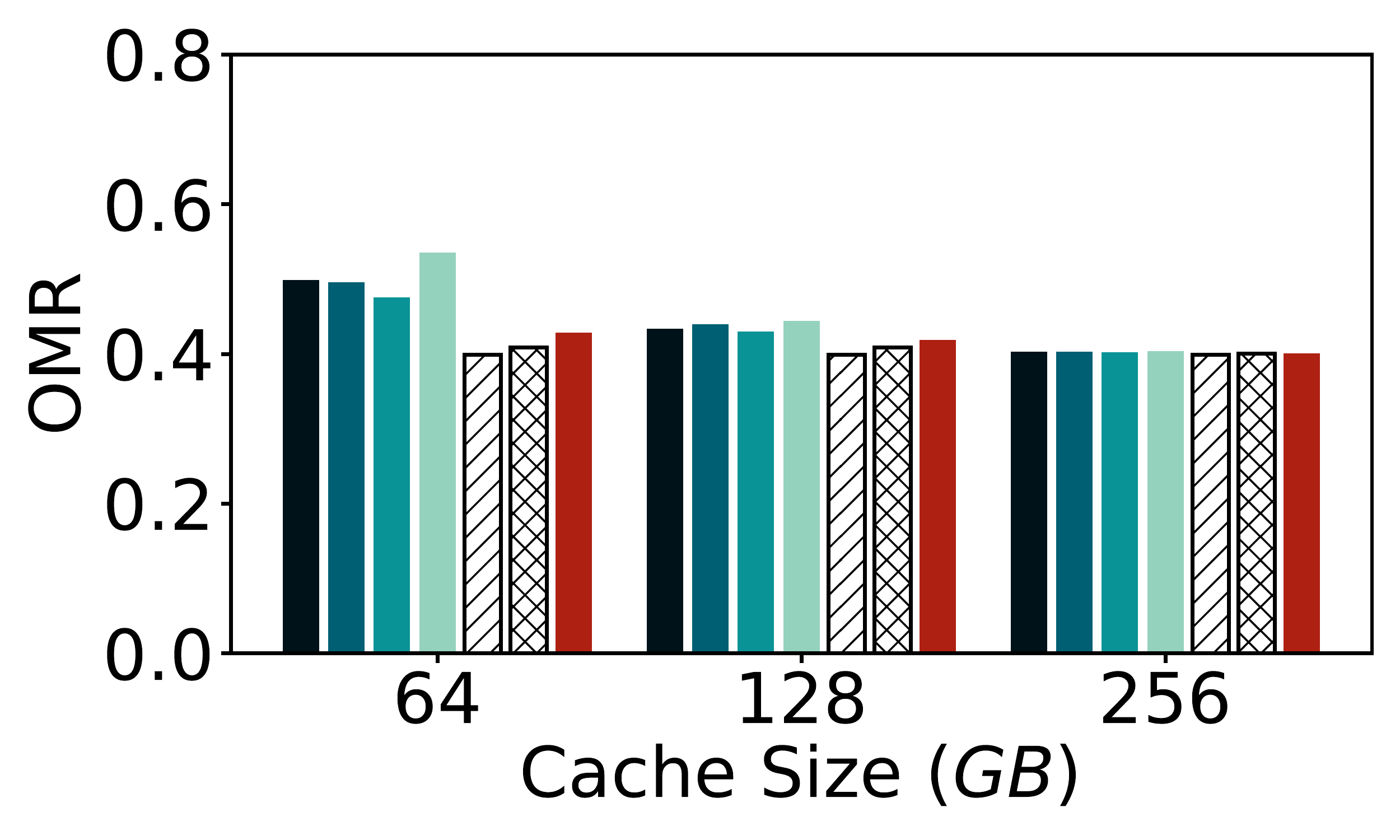}
	    \label{fig:cdn_q_omr3}
	}
	\subfigure[CDN-Q: CPU]{
	    \includegraphics[width=0.15\linewidth]{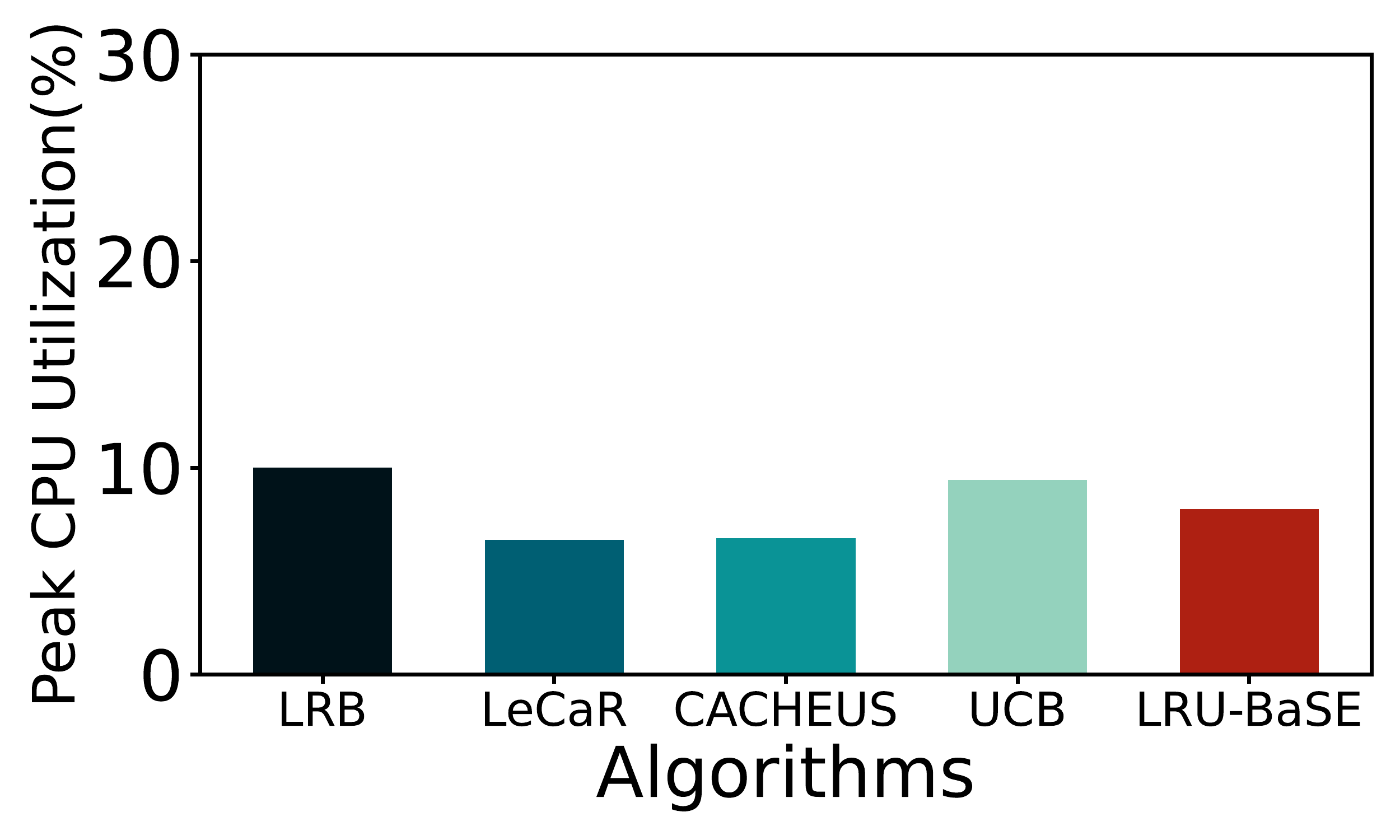}
	    \label{fig:cdn_q_cpu3}
	}
	\subfigure[CDN-Q: TPS]{
	    \includegraphics[width=0.15\linewidth]{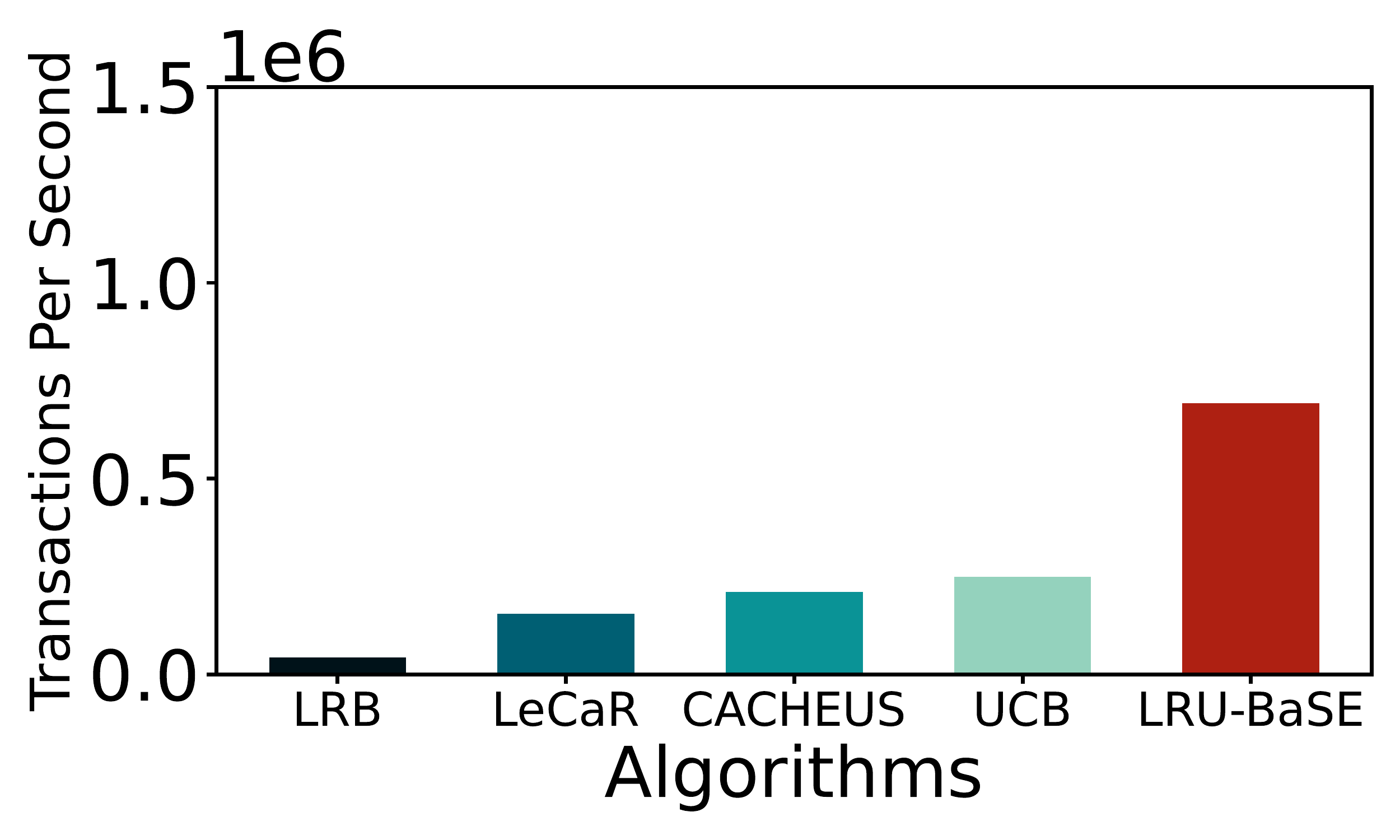}
	    \label{fig:cdn_q_throughput3}
	}
	\subfigure[CDN-Q: Memory]{
	    \includegraphics[width=0.15\linewidth]{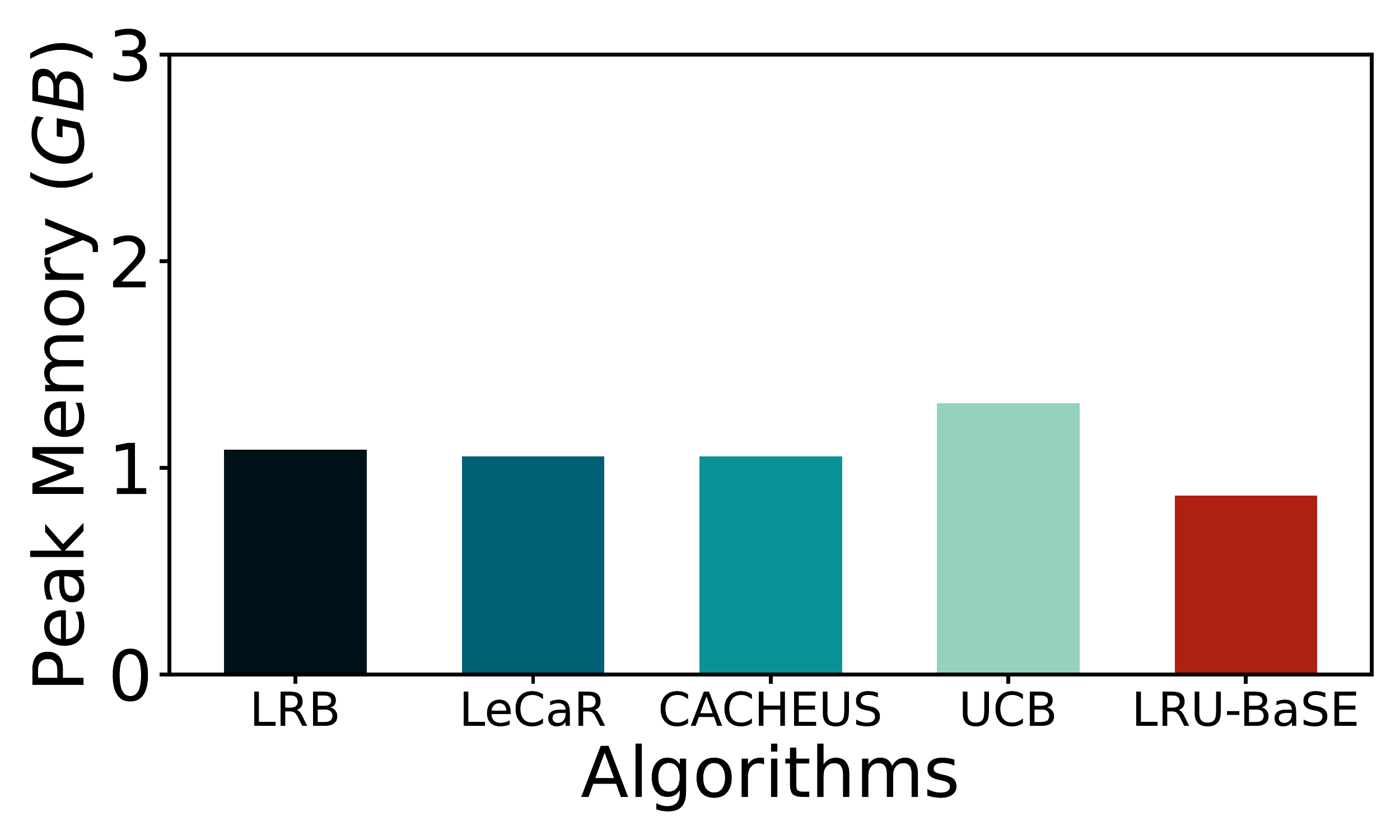}
	    \label{fig:cdn_q_memory3}
	}
	\subfigure[CDN-Q: D-Time]{
	    \includegraphics[width=0.15\linewidth]{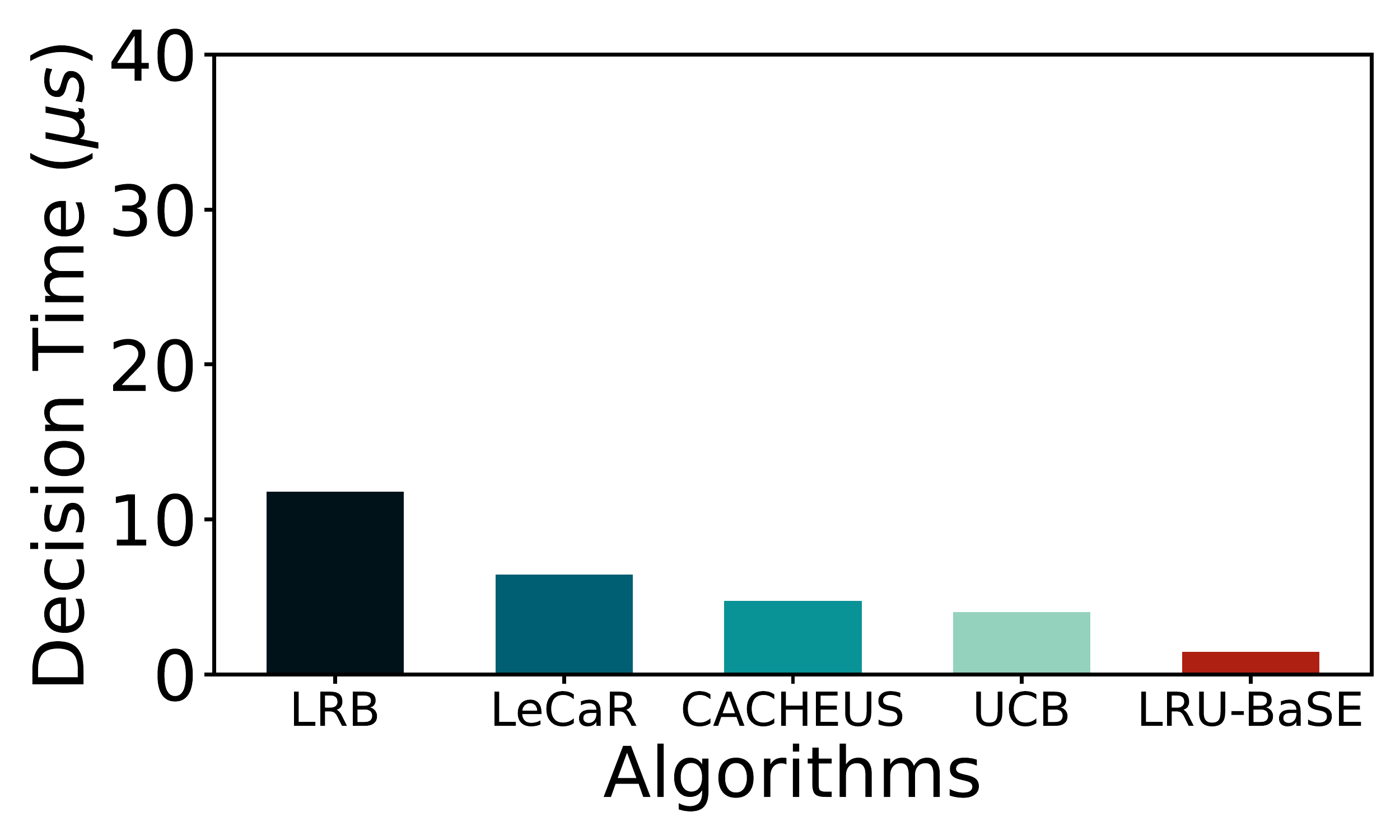}
	    \label{fig:cdn_q_dt}
	}
	\caption{Comparisons of Belady, PFOO, LRU-BaSE and four cache algorithms based on ML in terms of BMR, OMR, peak CPU utilization, transactions per second (TPS), peak memory and decision time (D-Time) at different cache sizes on three traces. The D-time here is the average value.}
	\vspace{-0.2cm}
	\label{fig:performance_pro3}
\end{figure*}

\noindent\textbf{Beyond the lower bound of BMR.} As shown in Figure~\ref{fig:performance_tro}, Figure~\ref{fig:performance_pro2}, and Figure~\ref{fig:performance_pro3}, the BMRs of LRU-BaSE are, by and large, superior to the lower bound, \ie, the BMRs of Belady and PFOO. When the cache size equals 64GB, the improvement over Belady and PFOO is 0.7\% and 0.7\%, respectively, on Trace-\textit{T}; 0.2\% and 0.1\%, respectively, on Wikipedia; and 1\% and 0.2\%, respectively, on CDN-Q. When the cache size equals 128GB, the corresponding improvements are 0.8\% and 0.8\%, 0.2\% and 0.06\%, 1\% and 0.2\%. When the cache size equals 256GB, the corresponding improvements are 1\% and 1\%, 0.3\% and 0.04\%, 1\% and 0.2\%. In addition, the average OMR of LRU-BaSE over the three cache sizes is 0.53\%, 0.48\%, and 1.74\% higher than that of Belady on Trace-\textit{T}, Wikipedia, and CDN-Q, respectively; is 0.54\%, 0.16\%, and 1.01\% lower than that of PFOO, respectively. We believe that the reason for LRU-BaSE's superiority is twofold. First, LRU-BaSE is extremely close to Belady in reducing BMR by hitting more data. Second, LRU-BaSE chooses a BMR-friendly eviction policy while ensuring OMR. Note that since both Belady and PFOO need all the information about the trace in advance, LRU-BaSE is the first algorithm that dynamically senses the trace to attain a seemingly unattainably low BMR, beyond the lower bound of Belady. 

\noindent\textbf{Compared to the classic algorithms.} As shown in Figure~\ref{fig:trace_t_bmr}, Figure~\ref{fig:wikipedia_bmr}, and Figure~\ref{fig:cdn_q_bmr}, LRU-BaSE outperforms others in both BMR and OMR. In BMR, the average BMR of LRU-BaSE over the three cache sizes is 2.52\%, 5.28\%, and 6.48\% lower than that of the best classic algorithm on Trace-\textit{T}, Wikipedia, and CDN-Q, respectively. The corresponding results in OMR are 1.26\%, 0.83\%, and 4.03\%.


In addition, as shown from Figure~\ref{fig:trace_t_cpu} to Figure~\ref{fig:trace_t_memory}, on average, LRU-BaSE overtakes other methods by 0.8\% in peak CPU utilization. As shown from Figure~\ref{fig:wikipedia_cpu} to Figure~\ref{fig:wikipedia_memory}, LRU-BaSE consumes 1.4GB more peak memory on average than LRU and S4LRU. As shown from Figure~\ref{fig:cdn_q_cpu} to Figure~\ref{fig:cdn_q_memory}, LRU-BaSE's peak memory consumption is only lower than LRU in TPS.

\noindent\textbf{Compared to the size-aware algorithms.} As shown in Figure~\ref{fig:trace_t_bmr2}, Figure~\ref{fig:wikipedia_bmr2}, and Figure~\ref{fig:cdn_q_bmr2}, LRU-BaSE outperforms other size-aware algorithms, where the average BMR of LRU-BaSE over the three cache sizes is 12.81\%, 12.3\%, and 14.76\% lower than that of the next best one on Trace-\textit{T}, Wikipedia, and CDN-Q, respectively.


As shown in Figure~\ref{fig:trace_t_cpu2}, Figure~\ref{fig:wikipedia_cpu2}, and Figure~\ref{fig:cdn_q_cpu2}, LRU-BaSE is comparable to AdaptSize and slightly higher than others in peak CPU utilization. As shown in Figure~\ref{fig:trace_t_memory2}, Figure~\ref{fig:trace_t_memory2}, and Figure~\ref{fig:trace_t_memory2}, LRU-BaSE is moderate among compared methods in terms of memory consumption. As shown in Figure~\ref{fig:trace_t_throughput2}, Figure~\ref{fig:wikipedia_throughput2}, and Figure~\ref{fig:cdn_q_throughput2}, LRU-BaSE is second only to ThLRU in transaction throughput.

\noindent\textbf{Compared to the ML cache algorithms.} As shown in Figure~\ref{fig:performance_pro3}, LRU-BaSE also outperforms the ML-based algorithms in both BMR and OMR. In BMR, the average BMR of LRU-BaSE over the three cache sizes is 9.12\%, 4.91\%, and 5.44\% lower than that of the best ML-based algorithm on Trace-\textit{T}, Wikipedia, and CDN-Q, respectively. In terms of peak CPU utilization, TPS, and peak memory consumption, the LRU-BaSE algorithm is the best performer. We attribute this improvement to two optimizations of LRU-BaSE. First, the decision space is greatly reduced by the discovery of the rear section. Second, the cache requirements for speed can be met by parallelizing the batch processing in the implementation. In addition, in terms of decision time, as shown in Figure~\ref{fig:trace_dt}, Figure~\ref{fig:wikipedia_dt}, and Figure~\ref{fig:cdn_q_dt}, since model training and decision making are performed on different threads in parallel, LRU-BaSE calculates only once when making a decision. It is a single calculation for multiple requests, thereby LRU-BaSE is also the best in terms of decision time for a single request.

\subsection{Robustness of LRU-BaSE}
\label{sec:robust}
\begin{figure}[t]
	\centering
	\includegraphics[width=0.9\linewidth]{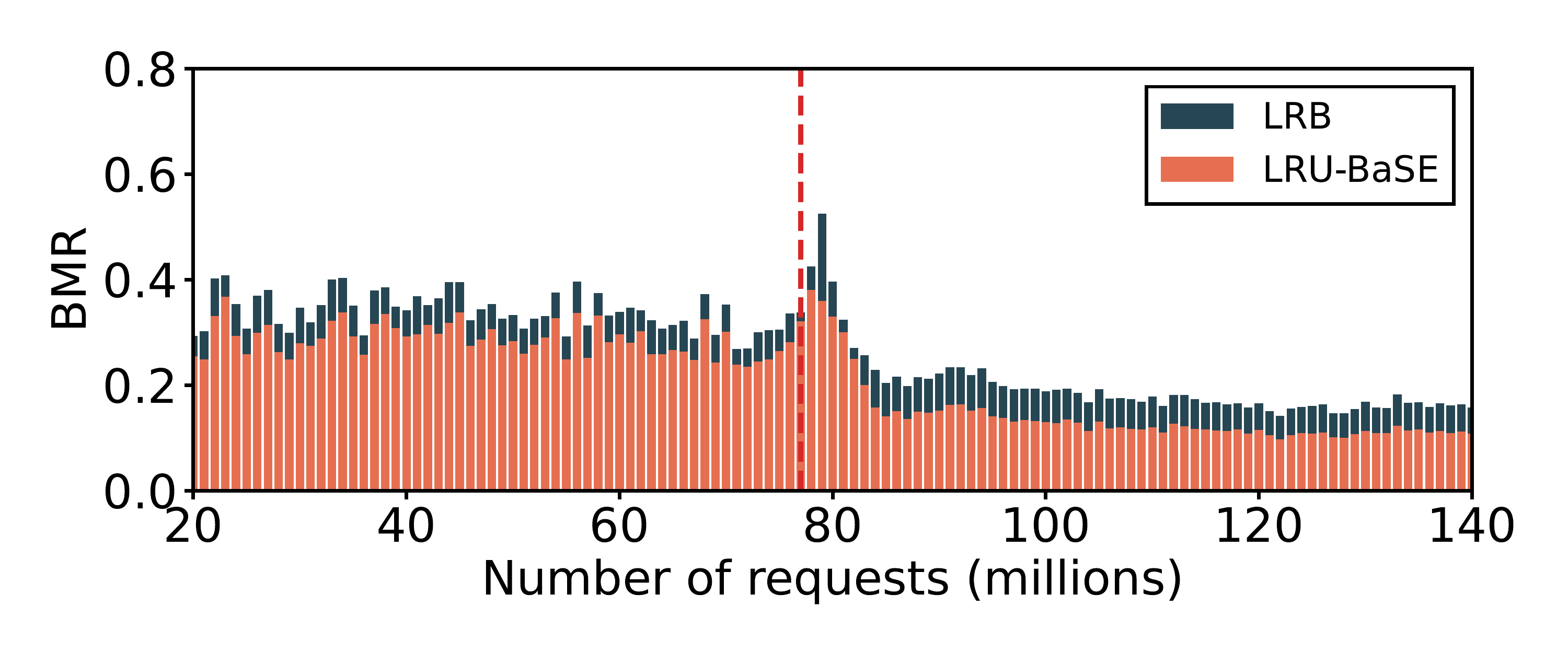}
	\caption{Compare the robustness of LRB and LRU-BaSE facing workload drift.}
	\vspace{-0.2cm}
	\label{fig:breakout}
\end{figure}

In this section, we show the robustness of LRU-BaSE by showing that our window-finding method is more resilient to workload drift than the pre-defined boundary method in LRB. 
We splice two traces (Trace-\textit{T} and Wikipedia from Table~\ref{table:trace}) with large variability to form a new trace. On this new trace, we run LRB and LRU-BaSE, and output the average BMR of each one million requests. As shown in Figure~\ref{fig:breakout}, data on the left of the red dashed line represent BMRs on Trace-\textit{T}, and the workload is changed to Wikipedia at the red dashed line, with the ensuing BMRs shown on the right of the red dashed line. According to the statistics, LRU-BaSE outperforms LRB by 3.9\% on average before the workload drift, while outperforming LRB by 5.5\% on average after the drift. Referring to their results in Figure~\ref{fig:trace_t_bmr3} and Figure~\ref{fig:wikipedia_bmr3}, we believe that LRB's customized window for Trace-\textit{T} inhibits its performance on Wikipedia, while LRU-BaSE can dynamically adapt to changes in workloads. Furthermore, LRU-BaSE exhibits smaller fluctuations at and right after the onset of the workload change.  


\section{Conclusion and Discussion}
With the wide development of CDNs, "back to original server" traffic will continue to surge, increasing the importance of studying byte miss ratio. Indeed, we believe that approximating Belady alone no longer provides enough room for lowering BMR, as the approximation for the reuse distance is close to the limit. Our work represents a first step towards comprehending the relationship between OMR and BMR and attempting to decrease BMR while maintaining OMR. Our discovery of the "flat" regions in the OMR (Eviction-Window) function curves offers theoretical possibilities for solutions with win-win BMR-OMR performance. Furthermore, we improve the RL model based on unique traits of CDN and LRU to tackle the feedback delay problem. Nonetheless, this is also a compromise solution due to the inability to find evident features corresponding to BMR. As our future work, we plan to analyze the most influential features from the choices that favor BMR to inform decisions for simple learning models.



\bibliographystyle{plain}
\bibliography{reference}

\begin{thebibliography}{10}

\bibitem{dqn_code}
Reinforcement learning: Dqn in pytorch.
\newblock
  \url{https://github.com/MorvanZhou/PyTorch-Tutorial/blob/master/tutorial-contents/405_DQN_Reinforcement_learning.py},
  2020.

\bibitem{lrb_code}
A simulator for cdn caching and web caching policies.
\newblock \url{https://github.com/sunnyszy/lrb}, 2020.

\bibitem{abrams1996removal}
Marc Abrams, Charles~R Standridge, Ghaleb Abdulla, Edward~A Fox, and Stephen
  Williams.
\newblock Removal policies in network caches for world-wide web documents.
\newblock In {\em ACM SIGCOMM}, pages 293--305, 1996.

\bibitem{abrams1995caching}
Marc Abrams, Charles~R Standridge, Ghaleb Abdulla, Stephen Williams, and
  Edward~A Fox.
\newblock Caching proxies: Limitations and potentials.
\newblock Technical report, 1995.

\bibitem{aggarwal1999caching}
Charu Aggarwal, Joel~L Wolf, and Philip~S. Yu.
\newblock Caching on the world wide web.
\newblock {\em IEEE TKDE}, 11(1):94--107, 1999.

\bibitem{arlitt2000evaluating}
Martin Arlitt, Ludmila Cherkasova, John Dilley, Rich Friedrich, and Tai Jin.
\newblock Evaluating content management techniques for web proxy caches.
\newblock {\em ACM SIGMETRICS Performance Evaluation Review}, 27(4):3--11,
  2000.

\bibitem{AtreSWB20}
Nirav Atre, Justine Sherry, Weina Wang, and Daniel~S. Berger.
\newblock Caching with delayed hits.
\newblock In {\em ACM SIGCOMM}, pages 495--513, 2020.

\bibitem{BahnKMN02}
Hyokyung Bahn, Kern Koh, Sang~Lyul Min, and Sam~H. Noh.
\newblock Efficient replacement of nonuniform objects in web caches.
\newblock {\em IEEE Computer}, 35(6):65--73, 2002.

\bibitem{bahn2002efficient}
Hyokyung Bahn, Kern Koh, Sam~H Noh, and SM~Lyul.
\newblock Efficient replacement of nonuniform objects in web caches.
\newblock {\em IEEE Computer}, 35(6):65--73, 2002.

\bibitem{beckmann2018lhd}
Nathan Beckmann, Haoxian Chen, and Asaf Cidon.
\newblock Lhd: Improving cache hit rate by maximizing hit density.
\newblock In {\em USENIX NSDI}, pages 389--403, 2018.

\bibitem{belady1966study}
Laszlo~A. Belady.
\newblock A study of replacement algorithms for a virtual-storage computer.
\newblock {\em IBM Systems journal}, 5(2):78--101, 1966.

\bibitem{berg2020cachelib}
Benjamin Berg, Daniel~S Berger, Sara McAllister, Isaac Grosof, Sathya
  Gunasekar, Jimmy Lu, Michael Uhlar, Jim Carrig, Nathan Beckmann, Mor
  Harchol-Balter, et~al.
\newblock The cachelib caching engine: Design and experiences at scale.
\newblock In {\em USENIX OSDI}, pages 753--768, 2020.

\bibitem{berger2018towards}
Daniel~S Berger.
\newblock Towards lightweight and robust machine learning for cdn caching.
\newblock In {\em ACM HotNets}, pages 134--140, 2018.

\bibitem{berger2018practical}
Daniel~S Berger, Nathan Beckmann, and Mor Harchol-Balter.
\newblock Practical bounds on optimal caching with variable object sizes.
\newblock {\em Proceedings of the ACM on Measurement and Analysis of Computing
  Systems}, 2(2):1--38, 2018.

\bibitem{berger2018robinhood}
Daniel~S Berger, Benjamin Berg, Timothy Zhu, Siddhartha Sen, and Mor
  Harchol-Balter.
\newblock Robinhood: Tail latency aware caching--dynamic reallocation from
  cache-rich to cache-poor.
\newblock In {\em USENIX OSDI}, pages 195--212, 2018.

\bibitem{berger2017adaptsize}
Daniel~S Berger, Ramesh~K Sitaraman, and Mor Harchol-Balter.
\newblock Adaptsize: Orchestrating the hot object memory cache in a content
  delivery network.
\newblock In {\em USENIX NSDI}, pages 483--498, 2017.

\bibitem{CaiLZZZLLCYX22}
Baoqing Cai, Yu~Liu, Ce~Zhang, Guangyu Zhang, Ke~Zhou, Li~Liu, Chunhua Li, Bin
  Cheng, Jie Yang, and Jiashu Xing.
\newblock {HUNTER:} an online cloud database hybrid tuning system for
  personalized requirements.
\newblock In {\em SIGMOD}, pages 646--659. {ACM}, 2022.

\bibitem{cao1997cost}
Pei Cao and Sandy Irani.
\newblock Cost-aware www proxy caching algorithms.
\newblock In {\em USENIX USITS}, volume~12, pages 193--206, 1997.

\bibitem{cherkasova2001role}
Ludmila Cherkasova and Gianfranco Ciardo.
\newblock Role of aging, frequency, and size in web cache replacement policies.
\newblock In {\em Springer HPCN}, pages 114--123, 2001.

\bibitem{costa2017mlcache}
Renato Costa and Jose Pazos.
\newblock Mlcache: A multi-armed bandit policy for an operating system page
  cache.
\newblock Technical report, Technical Report. University of British Columbia,
  2017.

\bibitem{einziger2021lightweight}
Gil Einziger, Ohad Eytan, Roy Friedman, and Benjamin Manes.
\newblock Lightweight robust size aware cache management.
\newblock {\em ACM TOS}, 2021.

\bibitem{einziger2017tinylfu}
Gil Einziger, Roy Friedman, and Ben Manes.
\newblock Tinylfu: A highly efficient cache admission policy.
\newblock {\em ACM TOS}, 13(4):1--31, 2017.

\bibitem{eytan2020s}
Ohad Eytan, Danny Harnik, Effi Ofer, Roy Friedman, and Ronen Kat.
\newblock It's time to revisit lru vs. fifo.
\newblock In {\em USENIX HotStorage}, 2020.

\bibitem{Jerome2001}
Jerome~H Friedman.
\newblock Greedy function approximation: a gradient boosting machine.
\newblock {\em Annals of statistics}, page 1189–1232, 2001.

\bibitem{GastH15}
Nicolas Gast and Benny~Van Houdt.
\newblock Transient and steady-state regime of a family of list-based cache
  replacement algorithms.
\newblock In {\em ACM SIGMETRICS}, pages 123--136, 2015.

\bibitem{hochreiter1997long}
Sepp Hochreiter and J{\"u}rgen Schmidhuber.
\newblock Long short-term memory.
\newblock {\em Neural computation}, 9(8):1735--1780, 1997.

\bibitem{huang2013analysis}
Qi~Huang, Ken Birman, Robbert Van~Renesse, Wyatt Lloyd, Sanjeev Kumar, and
  Harry~C Li.
\newblock An analysis of facebook photo caching.
\newblock In {\em ACM SOSP}, pages 167--181, 2013.

\bibitem{jain2018rethinking}
Akanksha Jain and Calvin Lin.
\newblock Rethinking belady's algorithm to accommodate prefetching.
\newblock In {\em ACM/IEEE ISCA}, pages 110--123, 2018.

\bibitem{jiang2002lirs}
Song Jiang and Xiaodong Zhang.
\newblock Lirs: an efficient low inter-reference recency set replacement policy
  to improve buffer cache performance.
\newblock {\em ACM SIGMETRICS}, 30(1):31--42, 2002.

\bibitem{JinB01}
Shudong Jin and Azer Bestavros.
\newblock Greedydual* web caching algorithm: exploiting the two sources of
  temporal locality in web request streams.
\newblock {\em Elsevier CC}, 24(2):174--183, 2001.

\bibitem{kaelbling1996reinforcement}
Leslie~Pack Kaelbling, Michael~L Littman, and Andrew~W Moore.
\newblock Reinforcement learning: A survey.
\newblock {\em AAAI JAIR}, 4:237--285, 1996.

\bibitem{ke2017lightgbm}
Guolin Ke, Qi~Meng, Thomas Finley, Taifeng Wang, Wei Chen, Weidong Ma, Qiwei
  Ye, and Tie-Yan Liu.
\newblock Lightgbm: A highly efficient gradient boosting decision tree.
\newblock {\em Advances in neural information processing systems},
  30:3146--3154, 2017.

\bibitem{li2015gd}
Conglong Li and Alan~L Cox.
\newblock Gd-wheel: a cost-aware replacement policy for key-value stores.
\newblock In {\em ACM EuroSys}, pages 1--15, 2015.

\bibitem{li2019beating}
Pengcheng Li, Colin Pronovost, William Wilson, Benjamin Tait, Jie Zhou, Chen
  Ding, and John Criswell.
\newblock Beating opt with statistical clairvoyance and variable size caching.
\newblock In {\em ACM ASPLOS}, pages 243--256, 2019.

\bibitem{lillicrap2015continuous}
Timothy~P Lillicrap, Jonathan~J Hunt, Alexander Pritzel, Nicolas Heess, Tom
  Erez, Yuval Tassa, David Silver, and Daan Wierstra.
\newblock Continuous control with deep reinforcement learning.
\newblock {\em arXiv preprint arXiv:1509.02971}, 2015.

\bibitem{maggs2015algorithmic}
Bruce~M Maggs and Ramesh~K Sitaraman.
\newblock Algorithmic nuggets in content delivery.
\newblock {\em ACM SIGCOMM}, 45(3):52--66, 2015.

\bibitem{megiddo2003arc}
Nimrod Megiddo and Dharmendra~S Modha.
\newblock Arc: A self-tuning, low overhead replacement cache.
\newblock In {\em USENIX FAST}, volume~3, pages 115--130, 2003.

\bibitem{mnih2016asynchronous}
Volodymyr Mnih, Adria~Puigdomenech Badia, Mehdi Mirza, Alex Graves, Timothy
  Lillicrap, Tim Harley, David Silver, and Koray Kavukcuoglu.
\newblock Asynchronous methods for deep reinforcement learning.
\newblock In {\em ACM ICML}, pages 1928--1937, 2016.

\bibitem{mnih2013playing}
Volodymyr Mnih, Koray Kavukcuoglu, David Silver, Alex Graves, Ioannis
  Antonoglou, Daan Wierstra, and Martin Riedmiller.
\newblock Playing atari with deep reinforcement learning.
\newblock {\em arXiv preprint arXiv:1312.5602}, 2013.

\bibitem{mnih2015human}
Volodymyr Mnih, Koray Kavukcuoglu, David Silver, Andrei~A Rusu, Joel Veness,
  Marc~G Bellemare, Alex Graves, Martin Riedmiller, Andreas~K Fidjeland, Georg
  Ostrovski, et~al.
\newblock Human-level control through deep reinforcement learning.
\newblock {\em Nature}, 518(7540):529--533, 2015.

\bibitem{mukerjee2016impact}
Matthew~K Mukerjee, Ilker~Nadi Bozkurt, Bruce Maggs, Srinivasan Seshan, and Hui
  Zhang.
\newblock The impact of brokers on the future of content delivery.
\newblock In {\em ACM HotNets}, pages 127--133, 2016.

\bibitem{o1993lru}
Elizabeth~J O'neil, Patrick~E O'neil, and Gerhard Weikum.
\newblock The lru-k page replacement algorithm for database disk buffering.
\newblock {\em ACM SIGMOD Record}, 22(2):297--306, 1993.

\bibitem{PeledMWE15}
Leeor Peled, Shie Mannor, Uri~C. Weiser, and Yoav Etsion.
\newblock Semantic locality and context-based prefetching using reinforcement
  learning.
\newblock In {\em ACM/IEEE ISCA}, pages 285--297, 2015.

\bibitem{PeledWE20}
Leeor Peled, Uri~C. Weiser, and Yoav Etsion.
\newblock A neural network prefetcher for arbitrary memory access patterns.
\newblock {\em ACM TACO}, 16(4):37:1--37:27, 2020.

\bibitem{rizzo2000replacement}
Luigi Rizzo and Lorenzo Vicisano.
\newblock Replacement policies for a proxy cache.
\newblock {\em IEEE/ACM TON}, 8(2):158--170, 2000.

\bibitem{rodriguez2021learning}
Liana~V Rodriguez, Farzana Yusuf, Steven Lyons, Eysler Paz, Raju Rangaswami,
  Jason Liu, Ming Zhao, and Giri Narasimhan.
\newblock Learning cache replacement with cacheus.
\newblock In {\em USENIX FAST}, pages 341--354, 2021.

\bibitem{sethumurugan2021designing}
Subhash Sethumurugan, Jieming Yin, and John Sartori.
\newblock Designing a cost-effective cache replacement policy using machine
  learning.
\newblock In {\em IEEE HPCA}, pages 291--303, 2021.

\bibitem{song2020learning}
Zhenyu Song, Daniel~S Berger, Kai Li, and Wyatt Lloyd.
\newblock Learning relaxed belady for content distribution network caching.
\newblock In {\em USENIX NSDI}, pages 529--544, 2020.

\bibitem{VaswaniSPUJGKP17}
Ashish Vaswani, Noam Shazeer, Niki Parmar, Jakob Uszkoreit, Llion Jones,
  Aidan~N. Gomez, Lukasz Kaiser, and Illia Polosukhin.
\newblock Attention is all you need.
\newblock In {\em MIT Press NeurIPS}, pages 5998--6008, 2017.

\bibitem{vietri2018driving}
Giuseppe Vietri, Liana~V Rodriguez, Wendy~A Martinez, Steven Lyons, Jason Liu,
  Raju Rangaswami, Ming Zhao, and Giri Narasimhan.
\newblock Driving cache replacement with ml-based lecar.
\newblock In {\em USENIX HotStorage}, 2018.

\bibitem{vitter1985random}
Jeffrey~S Vitter.
\newblock Random sampling with a reservoir.
\newblock {\em ACM TOMS}, 11(1):37--57, 1985.

\bibitem{wooster1997proxy}
Roland~P Wooster and Marc Abrams.
\newblock Proxy caching that estimates page load delays.
\newblock {\em Elsevier Computer Networks and ISDN Systems}, 29(8-13):977--986,
  1997.

\bibitem{yan2021learning}
Gang Yan, Jian Li, and Don Towsley.
\newblock Learning from optimal caching for content delivery.
\newblock In {\em ACM CoNEXT}, pages 344--358, 2021.

\bibitem{zhong2021lirs2}
Chen Zhong, Xingsheng Zhao, and Song Jiang.
\newblock Lirs2: an improved lirs replacement algorithm.
\newblock In {\em ACM SYSTOR}, 2021.

\bibitem{zhou2018demystifying}
Ke~Zhou, Si~Sun, Hua Wang, Ping Huang, Xubin He, Rui Lan, Wenyan Li, Wenjie
  Liu, and Tianming Yang.
\newblock Demystifying cache policies for photo stores at scale: A tencent case
  study.
\newblock In {\em ACM ICS}, pages 284--294, 2018.

\bibitem{zhu2017workloadcompactor}
Timothy Zhu, Michael~A Kozuch, and Mor Harchol-Balter.
\newblock Workloadcompactor: Reducing datacenter cost while providing tail
  latency slo guarantees.
\newblock In {\em ACM SoCC}, pages 598--610, 2017.

\end{thebibliography}

\end{document}